\documentclass[a4paper,11pt]{article}
\usepackage{jheppub} 
\usepackage{comment}
\usepackage{lineno}
\usepackage[dvipsnames]{xcolor}
\usepackage[utf8]{inputenc}  
\usepackage{pdflscape}
\usepackage{fix-cm}  
\usepackage{tikz}    
\usepackage{float}
\usepackage{longtable}
\usepackage{array}
\usetikzlibrary{arrows}
\usepackage{tikz-cd} 
\usepackage{tkz-euclide}
\usepackage{circuitikz}
\usepgfmodule{shapes}
\usetikzlibrary{decorations.pathmorphing}
\usetikzlibrary{backgrounds, arrows,calc,shapes,decorations.pathreplacing, automata,positioning}
\usepackage{cancel} 
\usepackage{graphicx}
\usepackage{mathdots}
\usepackage{slashed}
\usetikzlibrary{decorations.markings,backgrounds}
\usetikzlibrary{matrix}
\usetikzlibrary{shapes.misc}
\usepackage[most]{tcolorbox}
\usepackage{hyperref}
\usepackage{amsmath} 
\usepackage{amssymb}
\usepackage{graphicx}
\usepackage{caption}
\usepackage{subcaption}
\usepackage{mathrsfs}
\usepackage{mathtools}
\usepackage{multirow}

\graphicspath{ {images/} }

\usepackage{ifthen}
\definecolor{CScolor}{rgb}{1.0, 0.0, 0.0} 
\definecolor{BFcolor} {RGB} {0, 158, 96} 
\definecolor{FIcolor}{RGB}{250, 95, 85} 

\definecolor{bcolor}{RGB}{4, 55, 242} 
\definecolor{fcolor}{RGB}{255, 36, 0} 
\definecolor{labelcolor}{HTML}{808080} 

\def\with{\text{ with }}
\def\mH{m_H}
\def\Nt{\widetilde{N}}

\def\mon{\mathfrak{M}}

\newcommand{\zz}{\mathbb{Z}}

\newcommand{\NN}{\mathcal{N}}
\def\dualto{
	\qquad\leftrightarrow\qquad
}

\def\identityN[#1,#2]{
	_{#1}\mathbb{I}_{#2}^{[N]}
}

\def\identityOne[#1,#2]{
	_{#1}\mathbb{I}_{#2}^{[1^N]}
}

\def\DeltaN[#1]{
	\Delta^{[N]}\left( #1 \right)
}

\def\DeltaOne[#1]{
	\Delta^{[1^N]}\left( #1 \right)
}

\newcommand{\tikznode}[2]{\relax
\ifmmode%
  \tikz[remember picture,baseline=(#1.base),inner sep=0pt]\node(#1){$#2$};
\else
  \tikz[remember picture,baseline=(#1.base),inner sep=0pt]\node(#1){#2};%
\fi}

\tikzstyle{BFline}=[dashed,black,draw]
\tikzstyle{farrowstyle}=[fcolor,thick,draw]
\tikzstyle{barrowstyle}=[bcolor,thick,draw]

\tikzstyle{flavor}=[rectangle,draw=black,thick,inner sep = 0pt, minimum size = 6mm]
\tikzstyle{manifest}=[rectangle,draw=blue!50,thick,inner sep = 0pt, minimum size = 6mm]
\tikzstyle{gauge}=[circle,draw=black,thick,inner sep = 0pt, minimum size = 6mm] 
\tikzstyle{gauge2}=[circle,draw=black!50,thick,inner sep = 0pt, minimum size = 4.5mm] 
\tikzstyle{gauge3}=[rounded rectangle, draw=black!100, thick, minimum size=5mm] 

\tikzstyle{gaugefill}=[circle,draw=black, fill = black, thick,inner sep = 0pt, minimum size = 1.5mm]
\tikzstyle{flavorfill}=[rectangle,draw=black, fill = black,,thick,inner sep = 0pt, minimum size = 1.5mm]

\tikzset{->-/.style={decoration={
  markings,
  mark=at position .5 with {\arrow{>}}},postaction={decorate}}}
\tikzset{-<-/.style={decoration={
  markings,
  mark=at position .5 with {\arrow{<}}},postaction={decorate}}}
\tikzstyle{BFline}=[dashed]
  
\def\nodeCS(#1,#2)(#3,#4,#5){ 	
	\node at (#1,#2) (#3) [gauge,black] {#4};
	\draw[CScolor] (#3) ++(6pt,-7pt) node[anchor=west] {\tiny#5};
}

\def\flavorCS(#1,#2)(#3,#4,#5){ 	
	\node at (#1,#2) (#3) [flavor,black] {#4};
	\draw[CScolor] (#3) ++(6pt,-8pt) node[anchor=west] {\tiny#5};
}

\def\arrowBF(#1,#2)(#3){ 
	\begin{scope}[every node/.style={auto, outer sep=-1pt}]
	\path (#1) edge[->-] node[BFcolor,midway] {\tiny#3} (#2);
	\end{scope}
}

\def\arrowBFlr(#1,#2)(#3,#4){ 
	\begin{scope}[every node/.style={auto=#4, outer sep=-1pt}]
	\path (#1) edge[->-] node[BFcolor,midway] {\tiny#3} (#2);
	\end{scope}
}

\def\dottedBF(#1,#2)(#3){ 
	\begin{scope}[every node/.style={auto, outer sep=-1pt}]
	\path (#1) edge[BFline] node[BFcolor,midway] {\tiny#3} (#2);
	\end{scope}
}

\def\farrow(#1,#2,#3){
	\begin{scope}[every node/.style={auto=#3, outer sep=-1pt}]
	\path (#1) edge[->-,fcolor] node[fcolor,midway] {\tiny$\psi$} (#2);
	\end{scope}
}

\def\barrow(#1,#2,#3){
	\begin{scope}[every node/.style={auto=#3, outer sep=-1pt}]
	\path (#1) edge[->-,bcolor] node[bcolor,midway] {\tiny$\phi$} (#2);
	\end{scope}
}

\def\barrowBF(#1,#2,#3,#4){
	\begin{scope}[every node/.style={auto=#3, outer sep=-1pt}]
	\path (#1) edge[->-,bcolor] node[bcolor,midway] {\tiny$\phi$}  node[BFcolor,midway,swap] {\tiny#4} (#2);
	\end{scope}
}

\def\farrowBF(#1,#2,#3,#4){
	\begin{scope}[every node/.style={auto=#3, outer sep=-1pt}]
	\path (#1) edge[->-,fcolor] node[fcolor,midway] {\tiny$\psi$}  node[BFcolor,midway,swap] {\tiny#4} (#2);
	\end{scope}
}

\def\fQED(#1,#2){ 
\begin{tikzpicture}{baseline=(current bounding box).center}
	\node at (0,0) (g) [gauge,black] {$1$};
	\draw[CScolor] (g.south east) ++(-3pt,3pt) node[anchor=north west] {\tiny$#2$};
	\node at (1.5,0) (f) [flavor,black] {$#1$};
	\farrow(g,f,left)
\end{tikzpicture}
}

\def\sQED(#1,#2){ 
\begin{tikzpicture}{baseline=(current bounding box).center}
	\node at (0,0) (g) [gauge,black] {$1$};
	\draw[CScolor] (g.south east) ++(-3pt,3pt) node[anchor=north west] {\tiny$#2$};
	\node at (1.5,0) (f) [flavor,black] {$#1$};
	\barrow(g,f,left)
\end{tikzpicture}
}
\def\bQED(#1,#2){ \sQED(#1,#2) }

\tikzset{cross/.style={cross out, draw=black, minimum size=5*(#1-\pgflinewidth), inner sep=0pt, outer sep=0pt},
cross/.default={2pt}}

\tikzset{snake it/.style={decorate, decoration=snake}}

\tikzset{mid arrow/.style={postaction={decorate,decoration={
        markings,
        mark = at position .55 with {\arrow[#1]{Straight Barb[width=5pt]}}
      }}}}
      
\tikzset{mid arrowsm/.style={postaction={decorate,decoration={
        markings,
        mark = at position .55 with {\arrow[#1]{Straight Barb[width=3pt]}}
      }}}}
      
\tikzset{middx arrowsm/.style={postaction={decorate,decoration={
        markings,
        mark = at position .7 with {\arrow[#1]{Straight Barb[width=3pt]}}
      }}}}
      
\tikzset{midsx arrowsm/.style={postaction={decorate,decoration={
        markings,
        mark = at position .4 with {\arrow[#1]{Straight Barb[width=3pt]}}
      }}}}

\def\KFplane(#1)(#2,#3){ 
\begin{scope}[scale=#2]
\ifthenelse{\equal{#1}{1}}{
    \fill[fill=#3] (4,0)--(6,0)--(6,2)--cycle;
}{
\ifthenelse{\equal{#1}{1t}}{
    \fill[fill=#3] (4,0)--(2,0)--(2,2)--cycle;
}{
\ifthenelse{\equal{#1}{2old}}{
    \fill[fill=#3] (4,0)--(6,2)--(6,6)--(4,4)--cycle;
}{
\ifthenelse{\equal{#1}{2t}}{
    \fill[fill=#3] (4,0)--(4,4)--(2,2)--cycle;
}{
\ifthenelse{\equal{#1}{3old}}{
    \fill[fill=#3] (4,4)--(6,6)--(2,6)--cycle;
}{
\ifthenelse{\equal{#1}{3t}}{
    \fill[fill=#3] (0,4)--(2,2)--(4,4)--(2,6)--(0,6)--cycle;
}{
\ifthenelse{\equal{#1}{2}}{
    \fill[fill=#3] (4,0)--(6,2)--(6,6)--(4,4)--(2,2)--cycle;
}{
\ifthenelse{\equal{#1}{34}}{
    \fill[fill=#3] (4,4)--(6,6)--(2,6)--cycle;
     \fill[fill=#3] (0,4)--(2,2)--(4,4)--(2,6)--(0,6)--cycle;
}
\ifthenelse{\equal{#1}{3}}{
    \fill[fill=#3] (2,2)--(6,6)--(2,6)--cycle;
}{
\ifthenelse{\equal{#1}{4}}{
    \fill[fill=#3] (2,2)--(2,6)--(0,6)--(0,4)--cycle;
}{
}}}}}}}}}

\draw[->,thick] (0,0) -- (6,0) node[right] {$\scriptstyle F$};
\draw[->,thick] (0,0) -- (0,6) node[above] {$\scriptstyle -2k$};

\node[left] at (0,4) {$\scriptstyle 2N$};
\node[below] at (2,0) {$\scriptstyle N$};
\node[below] at (4,0) {$\scriptstyle 2N$};

\draw[thick] (2,0) -- (2,2);
\draw[thick] (4,0) -- (0,4);
\draw[thick] (2,2)--(2,6);
\draw[thick] (2,2)--(6,6);
\draw[thick] (4,0)--(6,2);

\ifthenelse{\equal{#1}{N4}}{
    \draw[#3,line width=3pt] (4,0)--(6,2);
}{
\ifthenelse{\equal{#1}{k0}}{
    \draw[#3,line width=3pt] (4,0)--(6,0);
}{
\ifthenelse{\equal{#1}{mchi}}{
    \draw[#3,line width=3pt] (2,2) -- (6,6);
}{
\ifthenelse{\equal{#1}{F0}}{
    \draw[#3,line width=3pt] (0,4)--(0,6);
}{
\ifthenelse{\equal{#1}{NOSUSYbdyLeft}}{
    \draw[#3,line width=3pt] (2,0)--(2,2);
}{
\ifthenelse{\equal{#1}{NOSUSYbdyTop}}{
    \draw[#3,line width=3pt] (2,2)--(0,4);
}{
\ifthenelse{\equal{#1}{cusp}}{
    \draw[#3,fill=#3] (2,2) circle (.2cm);
}{
\ifthenelse{\equal{#1}{bdy1t2}}{
    \draw[#3,line width=3pt] (2,2)--(4,0);
}{
}}}}}}}}
\end{scope}
}

\def\KFplaneUthree#1#2#3{ 
\begin{scope}[scale=#2]
\foreach \x in {0,...,8}
{
    \foreach \y in {0,...,8}
    {
        \pgfmathparse{int(\x+\y)}
        \ifodd\pgfmathresult
        \else
            \pgfmathparse{5-\x}
            \ifnum\x>2
                \draw[gray!70,fill] (\x*2/3,\y*2/3) circle (.05cm);
            \else
                \ifnum\y>\pgfmathresult
                    \draw[gray!70,fill] (\x*2/3,\y*2/3) circle (.05cm);
                \fi
            \fi
        \fi
    }
}
\draw[->,thick] (0,0) -- (7,0) node[right] {$\scriptstyle F$};
\draw[->,thick] (0,0) -- (0,6) node[above] {$\scriptstyle -2k$};
\node[left] at (0,4) {$\scriptstyle 6$};
\node[below] at (2,0) {$\scriptstyle 3$};
\node[below] at (4,0) {$\scriptstyle 6$};
\draw[thick] (2,0) -- (2,2);
\draw[thick] (4,0) -- (0,4);
\draw[thick] (2,2)--(2,6);
\draw[thick] (2,2)--(6,6);
\draw[thick] (4,0)--(6,2);
\foreach \x/\y in {#1}
{
    \draw[#3,fill] (\x*2/3,\y*2/3) circle (.1cm);
}
\end{scope}
}

\newcommand{\SB}[1]{{\textcolor{Red} {[SB: #1]}}}
\newcommand{\RC}[1]{{\textcolor{Purple} {[RC: #1]}}}
\newcommand{\GP}[1]{{\color{blue} {[GP: #1]}}}
\newcommand{\SR}[1]{{\color{orange} {[SR: #1]}}}

\newcommand{\AS}[1]{{\color{Blue} {[\textbf{AS}: #1]}}}

\author[a]{Sergio Benvenuti}
\author[b]{Riccardo Comi}
\author[a,c]{Gabriel Pedde Ungureanu}
\author[a,c]{Simone Rota}
\author[a,c]{Anant Shri}

\affiliation[a]{INFN, Sezione di Trieste, Via Valerio 2, I-34127 Trieste, Italy}
\affiliation[b]{Abdus Salam Centre for Theoretical Physics, Imperial College London, London SW7 2AZ, UK}
\affiliation[c]{SISSA, Via Bonomea 265, I-34136 Trieste, Italy}

\emailAdd{benve79@gmail.com} \emailAdd{rcomi@ic.ac.uk} 
\emailAdd{gpeddeun@sissa.it} \emailAdd{srota@sissa.it}  \emailAdd{ashri@sissa.it}

\title{\boldmath {Universal Planar Abelian Duals for 3d $\mathcal{N}=2$ Unitary CS-SQCD}}

\abstract{We provide an explicit planar Abelian dual for three-dimensional $\mathcal{N}=2$ $U(N)_k$ SQCD with $F$ fundamental chiral multiplets. This construction covers the entire $(N, F, k)$ parameter space (provided supersymmetry is unbroken), offering a unified framework for the infrared physics of these theories. Our results generalize a recently discovered class of chiral-planar dualities, which were previously limited to the locus $F = 2|k| + 2N$, which is a mass deformation of $\mathcal{N}=4$ mirror symmetry plus a restricted set of additional mass deformations. By developing a systematic algorithm to track the flow of the dual theory under generic mass deformations, we establish the planar Abelian quiver not merely as a specific dual description, but as a universal tool for analyzing 3d gauge dynamics.}

\begin{document}
\maketitle
\flushbottom

\newpage


\section{Introduction and Summary}

Infrared dualities are among the most distinctive and extensively studied features of supersymmetric quantum field theories. They relate seemingly different ultraviolet (UV) descriptions that flow to the same interacting fixed point in the infrared (IR) \cite{Seiberg_1995}. 

In three dimensions, infrared dualities exhibit rich structures and provide powerful probes of non-perturbative dynamics. They can be broadly organized into two classes: \emph{mirror} and \emph{Seiberg-like} dualities. Mirror dualities typically relate very different UV Lagrangian descriptions and, crucially, exchange the Higgs and Coulomb branches of the moduli space \cite{Intriligator_1996,Hanany_1997,Aharony:1997bx}. The Higgs branch is parameterized by gauge-invariant combinations of elementary fields (mesonic operators), while the Coulomb branch is described by monopole operators, namely local path-integral insertions carrying non-zero magnetic flux.

Seiberg-like dualities, in contrast, do not exchange these two branches and instead connect theories that are structurally similar. The differences typically lie in the rank of the gauge group and in the presence of additional gauge-singlet fields, while the overall matter content and interactions are largely preserved. The prototypical example is that of \emph{Aharony duality} \cite{Aharony:1997gp}, which has inspired numerous generalizations (see \cite{Closset_2023a} and references therein for a comprehensive overview), covering the full space of $\mathcal{N}=2$ (four supercharges) $U(N)_k$ SQCD theories with fundamental and anti-fundamental chiral multiplets.

On the other hand, mirror dualities are well understood primarily in theories with extended supersymmetry, namely $\mathcal{N}=4$ (eight supercharges). In this case, the Higgs and Coulomb branches are sharply distinguished, as they are hyper-K\"ahler manifolds that intersect only at the origin of the moduli space. When supersymmetry is reduced to $\mathcal{N}=2$, much of the mathematical structure is lost, although several isolated examples of mirror dualities have been proposed (see, for example, \cite{Aharony:1997bx,Kapustin:1999ha,Tong:2000ky,Benvenuti:2018bav,Giacomelli:2017vgk}).

In \cite{Benvenuti:2024seb,Benvenuti:2025a}, it was proposed that $\mathcal{N}=2$ mirror dualities can be obtained as supersymmetry-breaking deformations of $\mathcal{N}=4$ mirror dualities\footnote{A different proposal for general non-Abelian $\mathcal{N}=2$ mirror dualities was put forward in \cite{Benvenuti:2023qtv}, with a close relation to Type IIB brane constructions. These dualities are qualitatively different from those of \cite{Benvenuti:2024seb,Benvenuti:2025a}, which provide the starting point of this work.}. A distinctive feature of these dualities is that they relate non-Abelian gauge theories with Chern--Simons interactions to purely Abelian theories. In these constructions, the gauge and matter content of the dual theory are entirely encoded in a planar quiver diagram\footnote{The concept of planar quivers originates in the study of the $\text{AdS}_5/\text{CFT}_4$ correspondence with minimal supersymmetry \cite{Hanany:2005ve, Franco:2005sm, Benvenuti:2005cz, Benvenuti:2005ja, Franco:2005rj, Butti:2005sw}.}.

These dualities exhibit the characteristic mapping of monopole operators to mesonic operators, together with an exchange between flavor symmetries and enhanced topological symmetries. This makes them a promising framework for extending mirror symmetry to $\mathcal{N}=2$ non-Abelian gauge theories. At the same time, the Abelianization of non-Abelian dynamics is itself an interesting feature that merits further investigation. In \cite{Benvenuti:2024seb,Benvenuti:2025a}, the planar Abelian dual of $\mathcal{N}=2$ $U(N)_k$ SQCD with $F$ fundamental chiral multiplets was constructed explicitly along the locus $2|k| = F - 2N$ in the $(k,F)$ plane, and additional examples suggest that this duality extends more generally.

In this work, we develop a toolbox to explore the full $(N,k,F)$ hyperplane systematically. Real mass deformations provide a controlled bridge between the known planar duals on the special locus $2|k|=F-2N$ and generic theories. These deformations shift Chern--Simons levels, integrate out matter fields, and move the theory across the $(N,k,F)$ space along RG-flow trajectories, in a spirit similar to \cite{Benini_2011a}. By tracking these deformations simultaneously in the SQCD and in its planar Abelian mirror, we obtain \textbf{a constructive procedure to determine the planar Abelian dual of} $U(N)_k$ \textbf{SQCD with} $F$ \textbf{fundamental chiral multiplets for arbitrary} $(N,k,F)$.

From a technical perspective, tracking these deformations in the planar dual is achieved through the analysis of the $\mathbf{S}_b^3$ partition function \cite{Kapustin:2009kz,Hama:2010av,Hama_2011} in the presence of large real masses, following the methods of \cite{Aharony:2013dha,Benini_2011a} (see also Section 2.2.2 of \cite{Benvenuti:2025a}).

\subsubsection*{Future Directions}

In this work, we primarily focus on $U(N)_k$ SQCD with $F$ fundamental chiral multiplets. We also present examples of planar Abelian dualities for $U(N)_k$ SQCD with both fundamental and anti-fundamental matter. However, a complete classification of planar Abelian duals in this more general setting is still lacking.

A natural direction for future work is to extend the present analysis toward a systematic understanding of planar Abelian duals with generic chiral matter content.

Another interesting avenue is to clarify the relation between Aharony-like dualities and these more general planar quiver theories.

From a more technical perspective, it would also be valuable to study planar dual descriptions of pure $U(N)_k$ supersymmetric Yang--Mills theories. In particular, it would be interesting to analyze these dualities in the context of generalized global symmetries and their associated topological operators.

One may also consider further generalizations involving more intricate matter content, such as chiral multiplets in tensor representations, as well as theories with different gauge groups, including $USp(2N)$ and $SO(N)$. In the former case, preliminary results have already appeared in \cite{Benvenuti:2025a}, and a more systematic study is currently in progress \cite{Benvenuti:2026b}.

Finally, we anticipate that combining the results presented in this paper with the strategies employed in \cite{Benvenuti:2025qnq} would enable us to extend the scope of planar Abelianization to non-supersymmetric $SU(N)_k$ CS-QCD$_3$ with generic fermionic and bosonic flavors.

\subsubsection*{Organization of the Paper}
\begin{itemize}
    \item In Section \ref{sec: landscape}, we organize the space of theories in the $(k,F)$ parameter plane and identify distinct dynamical regimes of $\mathcal{N}=2$ $U(N)_k$ SQCD with fundamental chiral multiplets. 
    
    Building on this classification, we propose a planar Abelian theory that captures the infrared dynamics for arbitrary $(k,F)$, and study its behavior under real mass deformations. The detailed justification of this proposal is deferred to Section \ref{sec: planar abelian dual of SQCD}. We also conclude the section with illustrative examples of planar Abelian dualities and RG-flow trajectories in theories with both fundamental and anti-fundamental matter.
    
    \item In Section \ref{sec: mass deformations}, we analyze real mass deformations in the electric SQCD theory and map these flows to the corresponding Aharony-like dual descriptions.
    
    \item In Section \ref{sec: planar abelian dual of mass}, we study how these mass deformations are realized in the planar mirror dual theories.

    \item In Section \ref{sec: aharonyproof}, we describe how Aharony duality is related to the planar Abelian duality.
\end{itemize}
\subsubsection*{Notation}
We adopt a standard quiver notation, summarized below. Each round node represents a $U(1)$ gauge symmetry equipped with an $\NN=2$ $U(1)$ vector multiplet, while each arrow denotes a chiral multiplet charged under the two nodes it connects. Outgoing (ingoing) arrows correspond to fundamental (anti-fundamental) representations.

We use red, green, and orange labels to denote, respectively, CS levels, mixed CS levels, and FI parameters. Near flavor nodes we indicate their associated real mass parameter.
A dashed line between two nodes indicates the sole presence of a pure mixed CS interaction, whose level is specified by the adjacent green label. We summarize this notation, together with the contribution of each building block to the $\mathbf{S}_b^3$ partition function \cite{Kapustin:2009kz,Hama:2010av,Hama_2011}, as follows:
\begin{equation} 
\renewcommand{\arraystretch}{1.1}
\label{eq:quiv:notation_full}
    \begin{array}{|c|c|c|}
        \hline
        \text{\textbf{Quiver notation}} & \text{\textbf{Field theory}} &\mathbf{S}_b^3 \,\, \text{p.f.}
        \\ \hline
        \begin{tikzpicture}[baseline=(current bounding box).center]
        \node[gauge,black] at (0,0) (g) {$1$}; 
        \draw[black] (g) ++(-0.6,0) node {$u$};
        \draw[FIcolor] (g.north) ++(0,5pt) node {$\eta$};
        \draw[CScolor] (g.south east) ++(6pt,-2pt) node {\scriptsize$k$};
        \end{tikzpicture}
        &  \begin{array}{c} U(1) \; \NN=2 \text{ vector multiplet with CB } \\ \text{parameter $u$, FI parameter $\eta$ and CS level $k$} \end{array}
        & \int du \, e^{2\pi i \eta u - i \pi k u^2}
        \\ \hline
        \begin{tikzpicture}[baseline=(current bounding box).center]
        \node[flavor,black] at (0,0) (g) {$1$}; 
        \draw[black] (g) ++(-.6,0) node {$X$};
        \draw (0,-.4) {};
        \end{tikzpicture}
        &  \begin{array}{c} \text{$U(1)$ flavor node with real mass parameter $X$}\end{array}
        & -
        \\ \hline
        \begin{tikzpicture}[baseline=(current bounding box).center]
        \node[gauge,black] at (0,0) (g1) {$1$}; 
        \node[gauge,black] at (1.5,0) (g2) {$1$}; 
        \draw[->-,black] (g1) -- node[midway,below,black] {$r$} (g2);
        \draw[black] (g1) ++(-0.5,0) node {$u$};
        \draw[black] (g2) ++(0.5,0) node {$v$};
        \end{tikzpicture}
        &  \begin{array}{c} \text{Chiral multiplet with charge $(+1,-1)$} \\ \text{under $U(1)_{u} \times U(1)_{v}$ and R-charge $r$}  \end{array}
        & \Gamma (u - v + r \omega)
        \\ \hline
        \begin{tikzpicture}[baseline=(current bounding box).center]
        \node[gauge,black] at (0,0) (g1) {$1$}; 
        \node[gauge,black] at (1.5,0) (g2) {$1$}; 
        \draw[BFline,black] (g1) -- node[midway,above,BFcolor] {\scriptsize$k_{uv}$} (g2);
        \draw[black] (g1) ++(-0.5,0) node {$u$};
        \draw[black] (g2) ++(0.5,0) node {$v$};
        \end{tikzpicture}
        &  \begin{array}{c}
        \text{Mixed CS term between $U(1)_u$ and $U(1)_v$} \\ \text{at level $k_{uv}$ }
        \end{array}
        & e^{-\pi i k_{uv} uv}
        \\ \hline
    \end{array}
\end{equation}
The R-charges of chiral multiplets are assigned according the superpotential, and are likewise omitted unless needed.

In general, we adopt the convention that the CS action of a $U(N)_{(k,k+\ell N)}$ gauge group is
\begin{equation}
    -i\frac{k}{4\pi}\int tr(A\wedge dA)-i\frac{\ell}{4\pi}\int tr(A)\wedge tr(dA) + \text{SUSY completion},
\end{equation}
where $k$ is the level of the CS interaction for the $SU(N)$ factor, while $\ell$ is the level for the Abelian interaction. Whenever $\ell = 0$ we simply write $U(N)_k$.
One may also introduce mixed CS interactions at level $k_{ij}$ between two Abelian gauge fields as
\begin{equation}
    -i\frac{k_{ij}}{4\pi} \int A_i\wedge dA_j + \text{SUSY completion}.
\end{equation}
An FI term may, therefore, be interpreted as a BF coupling at level $-2$ between the gauge group and the $U(1)$ topological symmetry.

The theories considered in this work exhibit a common structure, described below \eqref{eq: Dual_k0}, allowing the superpotential and the CS interactions to be reconstructed directly from simple graphical rules based on the orientation of the arrows.
Moreover, since all theories discussed here are Abelian, we simplify the notation by leaving the gauge group implicit, with each round/square node understood to represent a gauge/flavor $U(1)$ factor.

The compact notation used for Abelian planar quivers, together with their contributions to the partition function, is summarized below in \eqref{eq:quiv:notation_compact}.

\begin{equation}    \label{eq:quiv:notation_compact}
    \begin{array}{|c|c|c|}
        \hline
        \text{\textbf{Quiver notation}} & \text{\textbf{Field theory}} &\mathbf{S}_b^3 \,\, \text{p.f.}
        \\ 
        \hline
        \begin{tikzpicture}[baseline=(current bounding box).center]
        \node[gaugefill,black] at (0,0) (g1) {}; 
        \draw[black] (g1) ++(-0.5,0) node {$u$};
        \end{tikzpicture}
        &  \begin{array}{c} U(1) \; \NN=2 \text{ vector multiplet}\end{array}
        & \int du
        \\ 
        \hline
        \begin{tikzpicture}[baseline=(current bounding box).center]
        \node[flavorfill,black] at (0,0) (g1) {}; 
        \draw[black] (g1) ++(-0.5,0) node {$X$};
        \end{tikzpicture}
        &  \begin{array}{c} \text{$U(1)$ Flavor node}\end{array}
        & -
        \\ 
        \hline
        \begin{tikzpicture}[baseline=(current bounding box).center,scale=0.7]
        \node[gaugefill,black] at (0,1) (g1) {}; 
        \node[gaugefill,black] at (1,0) (g2) {}; 
        \draw[->-,black] (g1) --  (g2);
        \draw[black] (g1) ++(-0.5,0) node {$u$};
        \draw[black] (g2) ++(0.5,0) node {$v$};
        \end{tikzpicture},
        \begin{tikzpicture}[baseline=(current bounding box).center,scale=0.7]
        \node[gaugefill,black] at (1,1) (g1) {}; 
        \node[gaugefill,black] at (0,0) (g2) {}; 
        \draw[->-,black] (g1) --  (g2);
        \draw[black] (g1) ++(-0.5,0) node {$u$};
        \draw[black] (g2) ++(0.5,0) node {$v$};
        \end{tikzpicture}
        &  \begin{array}{c} \text{bifundamental chiral with R-charge $r$} \\ \text{and mixed CS level $+1$}  \end{array}
        & \Gamma (u - v + r \omega) e^{-\pi i uv}
        \\ 
        \hline
        \begin{tikzpicture}[baseline=(current bounding box).center,scale=0.7]
        \node[gaugefill,black] at (0,1) (g1) {}; 
        \node[gaugefill,black] at (0,0) (g2) {}; 
        \draw[->-,black] (g1) --  (g2);
        \draw[black] (g1) ++(-0.5,0) node {$u$};
        \draw[black] (g2) ++(0.5,0) node {$v$};
        \end{tikzpicture}
        &  \begin{array}{c} \text{bifundamental chiral with R-charge $2r$} \\ \text{and mixed CS level $+1$}  \end{array}
        & \Gamma (u - v + 2r \omega) e^{-\pi i uv}
        \\ 
        \hline
        \begin{tikzpicture}[baseline=(current bounding box).center,scale=0.7]
        \node[gaugefill,black] at (0,1) (g1) {}; 
        \node[gaugefill,black] at (1,0) (g2) {}; 
        \draw[-<-,black] (g1) --  (g2);
        \draw[black] (g1) ++(-0.5,0) node {$u$};
        \draw[black] (g2) ++(0.5,0) node {$v$};
        \end{tikzpicture},
        \begin{tikzpicture}[baseline=(current bounding box).center,scale=0.7]
        \node[gaugefill,black] at (1,1) (g1) {}; 
        \node[gaugefill,black] at (0,0) (g2) {}; 
        \draw[-<-,black] (g1) --  (g2);
        \draw[black] (g1) ++(-0.5,0) node {$u$};
        \draw[black] (g2) ++(0.5,0) node {$v$};
        \end{tikzpicture}
        &  \begin{array}{c} \text{bifundamental chiral with R-charge $1-r$} \\ \text{and mixed CS level $-1$}  \end{array}
        & \Gamma (v - u + (1-r) \omega) e^{\pi i uv}
        \\ 
        \hline
        \begin{tikzpicture}[baseline=(current bounding box).center,scale=0.7]
        \node[gaugefill,black] at (0,1) (g1) {}; 
        \node[gaugefill,black] at (0,0) (g2) {}; 
        \draw[-<-,black] (g1) --  (g2);
        \draw[black] (g1) ++(-0.5,0) node {$u$};
        \draw[black] (g2) ++(0.5,0) node {$v$};
        \end{tikzpicture}
        &  \begin{array}{c} \text{bifundamental chiral with R-charge $2-2r$} \\ \text{and mixed CS level $-1$}  \end{array}
        & \Gamma (v - u + 2(1-r) \omega) e^{\pi i uv}
        \\ 
        \hline
        \begin{tikzpicture}[baseline=(current bounding box).center,scale=0.7]
        \node[gaugefill,black] at (0,0) (g1) {}; 
        \node[gaugefill,black] at (1.5,0) (g2) {}; 
        \draw[BFline,black] (g1) --  (g2);
        \draw[black] (g1) ++(-0.5,0) node {$u$};
        \draw[black] (g2) ++(0.5,0) node {$v$};
        
        \end{tikzpicture}
        &  \text{Mixed CS term at level $-2$}
        & e^{2\pi i uv}
        \\ \hline
    \end{array}
\end{equation}
For this shorthand to be effective, we distinguish between diagonal and vertical chirals, as well as their orientation (upward or downward). This graphical distinction encodes differences in trial R-charges and mixed CS couplings.

In almost all examples considered, the CS level of each gauge node satisfies
\begin{equation}
    k_i = -\frac{1}{2} \sum_{j \neq i} k_{ij}\,,
\end{equation}
where $k_i$ is the CS level of the $U(1)_i$ node and $k_{ij}$ is the mixed CS level between $U(1)_i$ and $U(1)_j$. The sum runs over all nodes in the quiver, including both gauge and flavor nodes. This relation allows one to reconstruct the CS levels entirely from the mixed CS couplings, and hence from the orientation of the chiral multiplets attached to each node. The superpotential can be also reconstructed from the quiver notation as explained below \eqref{eq: Dual_k0}.

\section{The Landscape of \texorpdfstring{$\mathcal{N}=2$ CS-SQCD}{N=2 SQCD3} and its Planar Mirror Duals}\label{sec: landscape}

In this section we describe the planar Abelian dual of the $U(N)_k$ SQCD with $F$ flavors for any choice of parameters $(N,k,F)$. Toward the end of the section we also present examples for the SQCD with both fundamental and anti-fundamental chirals.

\subsection{The Taxonomy of \texorpdfstring{$\mathcal{N}=2$ CS-SQCD}{N=2 SQCD3} with Fundamental Matter}
\label{subsec: zoology}

Three-dimensional $\mathcal{N}=2$ $U(N)_k$ SQCD theories with $F$ fundamental chiral multiplets can be organized in a two-dimensional parameter space spanned by $(k, F)$. Each point in this plane corresponds to a distinct theory. In this paper, we focus on SQCD with fundamentals, with CS level $ k\leq0$. The case $k>0$ is analogous. 

The results given in this section can be extended, in principle, to the most general case of SQCD with both fundamental and antifundamental fields by building on the results of \cite{Benvenuti:2025a}. 
The techniques needed for this generalization are analogous to the ones developed in this paper, but the taxonomy is more involved, and we defer this analysis to future work.

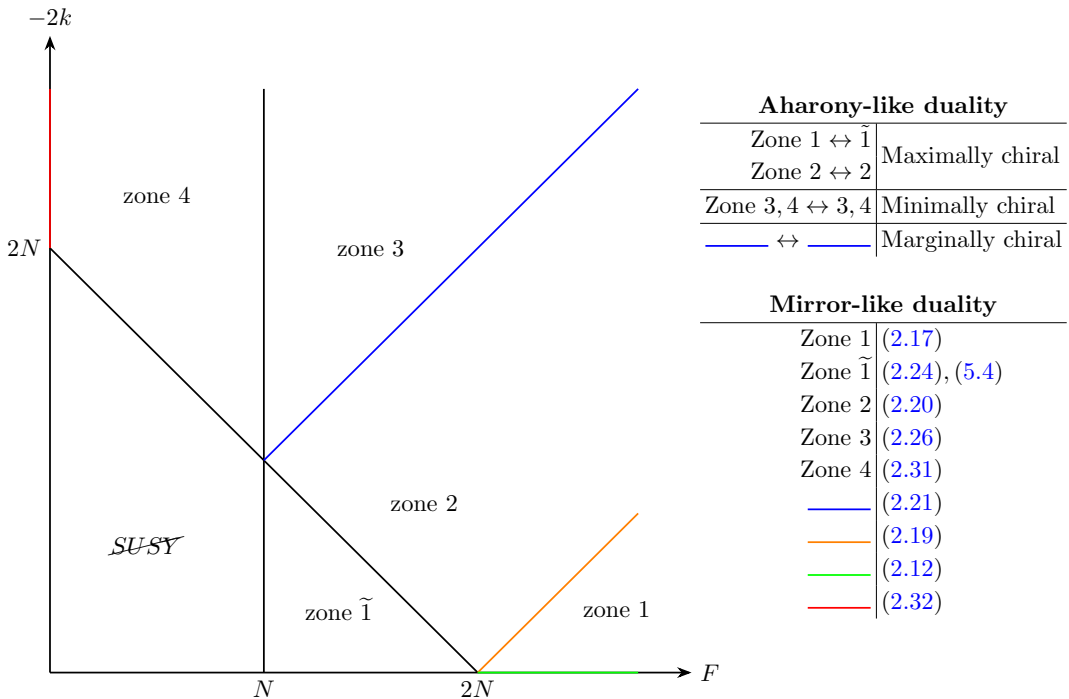
\begin{figure}[H]
    \centering
\resizebox{.95\hsize}{!}{ 
\begin{tikzpicture}[scale=1.7,>=Stealth]

\draw[->,thick] (0,0) -- (6,0) node[right] {$F$};
\draw[->,thick] (0,0) -- (0,6) node[above] {$-2k$};

\node[left] at (0,4) {$2N$};
\node[below] at (2,0) {$N$};
\node[below] at (4,0) {$2N$};

\draw[thick] (2,0) -- (2,2);
\draw[thick] (4,0) -- (0,4);
\draw[thick] (2,2) -- (2,5.5);

\draw[thick,blue] (2,2) -- (5.5,5.5);  
\draw[thick,orange] (4,0) -- (5.5,1.5); 
\draw[thick,green] (4,0) -- (5.5,0);
\draw[thick,red] (0,4) -- (0,5.5);

\node at (2.7,0.6) { zone $\widetilde{1}$};
\node at (5.3,0.6) {zone $1$};
\node at (3.5,1.6) { zone $2$};
\node at (3.,4) { zone $3$};
\node at (1.,4.5) { zone $4$};

\node[rotate=0] at (0.9,1.2) { $\cancel{SUSY}$};


\begin{scope}[shift={(6,3)}]
    
    
    


    \node[anchor=west] at (0,0) {
    $
    \begin{array}{r|l}
    \multicolumn{2}{c}{\textbf{Aharony-like duality}}
        \\
        \hline
        \text{Zone }1 \leftrightarrow \tilde 1 & \multirow{2}{*}{\text{Maximally chiral}}\\
        \text{Zone }2 \leftrightarrow  2 & \\ \hline
        \text{Zone }3,4 \leftrightarrow  3,4 & \text{Minimally chiral} \\ \hline
        \tikz{\path[draw,thick,blue](0,0)--(1,0);} \leftrightarrow \tikz{\path[draw,thick,blue](0,0)--(1,0);} & \text{Marginally chiral}
        \\ 
        \multicolumn{2}{c}{}\\
        \multicolumn{2}{c}{\textbf{Mirror-like duality}}\\
        \hline
       \text{Zone 1}  & \eqref{eq: Dual_zone1_schematic}  \\
       \text{Zone $\widetilde{1}$} & \eqref{eq: Zone1t_dual_raw}, \eqref{eq:dual_zone1t_alternative} \\
       \text{Zone 2}  & \eqref{eq: Zone22t_dual_raw}  \\
       \text{Zone 3}  & \eqref{eq: Zone33t_a_dual_raw}   \\
       \text{Zone 4}  & \eqref{eq: Zone33t_b_dual_raw}   \\
       \tikz{\path[draw,thick,blue](0,0)--(1,0);} & \eqref{eq: margchir_dual_raw} \\
       \tikz{\path[draw,thick,orange](0,0)--(1,0);} & \eqref{eq: Dual_N4line_schematic_nosquares} \\
       \tikz{\path[draw,thick,green](0,0)--(1,0);} & \eqref{eq:Dual_k0_schematic} \\
       \tikz{\path[draw,thick,red](0,0)--(1,0);} & \eqref{eq: tqft} \\
    \end{array}
    $
    };

\end{scope}

\end{tikzpicture}
}
    \caption{Phase diagram of $U(N)_k$ SQCD with $F$ fundamental fields in the $(F,\,k)$-plane for fixed $N$. The diagram is partitioned into various zones, within which the planar Abelian duals take qualitatively distinct forms. 
    }
    \label{fig:KF_plane}
\end{figure}

In Figure \ref{fig:KF_plane}, close to the origin, we identify the supersymmetry breaking region defined by:
\begin{equation}
\cancel{SUSY}: \;  2|k| < 2N - F \cap F < N.
\end{equation}
On the boundary of this region, namely for $F=N \,\cap\, 2|k| < N $ and $F < N \,\cap\, 2|k| = 2N-F$, the theory is trivial in the IR, with the exception of the cusp in $F=N$ and $2|k|=N$, where the theory reduces to a single free chiral. Indeed, for $F=0$, the red line in Figure \ref{fig:KF_plane}, the SQCD reduces to a $U(N)_k$ TQFT.

Outside the supersymmetry breaking region and for $F \neq 0$ the theory is interacting. As discussed in \cite{Benini_2011a}, the space of $\NN=2$ SQCD theories can be organized in three broad regimes distinguished by the relative magnitude of $F,\,k $ and $N$.
These three regions are distinguished by a qualitatively different Aharony-like dual\footnote{While the original example of a 3d $\mathcal{N}=2$ duality for $U(N)$ SQCD was presented in \cite{Aharony:1997gp}, generalizations to $U(N)_k$ theories with chiral matter content were subsequently developed by various authors. We refer to \cite{Closset_2023a} for a comprehensive overview and an extensive list of references.} and are denoted as minimally, marginally, and maximally chiral regions:
\begin{equation}
\begin{array}{rl}
\text{minimally chiral: }F < |2k|, & \qquad \Nt= \frac{F}{2} + |k| - N
\\
\text{marginally chiral: }F = |2k|, & \qquad\Nt=\frac{F}{2}  + |k| - N = F - N
\\
\text{maximally chiral: }F > |2k|, & \qquad\Nt=F - N
\end{array}
\end{equation}
where $\Nt$ is the rank of the Aharony-like dual theory. We point out that marginally chiral theories are characterized by the presence of a single gauge invariant monopole operator parameterizing a non-compact $\mathbb{C}$ moduli space of vacua. In general, for different values of $(k,F)$ the theory does not possess any chiral gauge invariant operator.
For the purpose of this paper we further divide these regions depending on the qualitative form the planar Abelian dual, as we will discuss in the rest of the section. 


A special role in the diagram in Figure \ref{fig:KF_plane} is played by the line $2|k|=F-2N$, that is highlighted in orange. The SQCD on this line can be derived starting from the $\mathcal{N}=4$ $U(N)$ SQCD with $F$ flavors and performing a SUSY breaking deformation that preserves the $SU(F)$ flavor symmetry\footnote{To be more precise, the $\NN=4$ SUSY breaking deformations results in an $\NN=2$ $U(N)_{2N-F,N-F}$ SQCD theory with $F$ fundamental chiral multiplets. By performing an $SL(2,\mathbb{Z})$ transformation \cite{witten2003sl2zactionthreedimensionalconformal}, implemented via gauging and ungauging the topological symmetry, one can shift the $U(1)$ Chern--Simons level so that it matches that of the $SU(N)$ factor.}. This deformation and the planar Abelian dual of these theories was studied in \cite{Benvenuti:2024seb, Benvenuti:2025a} (see Figure 6 of \cite{Benvenuti:2025a}), which is the conceptual starting point for this analysis.

Starting from this proposal, we can explore the entire $(k,F)$ plane by performing suitable real mass deformations, thereby arriving at a fully general construction of the planar Abelian dual of $U(N)_k$ SQCD with $F$ flavors. This analysis naturally partitions the parameter space into five distinct regions, shown in Figure \ref{fig:KF_plane}, in which the planar dual exhibits qualitatively different features.

For clarity, we adopt a slightly different presentation strategy. We first introduce the planar Abelian dual of $U(N)_0$ SQCD with $F \geq 2N$ fundamental fields, corresponding to the green line in Figure \ref{fig:KF_plane}. We then take this theory, rather than the $\mathcal{N}=4$ descendant, as the starting point to explore the full $(k,F)$ plane via real mass deformations. In Section \ref{sec: planar abelian dual of mass}, we explain how the $U(N)_0$ proposal itself can be obtained through a sequence of deformations starting from the $\mathcal{N}=4$ parent theory.

We emphasize that all dualities presented in this work follow from a systematic analysis of real mass deformations that break supersymmetry from $\mathcal{N}=4$ to $\mathcal{N}=2$. These flows are studied in detail using, for instance, the $\mathbf{S}^3_b$ partition function to provide exact mathematical identities between the partition functions of the two dual theories. As an additional non-trivial check, we can verify the dualities, for example, by matching the numerical expansion of the superconformal index for small values of $(N,k,F)$.

\subsubsection*{Motion in the \texorpdfstring{$(k,F)-$}{(k,F)-} Plane}
A convenient way to explore the $(k,F)$ parameter space is via real mass deformations of the matter fields. This will be the main tool employed throughout this paper. Here we briefly summarize how such deformations allow one to navigate this plane, while a more detailed discussion is deferred to Section \ref{sec: mass deformations}.

For each fundamental chiral multiplet, there are four basic possibilities:
\begin{enumerate}
    \item \textbf{Non-Higgsing Mass Flows}: Assigning a positive or negative real mass $\pm m$ removes one chiral field from the spectrum and shifts the CS level by $\pm\tfrac{1}{2}$. On the $(k,F)-$plane these correspond to diagonal moves toward the left:
    \begin{equation}
        \begin{tikzpicture}[baseline=(current bounding box).center]
        \node[label=right:{SQCD}] at (0,0) (SQCD) {$\bullet$};
        \node[] at (-1,1) (mm) {$-m$};
        \node[] at (-1,-1) (pm) {$+m$};
        \path[draw,->] (SQCD) -- (mm);
        \path[draw,->] (SQCD) -- (pm);
        \end{tikzpicture}
    \end{equation} 
    which can be stated more succinctly: 
    \begin{equation*}
        (N,F,k)\overset{\pm m}{\longrightarrow} (N,F-1, k\pm\tfrac{1}{2}).
    \end{equation*}

    \item \textbf{Higgsing Mass Flows}: Assigning a positive or negative real mass $\pm m_H$ removes one chiral field from the spectrum, shifts the CS level by $\pm\tfrac{1}{2}$ and Higgses the gauge group to $U(N-1)$. 
    These correspond to the following movement across two different $(k,F)$-planes\footnote{Notice that the mass deformations with Higgsing $\pm \mH$ are only available in certain regions of the $(F,k)$-plane; refer to Section \ref{sec: mass deformations} for more details.}:
    \begin{equation*}
        (N,F,k)\overset{\pm m_H}{\longrightarrow} (N-1,F-1, k\pm\tfrac{1}{2}).
    \end{equation*}
    
    We can interpret this operation as a movement to the right on the $(k,F)$-plane.
    As an example, starting from the pseudo-self-dual point $F=2N$, $k=0$, the deformation $-\mH$ produces a theory with $F' > 2N'$, which sits on the $\NN=4$-descendant line to the right of the pseudo-self-dual point $F' = 2N'$. The deformation $+\mH$ reduces a theory sitting on the  $\NN=4$-descendant line to a theory which lies in Zone 1:
    \begin{equation}
        \includegraphics[width=.7\linewidth]{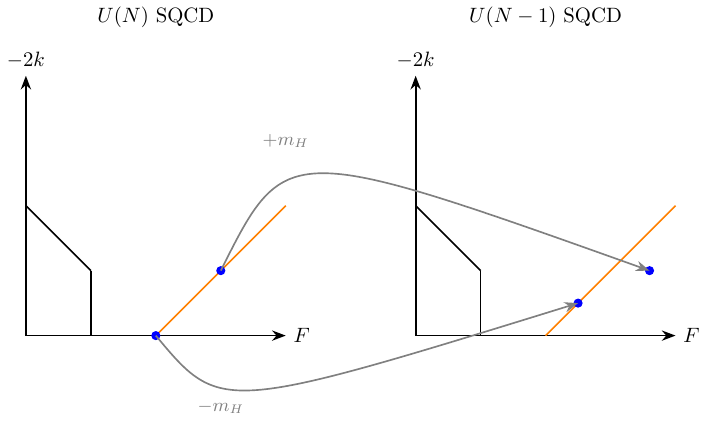}
    \end{equation}
\end{enumerate}

Therefore, with a slight abuse of notation, we depict the possible deformations in the $(k,F)$-plane: schematically as:
\begin{equation}    \label{eq:realmass_arrows_schematic}
\begin{tikzpicture}[baseline=(current bounding box).center]
\node[label=right:{SQCD}] at (0,0) (SQCD) {$\bullet$};
\node[] at (-1,1) (mm) {$-m$};
\node[] at (-1,-1) (pm) {$+m$};
\node[] at (1,1) (mmh) {$-\mH$};
\node[] at (1,-1) (pmh) {$+\mH$};
\path[draw,->] (SQCD) -- (mm);
\path[draw,->] (SQCD) -- (pm);
\path[draw,->] (SQCD) -- (mmh);
\path[draw,->] (SQCD) -- (pmh);
\end{tikzpicture}
\end{equation}

Taken together, these deformations allow us to probe the $(k,F)-$plane completely, starting from a given theory\footnote{A further consequence is that generic Aharony-like dualities for SQCD with only fundamentals can be derived assuming the pseudo-self-duality:
\begin{equation}
U(N)_{0} \with 2N \square \dualto U(N)_{0} \with 2N \overline{\square}
\end{equation}
for any $N$, as we discuss in more details in Appendix \ref{app: mass flow S3b}.}.
We emphasize that the Higgsing mass deformations $\pm\mH$ also reduce the gauge rank by one; consequently, when including such deformations, one must begin from an SQCD theory with gauge group 
$U(N+1)$. By systematically mapping these mass flows to the planar mirror description, we construct the planar Abelian dual of every $\mathcal{N}=2$ $U(N)_k$ SQCD theory with fundamental matter. 

In this section, we probe the $(k,F)$-plane using only $\pm m$ deformations, thereby avoiding Higgsing. Starting from the green line in Figure \ref{fig:KF_plane}, which lies along the $k$-axis, these deformations allow us to explore the entire plane.

To access the region corresponding to $U(N)_0$ with $F \geq 2N$ flavors, we instead employ massive deformations accompanied by Higgsing. In this case, we begin from the $\mathcal{N}=4$ descendant line (the orange line in Figure \ref{fig:KF_plane}), for which the planar Abelian duality was established in \cite{Benvenuti:2025a}.

The strategy employed to follow these deformations in the planar Abelian dual consists of studying the large mass asymptotic behavior of the $\mathbf{S}^3_b$ partition function, following the strategies of \cite{Aharony:2013dha,Benini_2011a} (see also Section 2.2.2 of \cite{Benvenuti:2025a}).


\subsection{A Planar Abelian Dual of \texorpdfstring{$U(N)_k$ SQCD with $F$ $\Box$}{U(N)k SQCD with F}}\label{sec: planar abelian dual of SQCD}

Having organized $\mathcal{N}=2$ $U(N)_k$ SQCD theories in the $(k,F)$–plane, we now turn to the construction of their planar Abelian duals. Our starting point is the case with vanishing CS level, $k=0$, for which the dual description takes a particularly symmetric form. We then show how real mass deformations, corresponding to motion in the $(k,F)$–plane, map naturally to Higgsing and mass flows in the planar quiver, generating duals for arbitrary $k$ and $F$. In this way, the planar Abelian dual provides a unifying framework that interpolates continuously between the $\mathcal{N}=4$–descendant line and more general $\mathcal{N}=2$ CS–SQCD theories.
\subsubsection{A Planar Abelian Dual of \texorpdfstring{$U(N)_0$ SQCD with $F\geq2N$ $\Box$}{U(N)0 SQCD with F}}\label{subsec: planar abelian dual of U(N)0}

In this section, we describe an Abelian theory that is infrared dual to $\mathcal{N}=2$ $U(N)_0$ SQCD with $F \geq 2N$ fundamental chiral multiplets. Notice that the number of chiral fields $F$ must be even due to the quantization of the effective Chern–Simons level. As mentioned above, the validity of this duality is supported by the fact that it can be derived starting from the mirror dual of $\mathcal{N}=4$ SQCD, followed by an appropriate supersymmetry-breaking real mass deformation and subsequent real mass deformations.

The Abelian dual exhibits close similarities to the duals discussed in \cite{Benvenuti:2024seb, Benvenuti:2025a} for $\mathcal{N}=2$ SQCD theories obtained via axial mass deformations of $\mathcal{N}=4$ SQCD. In particular, in Section \ref{sec: aharonyproof}, we show that a sequence of real mass deformations, combined with local Aharony dualities applied to the constructions of \cite{Benvenuti:2024seb, Benvenuti:2025a}, reproduces the duality presented below.

The dual of $U(N)_0$ SQCD with $F \geq 2N$ fundamental chiral multiplets is shown in \eqref{eq: Dual_k0}.
\begin{equation}
\label{eq: Dual_k0}
    \includegraphics[width=\linewidth]{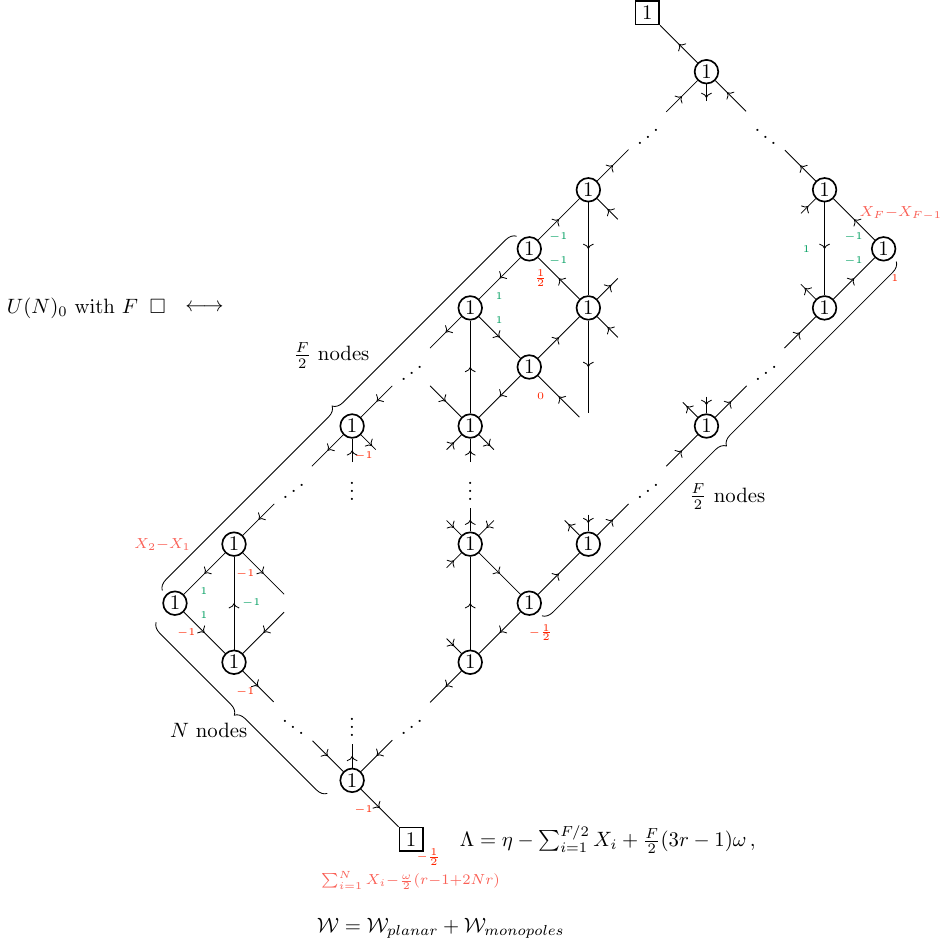}
\end{equation}

Before discussing the main features of the duality, we present in detail the Abelian dual theory in \eqref{eq: Dual_k0}. For clarity, we organize its defining data as follows.

\paragraph{Gauge structure and Chern--Simons couplings.}
The gauge group is $U(1)^{N \times (F - N)}$, with chiral multiplets whose charges are encoded in the quiver, following the notation of \eqref{eq:quiv:notation_full}. Each $U(1)$ factor carries a CS level, indicated by the red labels on the nodes, together with mixed CS interactions represented by green labels on the links.

We adopt the following convention for mixed CS terms. For any pair of nodes connected by a down-pointing arrow 
(\tikz{\draw[->-] (0,.5)--(.5,0);}, \tikz{\draw[->-] (.5,.5)--(0,0);}, \tikz{\draw[->-] (0,.5)--(0,0);}),
the mixed CS level is $+1$, while for upward-pointing arrows 
(\tikz{\draw[-<-] (0,.5)--(.5,0);}, \tikz{\draw[-<-] (.5,.5)--(0,0);}, \tikz{\draw[-<-] (0,.5)--(0,0);}),
it is $-1$. 

Moreover, the CS level $k_i$ of each $U(1)_i$ satisfies
\begin{equation}
    k_i = -\frac{1}{2} \sum_{j \neq i} k_{ij},
\end{equation}
where $k_{ij}$ denotes the mixed CS level between $U(1)_i$ and $U(1)_j$.
The sum runs over all nodes in the quiver, including both gauge and flavor nodes.
These rules allow one to reconstruct all CS interactions directly from the quiver.

\paragraph{Superpotential.}
The superpotential consists of two types of interactions:
\begin{itemize}
    \item $\mathcal{W}_{\text{planar}}$: it contains a cubic (quartic) term for each triangular (square) face of the quiver, with sign $-1$ ($+1$) if the arrows run clockwise (anticlockwise) around the face.
    \item $\mathcal{W}_{\text{monopole}}$: each vertical column of $k$ gauge nodes contributes with $k-1$ superpotential terms that are linear for the monopole with $-1/+1$ GNO fluxes for the top/bottom nodes connected by a vertical arrow. For the central column, where vertical arrows are absent, we have an analogous superpotential, that is linear for the monopole with $-1/+1$ GNO fluxes under the first/second gauge node of the column, also for the second/third nodes and so on.
\end{itemize}

\paragraph{Trial R-symmetry.}
On the SQCD side, we assign trial R-charge $r$ to the fundamental fields. 
On the quiver side, the chiral multiplets have R-charges:
\begin{equation}
    \tikz{\draw[->-] (0,.5)--(.5,0);}, \tikz{\draw[->-] (.5,.5)--(0,0);}: r 
    \quad,\quad 
    \tikz{\draw[->-] (0,.5)--(0,0);}: 2r 
    \quad,\quad 
    \tikz{\draw[-<-] (0,.5)--(.5,0);}, \tikz{\draw[-<-] (.5,.5)--(0,0);}: 1-r 
    \quad,\quad 
    \tikz{\draw[-<-] (0,.5)--(0,0);}: 2-2r \,,
\end{equation}
which are compatible with the planar superpotential.
The mixing parameter $r$ is to be fixed, in principle, via extremization \cite{Jafferis:2010un}. 
In this paper, we do not perform the extremization and keep $r$ generic instead.

\paragraph{Global symmetries.}
The Abelian theory has a $U(1)^F$ UV global symmetry. One factor, denoted $U(1)_\Lambda$, is a flavor symmetry compatible with $\mathcal{W}_{\text{planar}}$ and is not gauged. Without loss of generality, we assign charge $-1$ under $U(1)_\Lambda$ to the bottom-most chiral, with all other fields neutral. This symmetry corresponds to the bottom flavor node of the quiver.

In addition, the theory possesses a $U(1)^{F-1}$ topological symmetry that is preserved by $\mathcal{W}_{\text{monopole}}$. Effectively, each column of nodes carries one such symmetry. 
We encode the mapping in the FIs for the nodes in the $i$-th column which are, from top to bottom:
\begin{itemize}
\item $i=1,\ldots,\tfrac{F}{2}-1$:  $\{  X_{i+1} - X_{i}- \frac{\omega}{2}r,\;
                 X_{i+1} - X_{i}, \;
                 \dots \;, X_{i+1} - X_{i}, \;
                 X_{i+1} - X_{i}+ \frac{\omega}{2}r \}$
\item $i=\tfrac{F}{2}$: $\{ X_{\tfrac{F}{2}+1} - X_{\tfrac{F}{2}}+ \frac{\omega}{2}(r-1),\;\;
                 X_{\tfrac{F}{2}+1} - X_{\tfrac{F}{2}}, \;
                 \dots \;, X_{\tfrac{F}{2}+1} - X_{\tfrac{F}{2}}, \;
                 X_{\tfrac{F}{2}+1} - X_{\tfrac{F}{2}}+ \frac{\omega}{2}r \}$
\item $i=\tfrac{F}{2}+1,\ldots,F-1$: $\{ X_{i+1} - X_{i}+ \frac{\omega}{2}(r-1),\;
                 X_{i+1} - X_{i}, \;
                 \dots \;, X_{i+1} - X_{i}, \;
                 X_{i+1} - X_{i}- \frac{\omega}{2}(r-1) \}$
\end{itemize}
where $X_{i+1} - X_{i}$ parametrizes the corresponding Cartan of $SU(F)$ and the coefficient in $\omega$ parametrized the mixing between the topological and the R-symmetry. 
One of the $X_i$ is redundant, but we keep this overcomplete parametrization for later convenience.
One can check that the FIs are compatible with the monopole superpotential described above.

In the IR, the topological symmetry enhances to $SU(F)_X$, with parameters $X_i$ satisfying $\sum_{i=1}^F X_i = 0$. This enhancement can be seen, for instance, via the superconformal index and is also predicted by the duality as we explain in a moment.

\paragraph{Map of symmetries.}
Across the duality, flavor and topological symmetries are exchanged, as expected in mirror-like dualities. The SQCD side has an $SU(F)_X$ flavor symmetry under which the matter fields transform in the anti-fundamental representation. Its Cartan subgroup, parameterized by real masses $X_i$ with $\sum_i X_i = 0$, maps to linear combinations of the $U(1)^{F-1}$ topological symmetries of the Abelian theory, which enhance in the IR to $SU(F)_X$.

More precisely, the topological $U(1)$ symmetry associated with the $i$-th column maps to the anti-diagonal combination of $U(1)_{X_i}$ and $U(1)_{X_{i+1}}$.

The topological symmetry $U(1)_\eta$ of SQCD is mapped to a combination of $U(1)_\Lambda$, $U(1)_{X_i}$, and the R-symmetry in the Abelian theory. This map can be expressed at the level of real mass parameters as
\begin{equation}
    \Lambda = \eta - \sum_{i=1}^{F/2} X_i + \frac{F}{2}(3r-1)\omega \,,
\end{equation}
as shown in \eqref{eq: Dual_k0}. Replacing real masses by the corresponding generators yields the mapping of the $U(1)_\eta$ symmetry.

\paragraph{Background terms.}
Finally, the Abelian dual includes background CS terms, conveniently encoded in the $\mathbf{S}_b^3$ partition function as $\exp\!\left(\tfrac{\pi i}{2}\phi^{(\mathrm{mag})}\right)$, with
\begin{equation}
    \phi^{(\mathrm{mag})} =
    N\left(\sum_{i=1}^{\tfrac F2}(X_i^2 - 2\omega(1-3r) X_i)
    - \sum_{j=\tfrac F2+1}^{F}(X_j^2 +2\omega(1-r) X_j)\right)+\mathcal{O}(\omega^2).
\end{equation}
The quadratic terms encode CS couplings for the global symmetries $U(1)_{X_i}$, while the linear terms correspond to mixed CS couplings between $U(1)_{X_i}$ and the R-symmetry. These contributions are essential for matching the duality in the presence of background fields and for correctly implementing real mass deformations.

\paragraph{Compact Quiver Notation.}
All the data required to fully specify the planar Abelian theory are encoded in the quiver diagram. This will remain true throughout the paper, and we therefore adopt a compact notation to describe the duality in \eqref{eq: Dual_k0}, as well as in the rest of this work. This notation is summarized in \eqref{eq:quiv:notation_compact} and in the discussion below.

Within this compact notation, the duality in \eqref{eq: Dual_k0} can be expressed more succinctly as:
\vspace{-.5cm}
\begin{equation}    \label{eq:Dual_k0_schematic}
    \includegraphics[width=.8\linewidth]{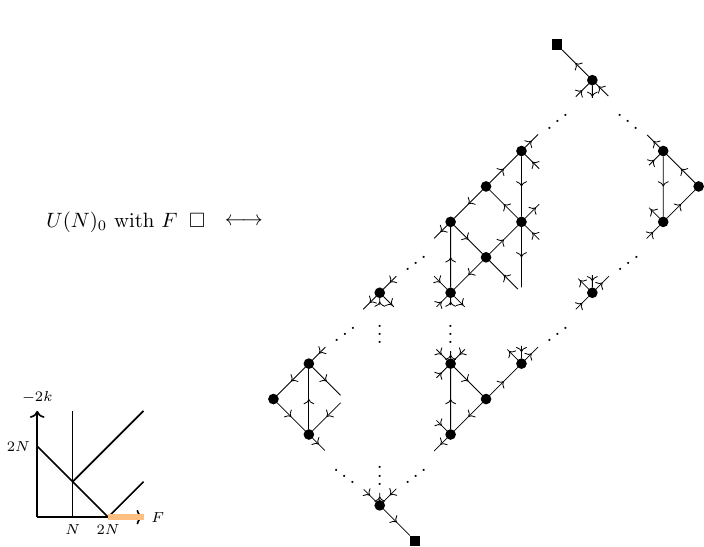}
\end{equation}
Notice that we avoid writing (mixed) CS levels explicitly. We remind the reader that these can be reconstructed with the rules in \eqref{eq:quiv:notation_compact}, which assign a standard mixed CS interaction to each pair of nodes connected by a chiral multiplet. We reiterate that the CS level of each node can be reconstructed from mixed CS levels through the relation:
\begin{equation}
    k_i = -\frac{1}{2}\sum_{j \neq i} k_{ij}
\end{equation}
where $k_i$ is the level of the $U(1)_i$ gauge node, and the sum runs over \textbf{all} the nodes of the quiver, including gauge and flavor nodes.
Furthermore, we avoid giving FI parameters explicitly to avoid clutter in the figures.

Here, and in the rest of the paper, we also include a diagram, on the bottom left, where we highlight in orange the regions in the $(k,F)$ plane where the duality is valid. These correspond to the various regions in Figure \ref{fig:KF_plane}. 

\subsubsection{Duality in Zone 1 (\texorpdfstring{$F \geq 2N+2|k|$}{Zone 1})}\label{subsec: negative mass deformations in mirror}

Starting from the duality between $U(N)_0$ SQCD with $F \geq 2N$ fundamental chiral multiplets and its planar Abelian dual in \eqref{eq: Dual_k0}, we now introduce negative real mass deformations and study the corresponding RG flows in both descriptions. 

On the “electric” side, this deformation drives the theory to $U(N)_{-\tfrac 12}$ SQCD with $F-1$ fundamental chiral multiplets. The analysis of the RG trajectory in the planar Abelian dual is more subtle. We propose that the corresponding deformation is implemented by Higgsing all the chiral fields along the bottom-left diagonal, as well as those connecting to the bottom flavor node. This Higgsing identifies all gauge nodes along the bottom-left diagonal, and the planar superpotential correspondingly reduces to mass terms.
Graphically the deformation is the following:
\begin{equation}\label{eq: K0_def_highlight_fields}
\resizebox{.8\hsize}{!}{ 
\begin{tikzpicture}[baseline=50pt] 

    \begin{scope}[xshift=-4.5cm, yshift = -2cm]
        \node at (-.5,0) {r.h.s.~ of \eqref{eq:Dual_k0_schematic}};
        \node[scale=1.5] at (2,0) {$\overset{ X_1\to+\infty}{\longrightarrow}$};
    \end{scope}

    \begin{scope}[xshift=-2cm, yshift = -5cm, scale=.8]
            \node at (0,0) (g11) [gaugefill] {};
            \node at (1,1) (g21) [gaugefill] {};
            \node at (1,-1) (g22) [gaugefill] {};
            \node at (2,2) (g31) {$\iddots$};
            \node at (2,0) (g32) [gaugefill] { };
            \node at (2,-2) (g33) {$\ddots$};
            \node at (3,3) (g41) [gaugefill] {};
            \node at (3,2) (g42) {$\vdots$};
            \node at (3,-2) (g43)  {$\vdots$};
            \node at (3,-3) (g44) [gaugefill] {};
            \node at (4,-4) (f1) [flavorfill,black] {};

            \node at (4,4) (g51) {$\iddots$};
            \node at (4,-2) (g52) {$\iddots$};

            \node at (5,5) (g61) [gaugefill] {};
            \node at (5,3) (g62) [gaugefill] {};
            \node at (5,2) (g63) {$\vdots$};
            \node at (5,1) (g64) [gaugefill] {};
            \node at (5,-1) (g65) [gaugefill] {};

            \node at (6,6) (g71) [gaugefill] {};
            \node at (6,4) (g72) [gaugefill] {};
            \node at (6,0) (g75) [gaugefill] {};
            
            \node at (7,7) (g81) [gaugefill] {};
            \node at (7,5) (g82) [gaugefill] {};
            \node at (7,3) (g83) [gaugefill,white]{};
            \node at (7,1) (g84) [gaugefill] {};

            \node at (8,8) (g91) {$\iddots$};
            \node at (8,2) (g92) {$\iddots$};

            \node at (9,9) (g101) [gaugefill] {};
            \node at (8,10) (f2) [flavorfill,black] {};
            \node at (9,3) (g102) [gaugefill] {};

            \node at (10,8) (g111) {$\ddots$};
            \node at (10,4) (g112) {$\iddots$};

            \node at (11,7) (g121) [gaugefill] {};
            \node at (11,5) (g122) [gaugefill] {};
            \node at (12,6) (g131) [gaugefill] {};

            \begin{scope}[on background layer]
        \draw[line width=6pt, blue!20, line cap=round]
    (g11.center) -- (g22.center) -- (g33.center) -- (g44.center) -- (f1.center);
   \draw[line width=6pt, orange!30, line cap=round]
   (g21.center) -- (g11);
    \draw[line width=6pt, orange!30, line cap=round]
   (g21.center) -- (g22);
    \draw[line width=6pt, orange!30, line cap=round]
   (g32.center) -- (g22);
    \draw[line width=6pt, orange!30, line cap=round]
   (g43.center) -- (g44);
      \end{scope}
            \draw[->-] (g11)--(g22);
            \draw[->-] (g21)--(g11);
            \draw[->-] (g22)--(g21);
            
            \draw[->-] (g22)--(g33);
            \draw[->-] (g32)--(g22);
            \draw[->-] (g31) -- (g21);
            \draw[->-] (g21)--(g32);

            \draw[->-] (g41)--(g31);
            \draw[->-] (g42)--(g41);
            \draw[->-] (g33)--(g44);
            \draw[->-] (g44)--(f1);
            \draw[->-] (g44)--(g43);
            \draw[->-] (g51)--(g41);
            \draw[->-] (g52)--(g44);
            \draw[-<-] (g41)++(.4,-.4) -- (g41);
            \draw[->-] (g61)--(g51);
            \draw[->-] (g62)--(g61);
            \draw[->-] (g51)--(g62);
            \draw[->-] (g63)--(g62);
            \draw[->-] (g64)--(g63);
            \draw[->-] (g65)--(g64);
            \draw[->-] (g65)--(g52);
            \draw[->-] (g65)++(-.4,.4) -- (g65);
            \draw[->-] (g64)++(-.4,.4) -- (g64);
            \draw[->-] (g64)++(-.4,-.4) -- (g64);
            \draw[->-] (g62)++(-.4,-.4) -- (g62);
            
            \draw[->-] (g71)--(g61);
            \draw[->-] (g61)--(g72);
            \draw[->-] (g72)--(g62);
            \draw[-<-] (g62)++(.4,.-.4) --(g62);
            \draw[->-] (g64)++(.4,.4) --(g64);
            \draw[->-] (g64)--(g75);
            \draw[->-] (g75)--(g65);
            \draw[->-] (g72)--(g82);
            \draw[->-] (g82)--(g71);
            \draw[->-] (g71)--(g81);
            \draw[->-] (g81)--(g82);
            \draw[->-] (g83)--(g72);
            \draw[->-] (g82)--(g83);
            \draw[->-] (g75)--(g84);
            \draw[-<-] (g84)++(-.4,.4) -- (g84);
            \draw[->-] (g84)++(0,.5) -- (g84);

            \draw[->-] (g81)--(g91);
            \draw[->-] (g81)++(.5,-.5) -- (g81);
            \draw[->-] (g82)++(.5,-.5) -- (g82);
            \draw[-<-] (g82)++(.5,.5) -- (g82);
            \draw[->-] (g84)--(g92);
            \draw[->-] (g91) --(g101);
            \draw[->-] (g101)--(f2);
            \draw[-<-] (g101)++(0,-.5) -- (g101);
            \draw[->-] (g102)++(0,.5) -- (g102);
            \draw[->-] (g92)--(g102);
            \draw[-<-] (g102)++(-.4,.4) -- (g102);
            \draw[->-] (g102)--(g112);
            \draw[-<-] (g101) -- (g111);
            \draw[->-] (g131)--(g121);
            \draw[->-] (g121)--(g122);
            \draw[->-] (g122)--(g131);
            \draw[->-] (g121)--(g111);
            \draw[->-] (g112)--(g122);
            \draw[->-] (g121)++(-.4,-.4) --(g121);
            \draw[-<-] (g122)++(-.4,.4) --(g122);
            \draw [decorate,decoration={brace,amplitude=5pt,raise=2ex}]
        (0,0)--(6,6) node[midway,xshift=-1.5em,yshift=2.4em]{$\tfrac{F}{2}$ nodes}; 

        \draw [decorate,decoration={brace,amplitude=5pt,raise=2ex}]
        (12,6)--(6,0) node[midway,xshift=3em,yshift=-1.4em]{$\tfrac{F}{2}$ nodes};

    \end{scope}
    \begin{scope}[shift={(3,-8)}]

    \node[anchor=west] at (0,0.55) { Higgsed:};
    \draw[blue!30, line width=6pt, opacity=0.8, line cap=round] (1.8,0.55) -- (2.8,0.55);

    
    \draw[orange!30, line width=6pt, opacity=0.8, line cap=round] (3.9,.2) -- (4.9,.2);
    \node[anchor=west] at (0,0.2) { Superpotential mass:};
    
\end{scope}

\end{tikzpicture}
}
\end{equation}
Here, the chiral fields shown in blue are those that acquire Higgs VEVs, consequently the fields in orange obtain complex masses from the planar superpotential as illustrated below:
\begin{equation}\label{eq: Higgsing_diag_separate}
\includegraphics[width=\linewidth]{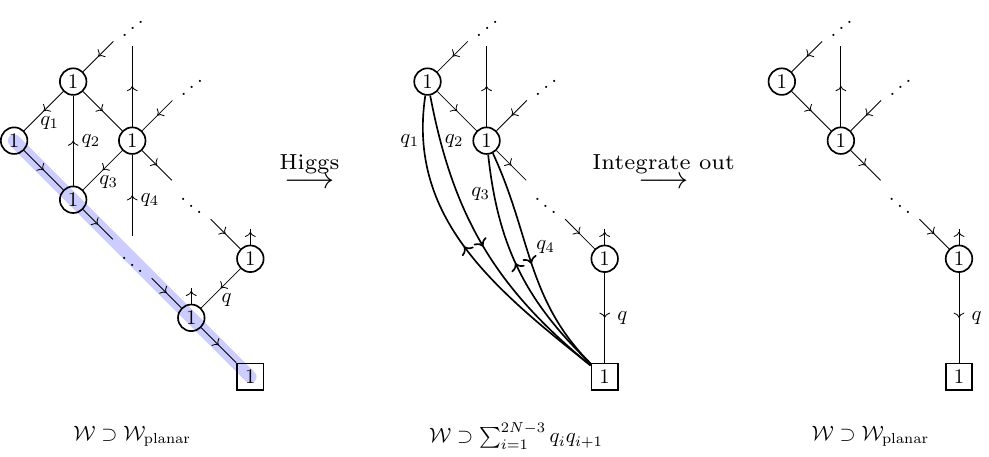}
\end{equation}

Integrating out the massive fields yields the planar Abelian dual for $U(N)_{-1/2}$ SQCD with $F-1$ fundamentals:
\begin{equation}\label{eq: First_def}
\includegraphics[width=.7\linewidth]{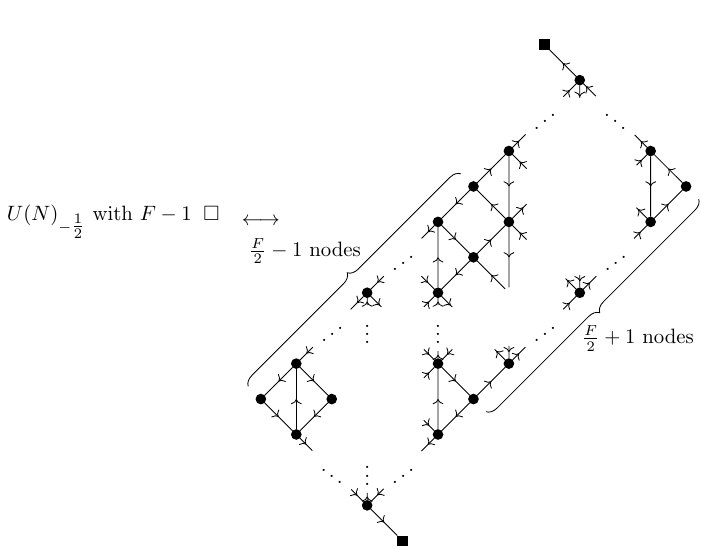}
\end{equation}
 
The resulting dual quiver closely resembles the Abelian dual before the mass deformation. The only differences are the number of columns and the position of the “column of squares”. In this case, the column of squares is displaced from the center of the quiver, as indicated by the sizes of the various quiver segments.

Further negative mass deformations are mapped analogously, leading to the following duality for SQCD with $F \geq 2N + 2|k|$ and $k\leq0$ (\textbf{Zone 1} of Figure \ref{fig:KF_plane}):
\begin{equation}\label{eq: Dual_zone1_schematic}
\includegraphics[width=.7\linewidth]{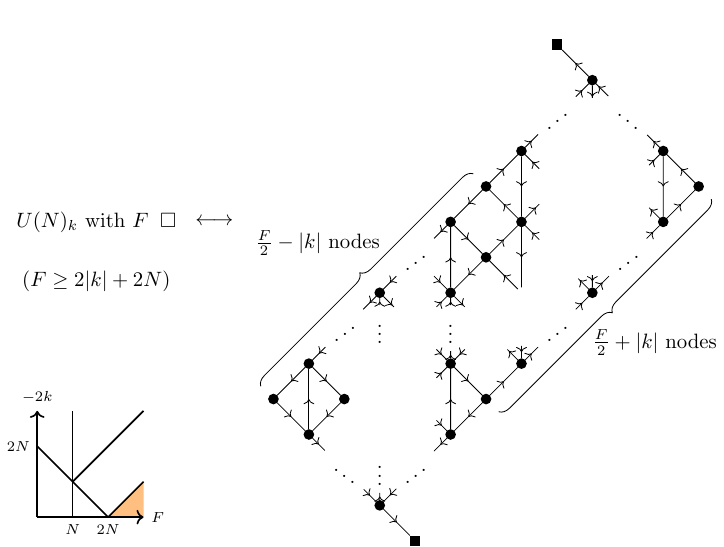}
\end{equation}

The three parameters $(N, F, k)$ are mapped to the sizes of the two edges of the quiver and to the position of the column of squares within it. The (mixed) CS interactions and the superpotential terms follow the same rule as in Section~\ref{subsec: planar abelian dual of U(N)0} (see the discussion below \eqref{eq: Dual_k0} or the summary in \eqref{eq:quiv:notation_compact}). 

This duality exhibits features very similar to those of \eqref{eq: Dual_k0}. In particular, the Abelian planar theory contains a superpotential with a cubic (quartic) term for each triangular (square) face of the quiver. In addition, there is a linear monopole superpotential for each monopole operator with $-1/+1$ GNO flux under the top/bottom gauge nodes connected by a vertical arrow. 

Overall, the theory possesses a single unbroken topological symmetry for each vertical column of chiral multiplets, giving a total $U(1)^{F-1}$ symmetry, which enhances in the IR to $SU(F)$. Moreover, there is a single $U(1)$ flavor symmetry that maps to the $U(1)$ topological symmetry of the SQCD\footnote{More precisely, a combination of the flavor $U(1)$, some of the topological $U(1)$ factors, and the $U(1)_R$ R-symmetry maps to the topological symmetry of the SQCD. We omit the precise combination here for brevity, but it can be determined by carefully tracking the RG flow on the $\mathbf S^3_b$ partition function.}.

We can systematically follow the first $\tfrac{F}{2} - N$ negative mass deformations in this way. The limiting case corresponds to theories obtained from supersymmetry-breaking real mass deformations of the $\NN=4$ dual pairs, shown in \eqref{eq: Dual_N4_line_schematic}.
\begin{equation}\label{eq: Dual_N4_line_schematic}
\includegraphics[width=.9\linewidth]{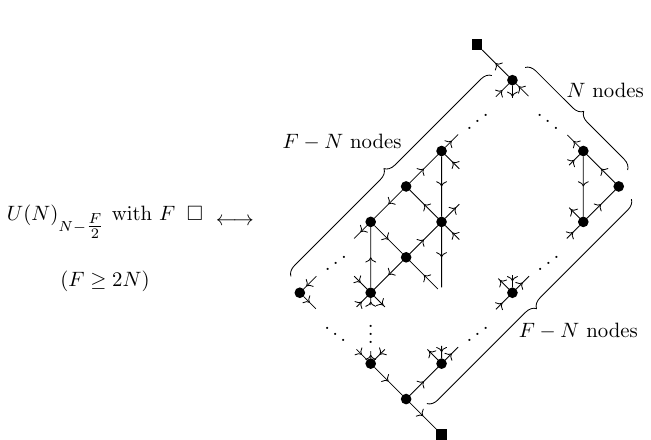}
\end{equation}

This proposal coincides with that of \cite{Benvenuti:2024seb,Benvenuti:2025a}, as can be shown through a sequence of local dualizations following the procedure described in detail in Appendix~B of \cite{Benvenuti:2025a}. In short, starting from the quiver in~\eqref{eq: Dual_N4_line_schematic}, one performs Aharony duality on all the gauge nodes in each column, from left to right, stopping at the $(N-1)$-th column, then repeats the process up to the $(N-2)$-th column, and so on. All the dualities utilized in this procedure are described in detail in Appendix \ref{app: rel_duals}, along with an example of such sequence of dualizations.
After performing this sequence $N-1$ times, the resulting quiver matches the planar Abelian dual of SQCD on the $\NN=4$–descendant line proposed in~\cite{Benvenuti:2025a}, shown in \eqref{eq: Dual_N4line_schematic_nosquares}.
\begin{equation}\label{eq: Dual_N4line_schematic_nosquares}
\includegraphics[width=.9\linewidth]{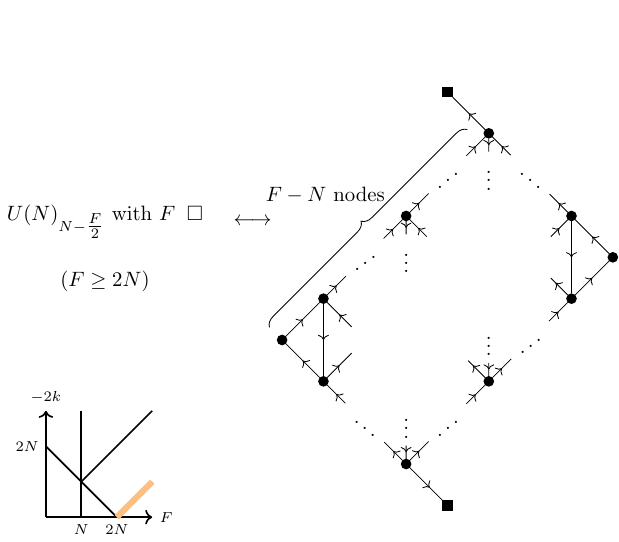}
\end{equation}

\subsubsection{Duality in Zone 2 (\texorpdfstring{$2N+2|k|>F\geq 2|k|$}{Zone 2})}

We can further probe the region $2N + 2|k| > F \geq 2|k|$ (\textbf{Zone 2} of Figure \ref{fig:KF_plane}) by introducing additional negative real masses in the electric theory and studying the corresponding mirror RG flow trajectory. The associated deformation on the planar Abelian dual is similar to that described at the beginning of Section \ref{subsec: negative mass deformations in mirror}, with the difference that they now affect fields along the bottom-right diagonal instead of the bottom-left one. A more detailed analysis of this deformation is presented in Section \ref{subsec: deformations of N4 sqcd}; for now, we focus on describing the resulting theory.
\newpage

The resulting duality for SQCD with $2N + 2|k| > F \geq 2|k|$ is shown in \eqref{eq: Zone22t_dual_raw}.
\begin{equation}\label{eq: Zone22t_dual_raw}
\includegraphics[width=.9\linewidth]{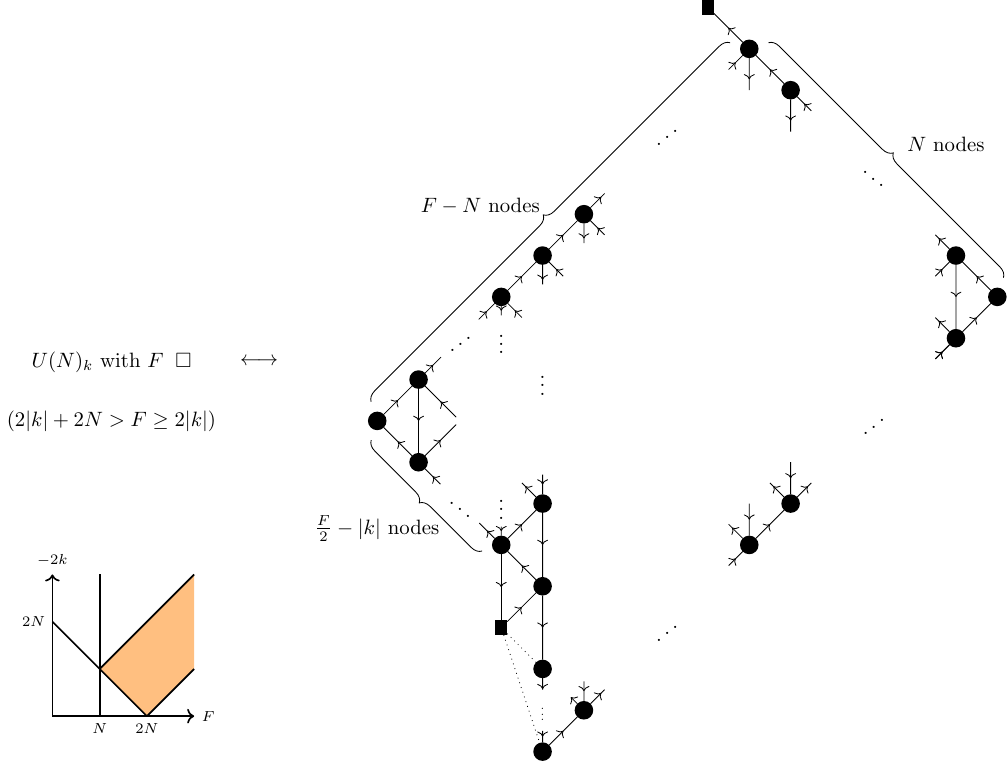}
\end{equation}

Turning on exactly $N$ negative real masses results in a planar Abelian dual for marginally chiral SQCD \eqref{eq: margchir_dual_raw}.
\begin{equation}\label{eq: margchir_dual_raw}
\includegraphics[width=.65\linewidth]{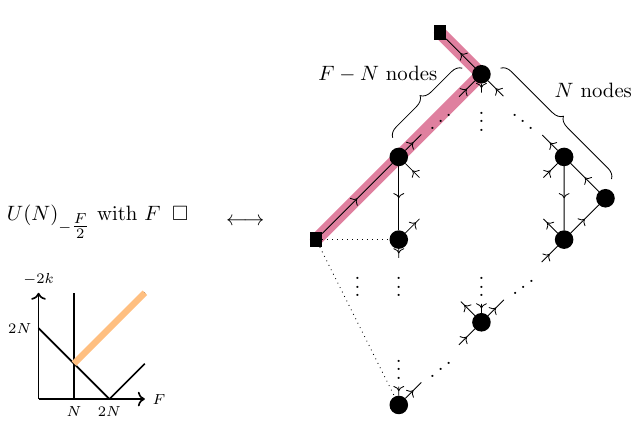}
\end{equation}
In this case, we observe that the planar Abelian quiver contains a long gauge-invariant mesonic operator (highlighted in maroon) in the chiral ring that parametrizes a non-compact moduli space. Consistently, the dual SQCD features a gauge-invariant monopole operator parametrizing a $\mathbb C$ moduli space of vacua.

Notice that Zone 2 includes, on its boundary, the case of $U(N)_{-\tfrac N 2}$ SQCD with $N$ flavors, which corresponds to the cusp of the supersymmetry-breaking region. The planar dual of this theory cannot be obtained as a simple limiting case of the dualities in \eqref{eq: Zone22t_dual_raw} and \eqref{eq: margchir_dual_raw}. Indeed, in this limit, certain parameters controlling the size of the quiver degenerate, indicating that the planar dual takes a qualitatively different form.

Nevertheless, the dual description can be accessed by following real mass deformations in the Abelian theory. For instance, starting from the dual of $U(N)_{-\tfrac N 2 + \tfrac 1 2}$ with $N+1$ fundamental chiral multiplets and integrating out one field, we are led to the duality shown in \eqref{eq: freechir_dual}.
\begin{equation}\label{eq: freechir_dual}
\includegraphics[width=.75\linewidth]{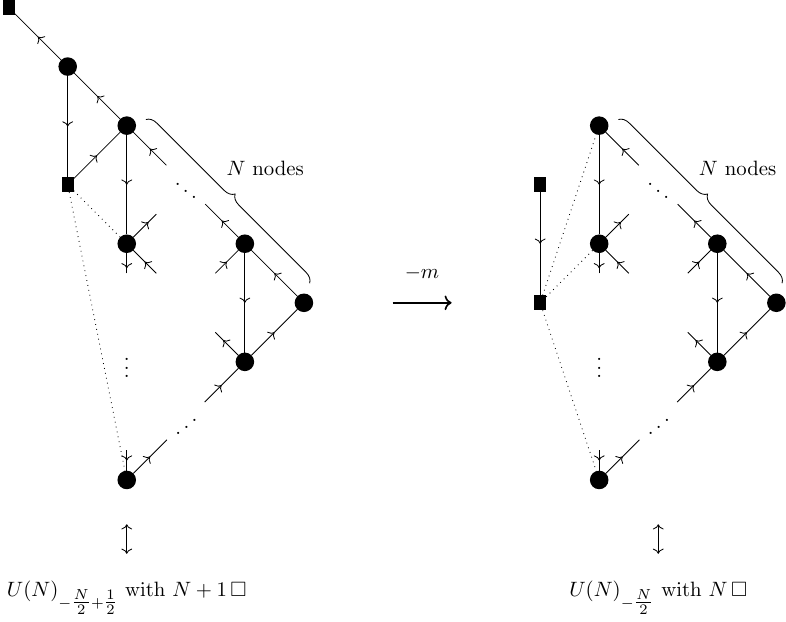}
\end{equation}

By applying a sequence of Aharony dualities on the gauge nodes of the Abelian quiver, one finds that the theory confines to a single free chiral multiplet. All the dualities involved in this procedure and a similar example is described in detail in Appendix \ref{app: rel_duals}.
This reproduces the expected infrared duality
\begin{equation}
    U(N)_{-\tfrac N 2} \quad \text{with} \quad N\, \Box 
    \qquad \longleftrightarrow \qquad 
    \text{free chiral}\,,
\end{equation}
where the free chiral is identified with the gauge-invariant monopole operator of the SQCD theory.

\subsubsection{Duality in Zone \texorpdfstring{$\tilde{1}$ ($2N-2|k|>F>N$)}{1t}}
Starting from a theory in Zone 2, we can consider further positive mass deformations to probe the regime $2N - 2|k| > F > N$ (\textbf{Zone} $\mathbf{\widetilde{1}}$ of Figure \ref{fig:KF_plane}). 
For the mirror theory, this deformation is analogous to that described at the beginning of Section \ref{subsec: negative mass deformations in mirror}, which has simply the effect of shortening the quiver by ``removing" the top-right diagonal of the quiver. As in the previous section, we refer to Section \ref{subsec: deformations of N4 sqcd} for more details.

The corresponding duality for SQCD is shown in \eqref{eq: Zone1t_dual_raw}.
\begin{equation}\label{eq: Zone1t_dual_raw}
\includegraphics[width=\linewidth]{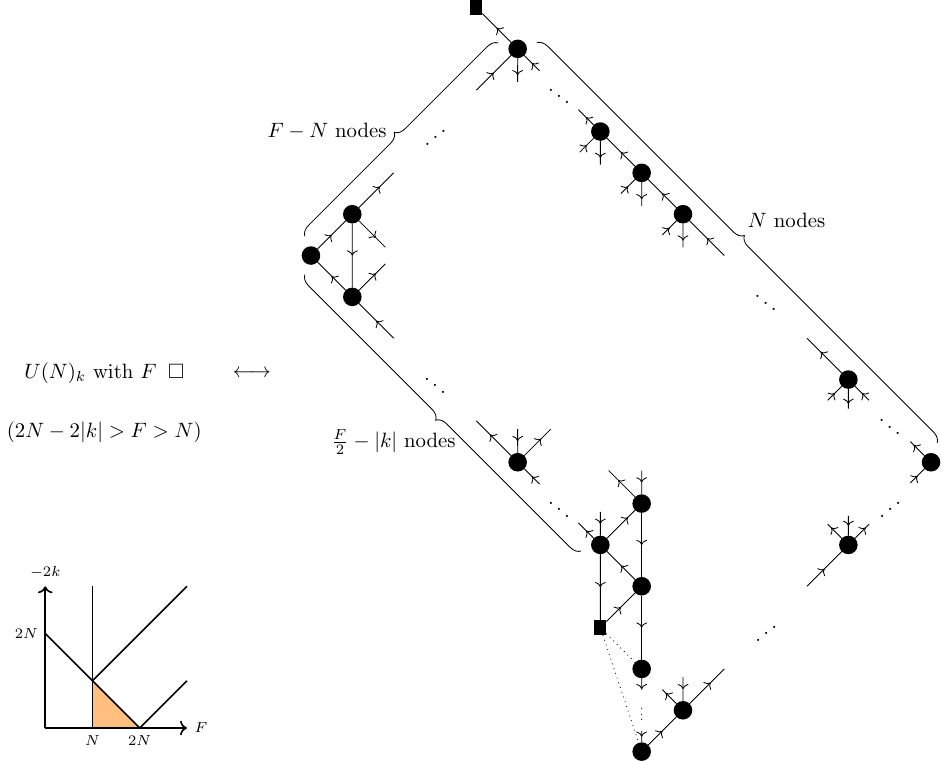}
\end{equation}
The dual quiver obtained is similar to those considered in \eqref{eq: Zone22t_dual_raw}. 
In Section~\ref{sec: aharonyproof}, we demonstrate that, through a sequence of local dualizations, the planar Abelian dual of theories in \textbf{Zone $\widetilde{1}$} admits an alternative representation, reported in \eqref{eq:dual_zone1t_alternative}, which more closely resembles the planar Abelian dual of \textbf{Zone 1} in \eqref{eq: Dual_zone1_schematic}. 
Indeed, theories in \textbf{Zone $\widetilde{1}$} and \textbf{Zone 1} are expected to be related by Aharony duality, and the sequence of local dualizations explicitly relates the planar Abelian duals of the two zones, consistent with this expectation.

This duality also closely parallels that of \eqref{eq: Dual_k0}. 
The superpotential contains both mesonic and monopole terms, following the rules outlined below \eqref{eq: Dual_k0}. 
As a result, the theory has a single $U(1)$ flavor symmetry, while each column of vertical chiral fields contributes one $U(1)$ topological symmetry due to the monopole superpotential. 
This yields a total $U(1)^{F-1}$ symmetry, which enhances in the IR to $SU(F)$. 
These symmetries map naturally to the $U(1) \times SU(F)$ topological and flavor symmetries of the original SQCD.

On the boundary between Zone $\widetilde{1}$ and the SUSY-breaking region, corresponding to $F=N$ and $2|k| \leq 2N - F$, the theory is expected to become trivial. This behavior cannot be directly inferred from \eqref{eq: Zone1t_dual_raw}, since reaching this point would require setting one of the quiver size parameters to zero, leading to a degenerate description. 

Nevertheless, this limit can be accessed by considering theories that flow to this boundary via a single real mass deformation. For example, one may start from theories along the vertical line with $F=N+1$ and $2|k| \leq 2N - F$, and follow the deformation. In this way, one finds that the planar Abelian dual reduces to a quiver that fully confines to a $U(1)_1$ theory, which is an almost-trivial theory.

In contrast, the boundary between Zone $\widetilde{1}$ and Zone 2 can be straightforwardly obtained by setting $2|k| = 2N - F$ in either \eqref{eq: Zone22t_dual_raw} or \eqref{eq: Zone1t_dual_raw}, which is a well-defined operation at the level of the quiver.

Similarly, theories along the horizontal line $k=0$, with $N < F < 2N$, can be obtained by directly setting $k=0$ in \eqref{eq: Zone1t_dual_raw}. The special case $k=0$ and $F=2N$ is instead described by \eqref{eq:Dual_k0_schematic}. This point is not captured by a naive limit of \eqref{eq: Zone1t_dual_raw}, as expected, since it cannot be reached via positive real mass deformations starting from theories in Zone 2.

\subsubsection{Duality in Zone \texorpdfstring{3 ($2|k|>F\geq N$)}{3}}

We can also probe the region $2|k| > F \geq N$ (\textbf{Zone 3} of Figure \ref{fig:KF_plane}). 
This is achieved by starting from the dualities on the boundary between Zone 2 and Zone 3 in \eqref{eq: margchir_dual_raw} and performing more negative mass deformations. The effect of the first deformation on the planar quiver in \eqref{eq: margchir_dual_raw} is to simply give a negative real mass to the left flavor chiral. The following real masses have instead the effect of ``integrating out columns", meaning that a positive/negative real mass is acquired by all the chirals represented as diagonal lines connecting two columns of vertical chirals. In this way, we are left simply by column of vertical chirals of height $N$ that are ``connected" only through mixed CS interactions between gauge nodes in the same relative position inside the nearest columns.
This effect occurs consecutively on each column, starting from the leftmost one and proceeding to the right, as shown in \eqref{eq:columns_howto}.

\begin{equation}
\label{eq:columns_howto}
    \includegraphics[width=.9\textwidth]{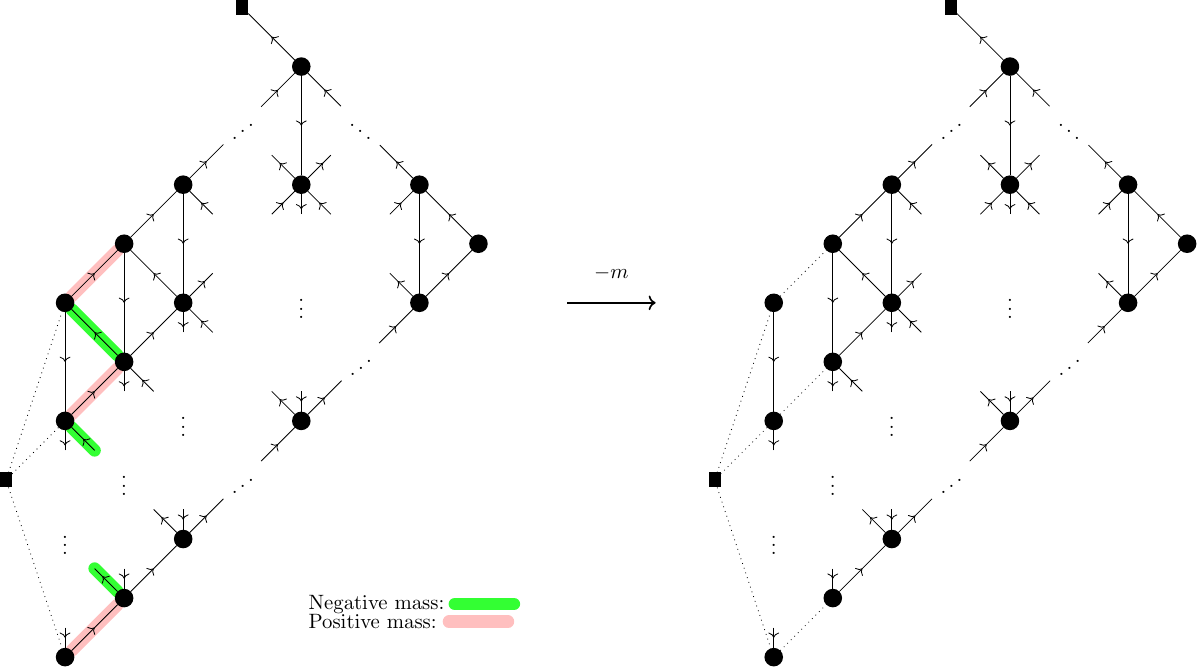}
\end{equation}

This description is indeed valid so long as we do not cross the boundary between Zone 3 and Zone 4.

We thus obtain the following duality \eqref{eq: Zone33t_a_dual_raw}:
\begin{equation}\label{eq: Zone33t_a_dual_raw}
\includegraphics[width=.8\linewidth]{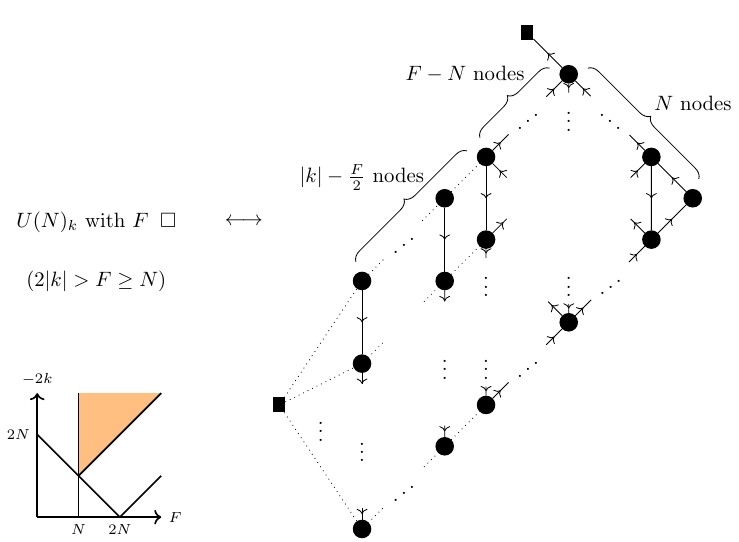}
\end{equation}

Let us comment on the monopole superpotential in the quiver theories above. There is a monopole superpotential term for each pair of consecutive nodes in a column with GNO flux $-1/+1$ under the top/bottom nodes. In the ``bulk" of the quiver, where there are diagonal arrows on both sides of the column, these monopoles are gauge-invariant BPS operators and, hence, enter the superpotential as described below \eqref{eq: Dual_k0}. When the gauge nodes are missing diagonal chirals on at least one side then the monopole superpotential follows these two new rules. 
\begin{itemize}
    \item In the column that separates the ``bulk" of the quiver and the region with BF interactions, as shown in \eqref{eq: mon_W1}, these monopoles are dressed with a vertical chiral field $\phi$ as following:
    \begin{equation}\label{eq: mon_W1}
    \begin{tikzpicture}[baseline=(current bounding box).center, scale=.6]
        \foreach \x in {1,...,3}
        {
            \foreach \y in {1,2}
            {
                \node at (2*\x,-2*\y+\x) (g\x\y) [gaugefill] {$1$};
            }
        \path[->-,draw] (g\x1) -- (g\x2); 
        \path[->-,draw] (g\x1)++(0,1) -- (g\x1);
        \path[-<-,draw] (g\x2)++(0,-1) -- (g\x2);
        }
        
        \path[-<-,draw] (g31) -- (g21) ;
        \path[-<-,draw] (g21) -- (g32);
        \path[-<-,draw] (g32) -- (g22);
        \draw[->-] (g22)++(1,-.5) -- (g22);
        \draw[->-] (g31)++(1,-.5) -- (g31);
        \draw[->-] (g32)++(1,-.5) -- (g32);
        \draw[-<-] (g31)++(1,.5) -- (g31);
        \draw[-<-] (g32)++(1,.5) -- (g32);
        \draw[-<-] (g31)++(-1,.5) -- (g31);
        
        \draw[dotted] (g11)++(-1,-.5) --(g11) -- (g21);
        \draw[dotted] (g12)++(-1,-.5) --(g12) -- (g22);
        \path[] (g11) -- node[midway,left] {$\phi$} (g12);
    \end{tikzpicture}
    \qquad\qquad
    \mathcal{W} \ni \mathfrak M^{\left( \begin{array}{ccc} 0&-1&0\\0&1&0\end{array} \right)} \phi
\end{equation}
    The FIs of the nodes in this type of column are, from top to bottom:
    \begin{equation}
        \text{FIs: } X_1 -\frac{\omega}{2}, \;\;
                    X_1 - r \omega, \;\;
                    \dots,  X_1 - r\omega, \;\;
                    X_1 + \frac{\omega}{2}(1 - 3r)
    \end{equation}
    where $X_1$ is a Cartan of the $SU(F)$ flavor symmetry and the coefficient in front of $\omega$ encodes the mixing between the topological and the R-symmetry. We remind the reader that we assign trial R-charge $r$ to the fundamental fields of the SQCD.
    One can check that the FIs above are compatible with the monopole superpotential.
\item In the columns with BF interactions on both sides, as shown in \eqref{eq: mon_W2}, the monopoles are dressed with two vertical chiral fields $\phi_1$ and $\phi_2$: 
\begin{equation}
\label{eq: mon_W2}
    \begin{tikzpicture}[baseline=(current bounding box).center, scale=.6]
        \foreach \x in {1,...,3}
        {
            \foreach \y in {1,2}
            {
                \node at (2*\x,-2*\y+\x) (g\x\y) [gaugefill] {$1$};
            }
        \path[->-,draw] (g\x1) -- (g\x2); 
        \path[->-,draw] (g\x1)++(0,1) -- (g\x1);
        \path[-<-,draw] (g\x2)++(0,-1) -- (g\x2);
        }

        \draw[dotted] (g11)++(-1,-.5) --(7,1.5);
        \draw[dotted] (g12)++(-1,-.5) --(7,-.5);
        \path[] (g11) -- node[midway,left] {$\phi_1$} (g12);
        \path[] (g31) -- node[midway,right] {$\phi_2$} (g32);
    \end{tikzpicture}
    \qquad\qquad
    \mathcal{W} \ni \mathfrak M^{\left( \begin{array}{ccc} 0&-1&0\\0&1&0\end{array} \right)} \phi_1 \phi_2
\end{equation}
    The FIs of the nodes in this type of column are, from top to bottom:
    \begin{equation}
        \text{FIs: } \frac{\omega}{2}(2r-1), \;\;
                    0, \;\;
                    \dots,  0, \;\;
                    \frac{\omega}{2}(1 - 2r)
    \end{equation}
    One can check that the FIs above are compatible with the monopole superpotential.
\item We further notice that in a column with no diagonal chirals on both sides, the single unbroken topological symmetry can be reabsorbed by a gauge transformation due to the presence of non-zero CS-levels. Therefore, only the columns in the ``bulk" of the quiver and at the edge of the ``bulk" contribute to the Cartan of the global symmetry, thus matching the rank of the global symmetry predicted by the dual SQCD.
\end{itemize}

The same discussion also applies to the planar Abelian quivers in Zone 4 discussed below.

\subsubsection{Duality in Zone \texorpdfstring{4 ($2N-2|k|<F<N$)}{4}}

We can continue to perform negative real mass deformations for the SQCD to probe the region $2N-2|k|<F < N$ (\textbf{Zone 4} of Figure \ref{fig:KF_plane}). Starting from the planar quiver in Zone 3 describing $F=N$, in \eqref{eq: Zone33t_a_dual_raw}, the effect of the deformation is now mapped to the chiral fields in the top-right diagonal of the quiver. Qualitatively, the effect of the deformation is the same as the one that we performed to cross the boundary between Zone 1 and Zone 2, but now mapped to the top-right diagonal instead of the bottom-left. By performing multiple such deformations, we can fully explore the entirety of Zone 4.

The resulting dualities are shown in \eqref{eq: Zone33t_b_dual_raw}.
\begin{equation}\label{eq: Zone33t_b_dual_raw}
\includegraphics[width=.9\linewidth]{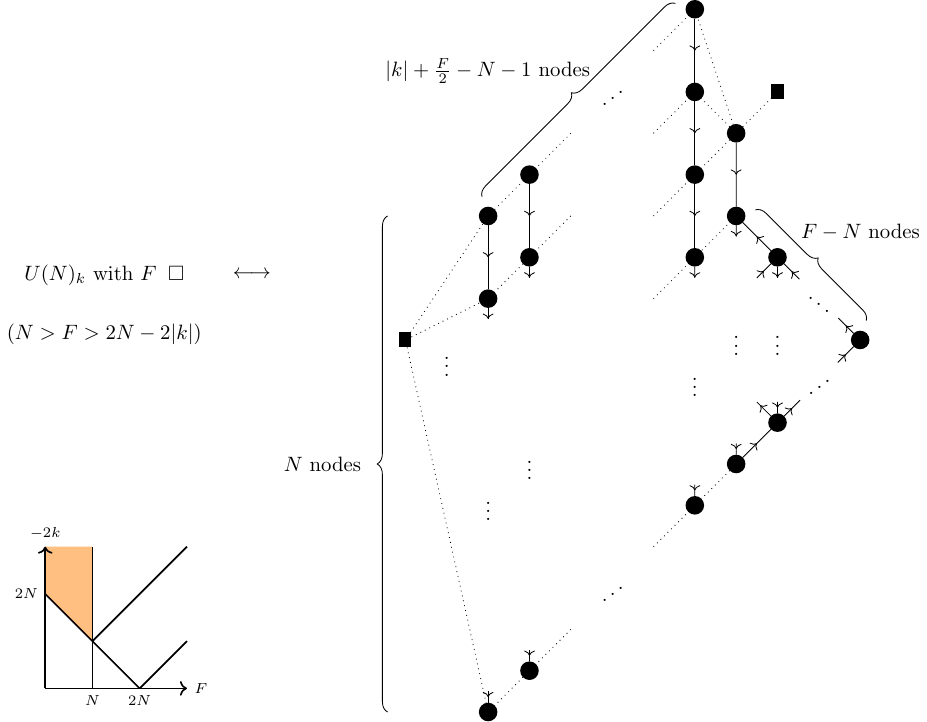}
\end{equation}

On the boundary between Zone 4 and the SUSY-breaking region, namely for $2|k| = 2N - F$ with $F < N$, the planar quiver takes a qualitatively different form and cannot be obtained as a well-defined zero-size limit of \eqref{eq: Zone33t_b_dual_raw}. 

This duality also describes the line separating Zone 3 and Zone 4, that is, for $F=N$ and $2|k| > N$.

However, using the fact that the theory at $F = |2k| = N$ reduces to a single free chiral multiplet, one can immediately infer the behavior along this boundary. Indeed, a negative real mass deformation of $U(N)_{-N/2}$ SQCD with $N$ flavors maps, in the dual description, to a mass term for the free chiral. Integrating out this field yields a trivial theory, implying that the boundary theory is trivial.

\subsubsection{Dualities for \texorpdfstring{$U(N)_k$ Super-Yang-Mills Theory}{TQFTs}}

Starting from Zone 4, we can add more negative massive deformations to integrate out all matter and finally reach the case of $F=0$. In such a situation, the SQCD reduces to a $U(N)_k$ TQFT. We can follow this sequence of deformations in the planar Abelian theory describing Zone 4, in \eqref{eq: Zone33t_b_dual_raw}, all the way down to $F=0$. The resulting duality that we obtain is:

\begin{equation}
\label{eq: tqft}
    \includegraphics[width=.8\linewidth]{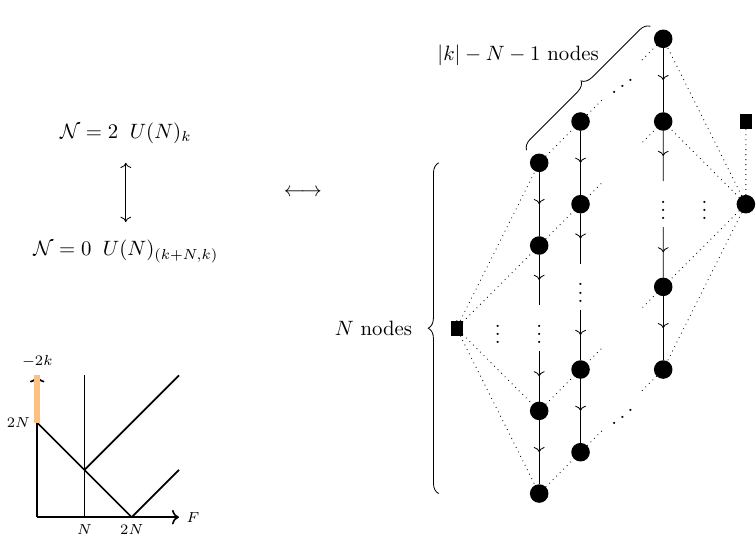}
\end{equation}


Interestingly, the resulting planar dual is not manifestly topological, as it still contains massless charged chiral fields. 
The duality, therefore, implies that the planar Abelian quiver shown in \eqref{eq: tqft} is in fact a gapped theory, flowing in the infrared to a $U(N)$ Chern–Simons theory with levels $(k+N,k)$.

Here we consider two limiting cases corresponding to $k=-N-1$ and $k=-N$. 
In these cases the planar Abelian dual does not involve any chiral field and is manifestly a TQFT.
In the first case we find the following duality\footnote{Here and in the rest of the paper we do not keep track of background gravitational CS terms.}:
\begin{equation}
    \begin{tikzpicture}[baseline=(current bounding box).center]
        \node at (0,0) (N2) {$\NN=2 \qquad U(N)_{-N-1}$};
        \node at (0,-2) (N0) {$\NN=0 \qquad U(N)_{-1,-N-1}$};
        \draw[->] (N2) -- (N0);
    \end{tikzpicture}
    \qquad \longleftrightarrow \qquad
    \begin{tikzpicture}[baseline=(current bounding box).center,scale=.8]
        \node[flavorfill] at (0,0) (f1) {} ;
        \node[gaugefill] at (3,0) (g) {} ;
        \node[flavorfill] at (3,2) (f2) {} ;
        \draw[dotted] (g) -- (f2);
        \draw[dotted] (f1) edge[out=60, in=120] (g);
        \draw[dotted] (f1) edge[out=-60, in=-120] (g);
        \draw[dotted] (f1) edge[out=30, in=150] (g);
        \draw[dotted] (f1) edge[out=-30, in=-150] (g);
        \path[] (f1) -- node[midway] {$\vdots$} (g);

        \draw [decorate,decoration={brace,amplitude=5pt,raise=2ex}]
             (-.5,-1)--(-.5,1) node[midway,xshift=-4em]{$\begin{array}{c}N \text{ dotted} \\ \text{lines}\end{array}$}; 
    \end{tikzpicture}
    \qquad = U(1)_{N+1}
\end{equation}

In the case $k=-N$ the planar dual does not involve any gauge nodes and is trivially gapped in the IR.
In both cases the planar Abelian dual reproduces the  generalized level-rank duality discussed in \cite{Hsin:2016blu,Radicevic:2016wqn,Closset_2023b} for $U(N)_{k,k-N}$ with $k=-1, 0$.

It would be interesting to study this duality further, in particular its possible connection with level–rank duality in the cases with $|k|>N+1$. We defer this to future work.
\\

To summarize, we find a planar Abelian dual for every SQCD theory with gauge group $U(N)$, CS level $k$, and $F$ fundamental fields. All these dualities can be obtained from the duality \eqref{eq:Dual_k0_schematic} by introducing large real mass deformations for some of the fundamental fields.
In the following sections, we analyze these real mass deformations in detail, both on the SQCD side and on the mirror side. 
We also show that the original duality \eqref{eq:Dual_k0_schematic} can be obtained from the dualities discussed in \cite{Benvenuti:2025a}.



\subsection{An Example: The Zoology of \texorpdfstring{$U(3)_k$ SQCD with $F\, \Box$}{U3 with 7}}\label{subsec: example u4}

Having established the framework for mapping real mass deformations across mirror duals, we now present a set of illustrative examples for $U(3)_k$ SQCD with $F$ flavors. These examples highlight how the planar Abelian dual realizes qualitatively different structures across the various regions of the $(k,F)$ parameter space.

We begin with the planar Abelian dual of $U(3)_0$ SQCD with $F=8$, as described by the general proposal in \eqref{eq: Dual_k0}. In standard quiver notation, the duality takes the form
\begin{equation}
\label{eq: u3_k=0_F=8}
    \includegraphics[width=.8\linewidth]{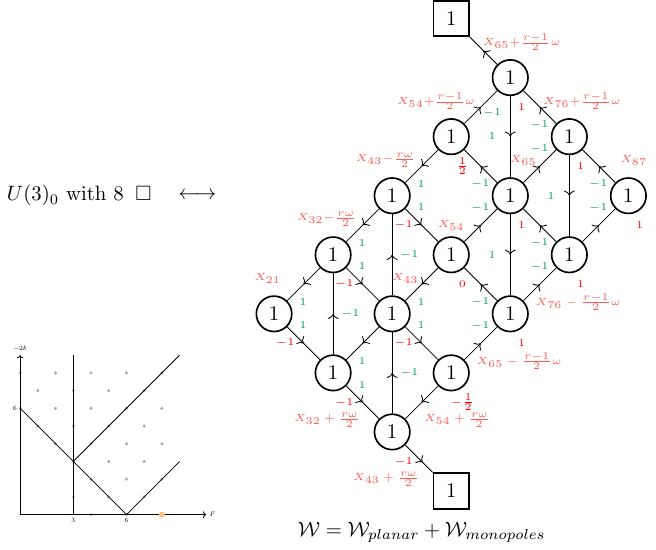}
\end{equation}
where we use the shorthand $X_{ij}:= X_i-X_j$ to parametrize the FI terms of the gauge nodes in the planar theory. In the planar theory, there are 14 mesonic superpotential terms, one associated with each face of the quiver, and a total of 8 monopole terms. These correspond to $k-1$ terms for each vertical column of gauge nodes of height $k$, as described below \eqref{eq: Dual_k0}. The CS levels are equal to 1 for all nodes except those in the central column, as indicated by the red labels in the figure. The BF couplings follow the rules described below \eqref{eq: Dual_k0}.

Starting from this duality, we can turn on a negative real mass for one flavor in the SQCD, flowing to $U(3)_{-1/2}$ with 7 flavors. This theory lies on the $\mathcal{N}=4$-descendant line of Figure \ref{fig:KF_plane}. The resulting duality for $U(3)_{-1/2}$ SQCD with 7 fundamental chiral multiplets is shown in \eqref{eq: u3_k=1_2_F=7}.

\begin{equation}
\label{eq: u3_k=1_2_F=7}
    \includegraphics[width=.85\linewidth]{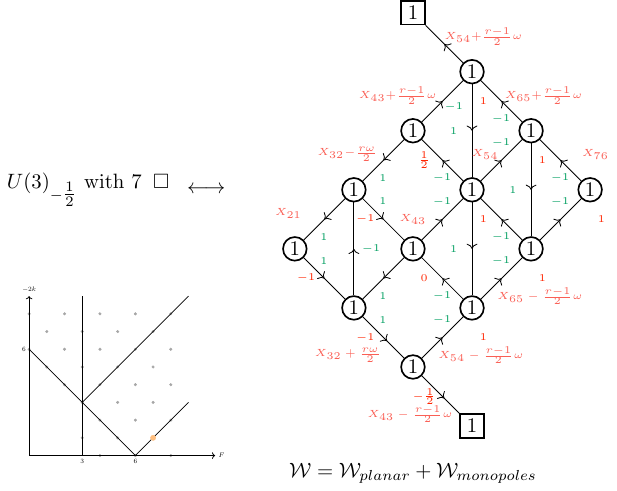}
\end{equation}

The resulting theory is shown in \eqref{eq: Dual_N4_line_schematic}. To recast it in the same form as \eqref{eq: Dual_N4line_schematic_nosquares}, we perform a sequence of Aharony dualities on the $U(1)$ gauge nodes, as shown in \eqref{eq: u3_k=1_2_F=7_dual}.

\begin{equation}
\label{eq: u3_k=1_2_F=7_dual}
    \includegraphics[width=\linewidth]{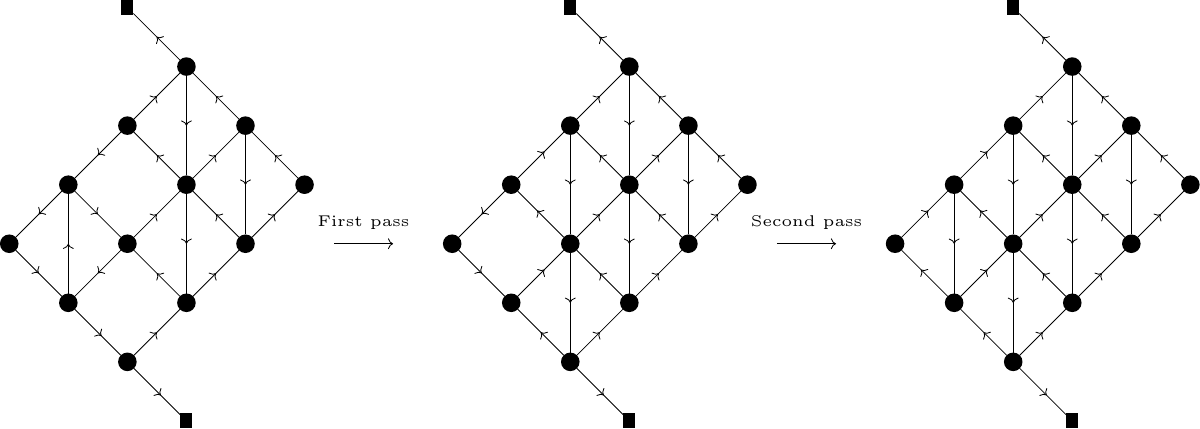}
\end{equation}
In a first pass, we apply Aharony duality to all nodes in the first vertical column, and then to those in the second column. In a second pass, we apply Aharony duality only to the left-most gauge node. After this sequence of dualizations, we recover the general proposal in \eqref{eq: Dual_N4line_schematic_nosquares}, which matches the result first proposed in \cite{Benvenuti:2025a}.

For completeness, we also present the dual of $U(3)_{-1/2}$ with 9 flavors, providing an example of a duality for SQCD in \textbf{Zone 1}. From this point onward, we will present examples using the compact quiver notation summarized in \eqref{eq:quiv:notation_compact}.
\begin{equation}
        \includegraphics[width=.7\textwidth]{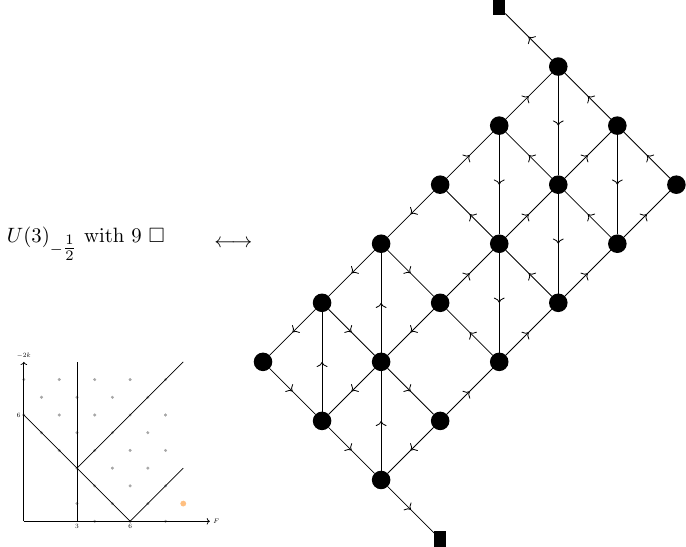}
\end{equation}

To provide further examples, we now return to the case of $U(3)_{-1/2}$ with 7 flavors and study all theories that can be reached via positive or negative real mass deformations. The resulting dualities are shown in Figures \ref{fig: zoology}, \ref{fig: zoo3}, and \ref{fig: zoology_2_half}, and exhaustively catalog the theories generated by $\pm m$ real mass deformations.

\begin{landscape}
\thispagestyle{empty}
\begin{figure}
\vspace{-2cm}
    \centering
    \includegraphics[width=.8\linewidth]{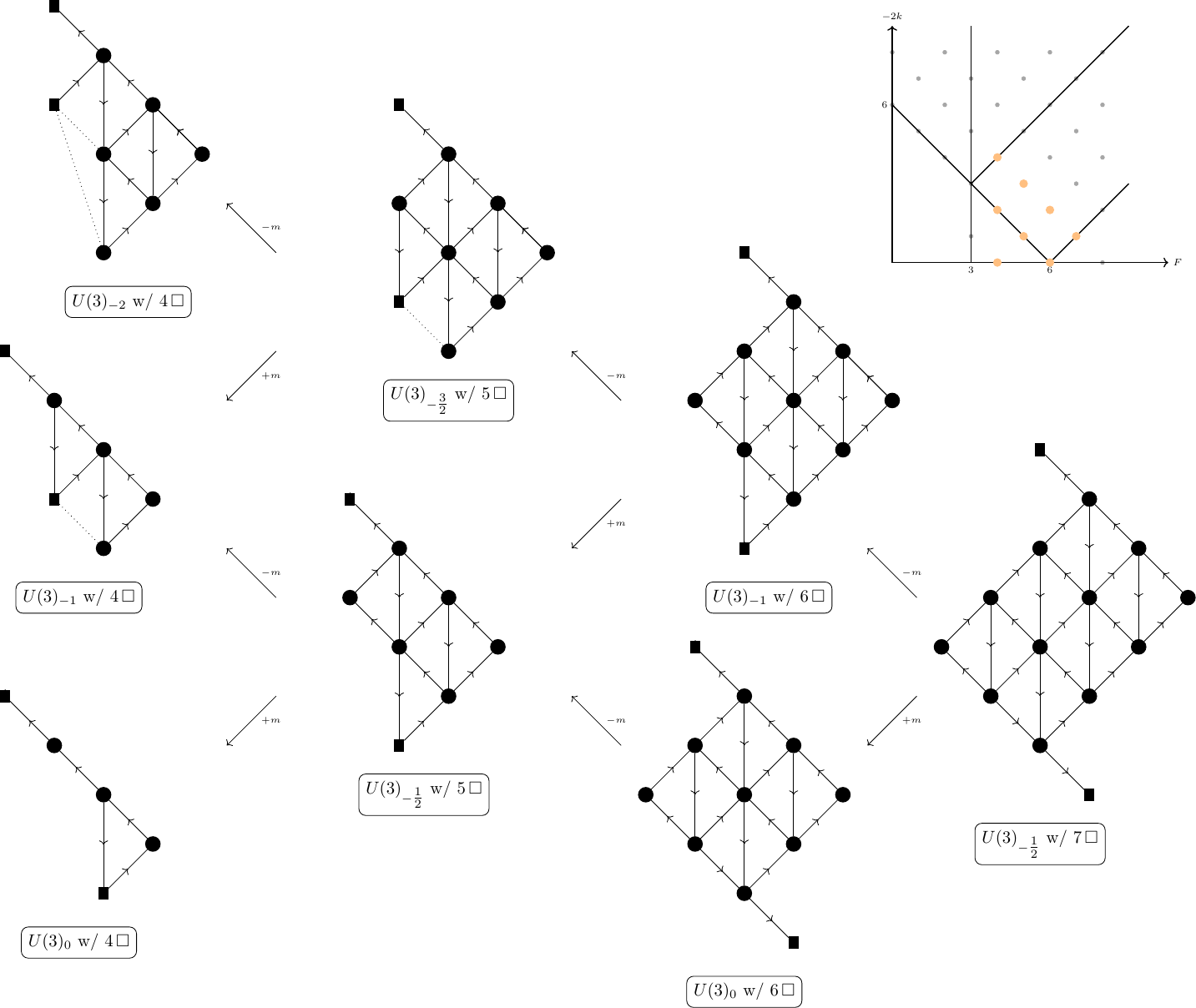}
    \caption{The planar mirror dual of $U(3)_{-1/2}$ SQCD with $7$ fundamental chiral multiplets is an $\NN=4$ descendant theory. Starting from this UV theory, we study the RG flow trajectories induced by $\pm m$ real mass deformations. In this figure, we display the first three levels of this deformation sequence. Specifically, we begin with an electric theory containing 7 fundamental chirals and follow the flow to theories with 4 fundamental chirals. 
    We also highlight the location of these theories on the $(k,F)-$plane in the top-right corner. }
    \label{fig: zoology}
\end{figure}
\end{landscape}

\thispagestyle{empty}
\begin{figure}
    \centering
    \includegraphics[width=.8\linewidth]{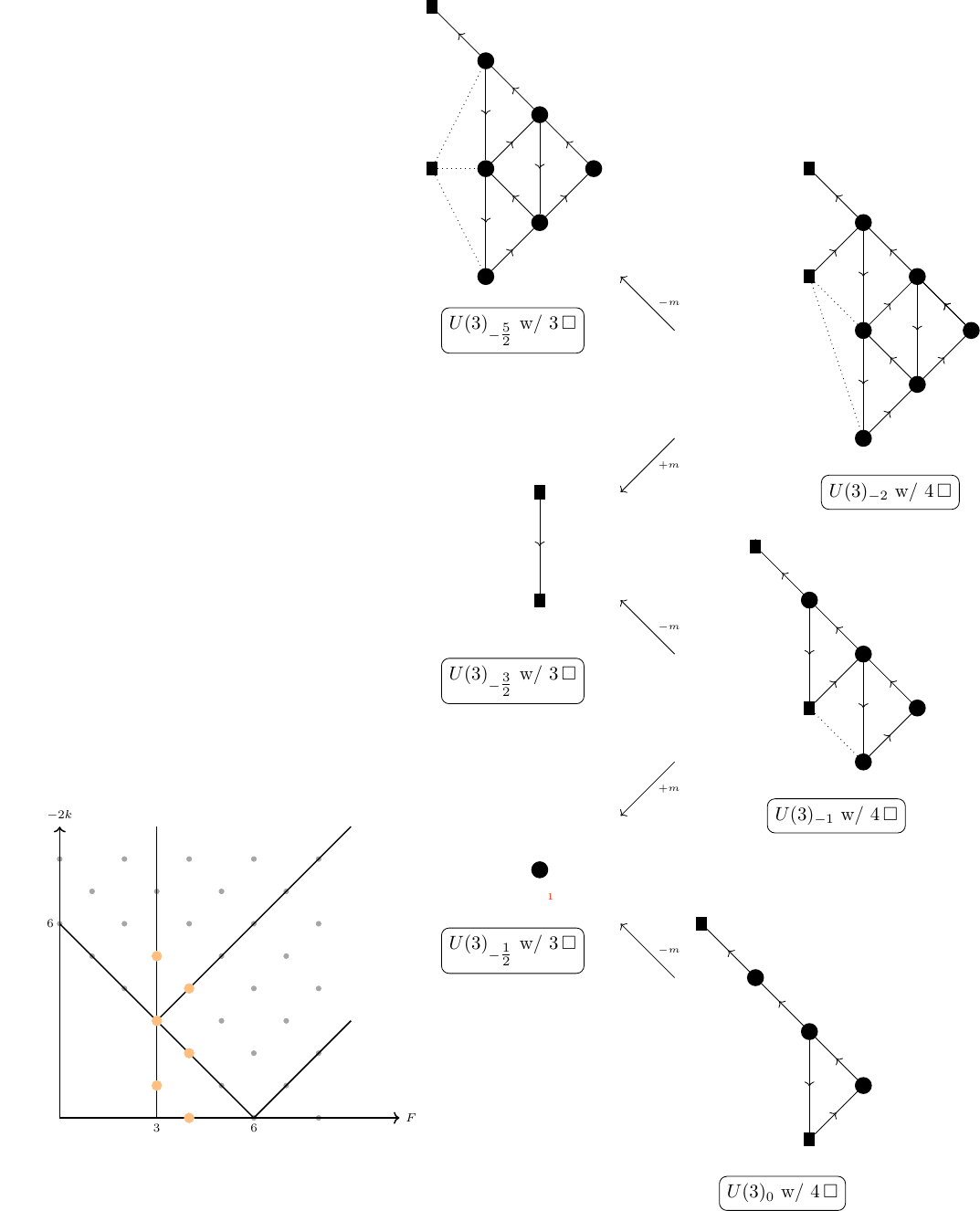}
    \caption{We continue analyzing the RG flow illustrated in Figure~\ref{fig: zoology}. Starting from $U(3)_k$ with 4 fundamental multiplets, we study a single mass deformation down to 3 fundamental multiplets. Note that the dual of $U(3)_{-1/2}$ with 3 flavors is a trivial TQFT, while that of $U(3)_{-3/2}$ with 3 flavors is a free chiral. We highlight the location of these theories in the $(k,F)-$plane in the bottom-left corner.}
    \label{fig: zoo3}
\end{figure}

\begin{landscape}
\thispagestyle{empty}
\begin{figure}
\vspace{-2cm}
    \centering
    \includegraphics[width=.8\linewidth]{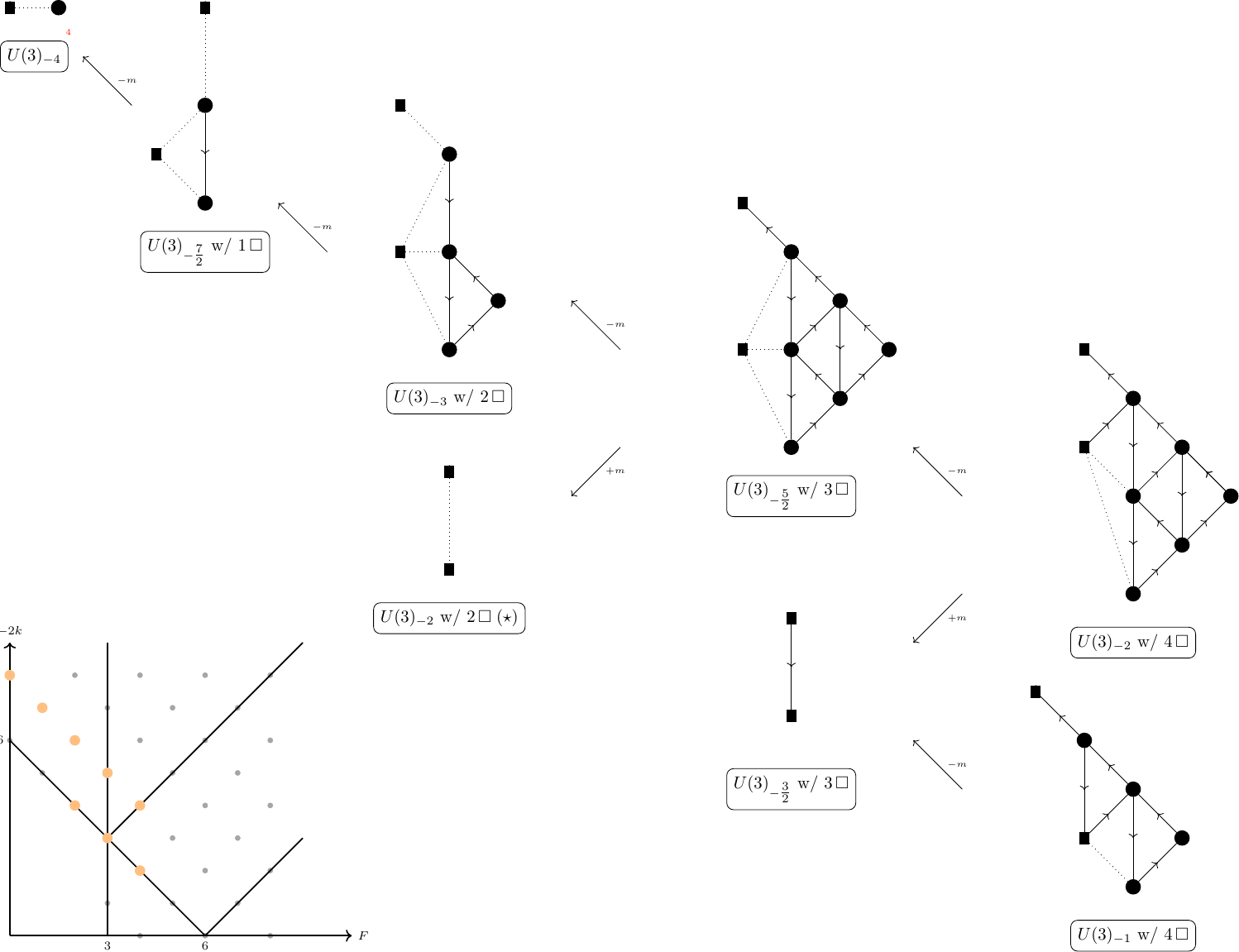}
    \caption{We continue the analysis of RG flow trajectories induced by $\pm m$ deformations, extending those considered in Figure~\ref{fig: zoo3}. We focus on the RG flow trajectories of $U(3)_{-5/2}$ and $U(3)_{-3/2}$ SQCD with 3 fundamental chirals and follow these flows until all matter is integrated out. Notice that since the dual of $U(3)_{-3/2}$ with 3 fundamentals is a single free chiral, when we perform a mass deformation, we flow to a trivial theory. The theory with the $(\star)$ label is a gapped theory; in fact, the dual theory is simply a BF coupling. We highlight the location of these theories in the $(k,F)-$plane in the bottom-left corner.}
    \label{fig: zoology_2_half}
\end{figure}
\end{landscape}

For completeness, we also present two additional examples. As a representative of \textbf{Zone 3}, we display the planar Abelian dual of $U(3)_{-3}$ with 4 fundamental chiral multiplets:
\begin{equation}
    \includegraphics[width=.7\linewidth]{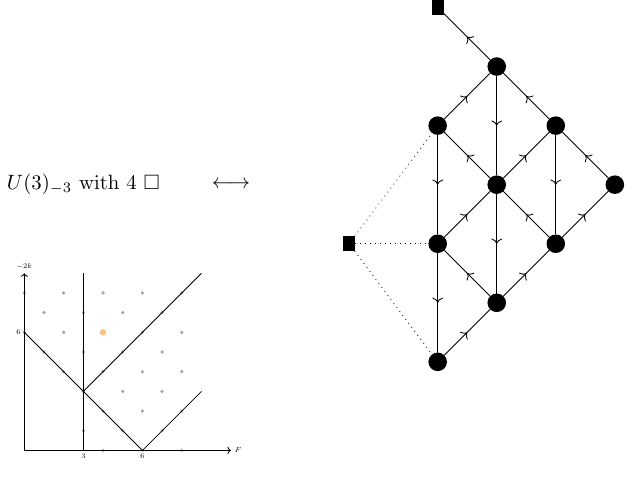}
\end{equation}

As a representative of the dualities along the vertical line $F=0$, we present the planar Abelian dual of $\mathcal{N}=2$ $U(3)_{-5}$ SYM.
\begin{equation}
    \includegraphics[width=.5\linewidth]{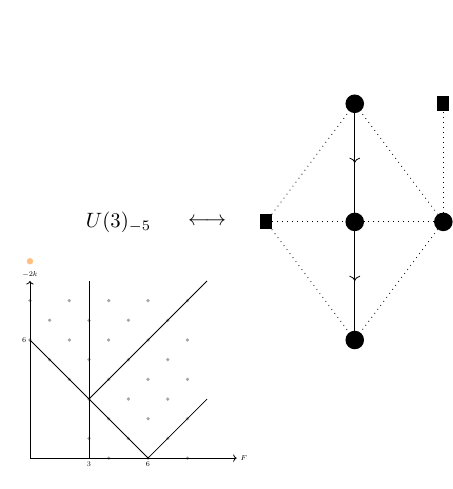}
\end{equation}

\subsection{Further Examples: CS-SQCD with Fundamentals and Anti-Fundamentals}

As mentioned at the beginning, planar Abelian duals can also be constructed for $U(N)_k$ SQCD with both $F$ fundamental and $A$ anti-fundamental chiral multiplets \cite{Benvenuti:2025a}. A natural starting point for analyzing these theories is the class of $U(N)_k$ SQCD models with $[F,F]$ flavors satisfying the constraint $2|k| = 2F - 2N$. 

The planar Abelian dual for this class can be obtained by starting from the mirror duality of the $\mathcal{N}=4$ $U(N)$ SQCD with $2F$ flavors and performing a real mass deformation that breaks supersymmetry to $\mathcal{N}=2$ while preserving the $S[U(F)\times U(F)]$ subgroup of the $SU(2F)$ flavor symmetry. Alternatively, the same dual can be derived using the algorithmic procedure introduced in \cite{Benvenuti:2025a} (see Section 5.1.1), obtained via real mass deformations from the $\NN=4$ algorithm developed in \cite{Bottini:2021vms,Hwang:2021ulb,Comi:2022aqo}.

Starting from these theories, one can explore the full three-dimensional parameter space $(k,F,A)$ by turning on real mass deformations and tracking their effects in the planar Abelian dual. These deformations can be analyzed using the techniques developed in this work and follow the same general logic as in the case of $U(N)_k$ SQCD with $[F,0]$ matter content.

We will not provide a detailed analysis of these deformations, but instead present an illustrative example given by $U(2)_k$ SQCD. We consider $U(2)_{-1}$ SQCD with $[3,3]$ chiral multiplets (described in Section 5.1.1 of \cite{Benvenuti:2025a}) and turn on positive/negative real mass deformations to explore theories with $4 \leq F + A \leq 6$. The resulting set of dualities is collected in Figure \ref{fig:u2w33flows}.

For some of the theories in Figure \ref{fig:u2w33flows}, the planar Abelian dual can also be derived starting from the mirror duality of the $\mathcal{N}=4$ $U(2)$ SQCD with an appropriate number of flavors, possibly supplemented by real mass deformations following the strategy of \cite{Benvenuti:2025a}. This applies, for example, to $U(2)_{-1}$ with $[3,3]$, as already mentioned, as well as to $U(2)_{-1/2}$ with $[2,3]$, $U(2)_0$ with $[2,2]$, and $U(2)_0$ with $[1,3]$.

We emphasize that all these dualities can be viewed, in a broad sense, as instances of mirror symmetry. A key feature is that the planar theories exhibit a single $U(1)$ flavor symmetry together with a $U(1)^{F+A-1}$ topological symmetry, which enhances in the infrared to $S[U(F)\times U(A)]$. This structure arises from the presence of both mesonic and monopole superpotentials: the former reduces the flavor symmetry to a single $U(1)$, while the latter ensures that each vertical column of gauge nodes contributes only one unbroken topological symmetry. The superpotential, and more generally the structure of the theory, follows the same rules described below \eqref{eq: Dual_k0}. 

In the first theory shown at the top of Figure \ref{fig:u2w33flows}, the topological symmetries enhancing to the $SU(F)$ and $SU(A)$ factors originate from the columns to the left and right of the central column, respectively. As fundamentals or anti-fundamentals are integrated out, the number of such columns on the left/right-hand side decreases accordingly. Moreover, the planar Abelian theory admits a single mesonic gauge-invariant operator only when $2|k| = |F-A|$. Since this operator is expected to map to a gauge-invariant monopole parametrizing a $\mathbb C$-moduli space of vacua in the SQCD theory, this observation allows one to predict when such operators are present in the non-Abelian theory.

These examples illustrate how real mass deformations can, in principle, be used to construct the planar Abelian dual of a general $U(N)_k$ SQCD with $[F,A]$ flavors. A complete classification of these theories, however, is more intricate and is left for future work.

\begin{landscape}
\thispagestyle{empty}
\begin{figure}
\vspace{-2cm}
    \centering
    \includegraphics[width=1.1\linewidth]{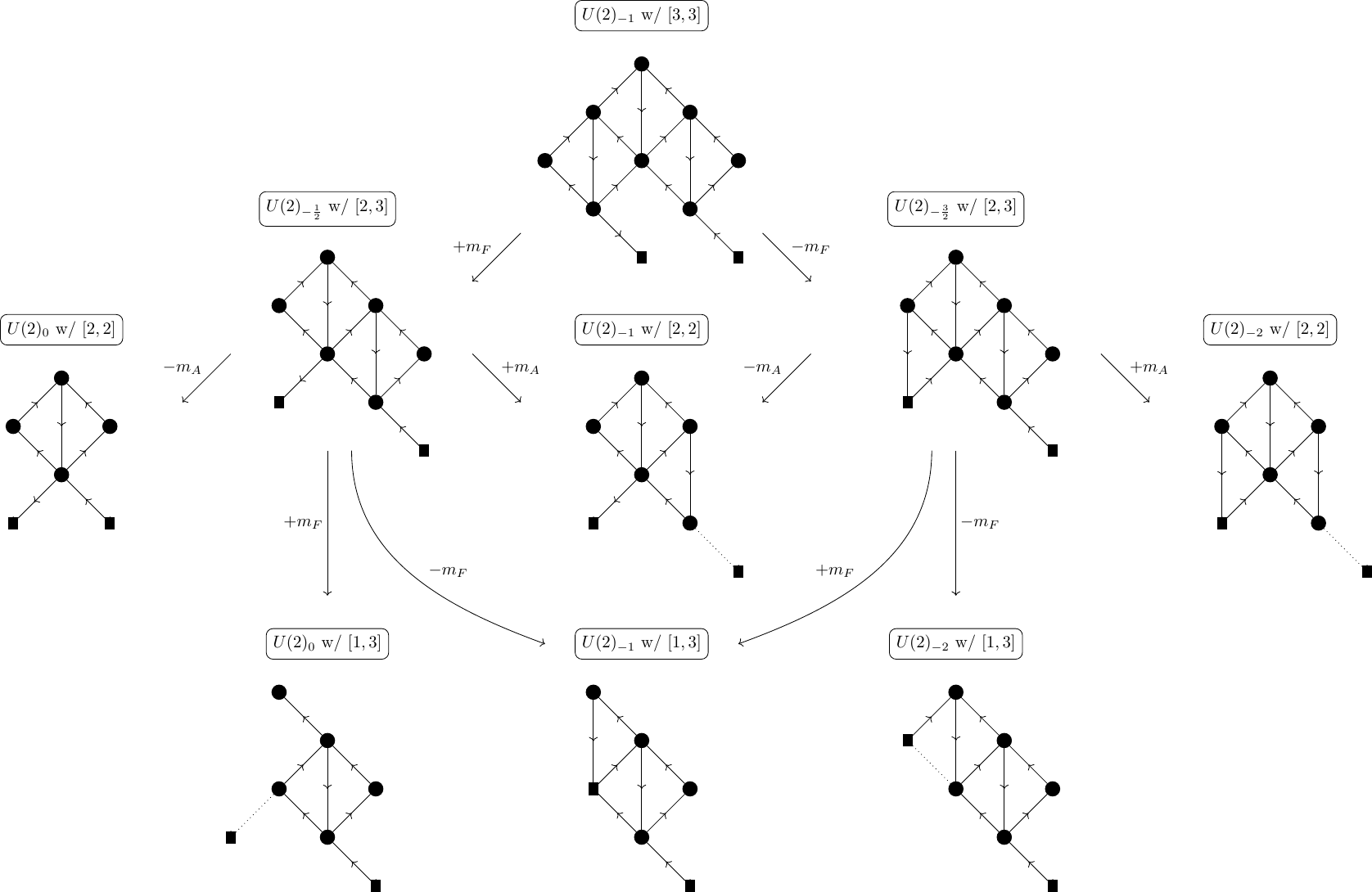}
    \caption{Examples of planar Abelian duals for $U(2)_k$ SQCD with matter content $[3,3]$, $[2,3]$, $[2,2]$, and $[1,3]$, for various values of $k$ are shown here. All dualities are obtained by systematically tracking real mass deformations in the planar Abelian theory, following the same strategy outlined in Section \ref{sec: landscape} for the case of $[F,0]$ flavors. A positive (negative) real mass for a fundamental chiral multiplet shifts the CS level by $+1/2$ ($-1/2$), while a positive (negative) real mass for an anti-fundamental shifts it by $-1/2$ ($+1/2$). The compact quiver notation is summarized in \eqref{eq:quiv:notation_compact}. Further details of the theories, including the superpotential and R-charge assignments, follow the same rules described below \eqref{eq: Dual_k0}.}
    \label{fig:u2w33flows}
\end{figure}
\end{landscape}






\subsection{An \texorpdfstring{$\mathcal{N}=4$}{N=4} Surprise}

From the analysis carried out in this section, we uncover that $U(N)_{(-N/2,-3N/2)}$ SQCD with $N$ fundamental chirals exhibits an accidental enhancement to $\mathcal{N}=4$ supersymmetry. The argument proceeds through a chain of dualities generated by gauging $U(1)$ global symmetries, which we now describe.

We begin with $U(N)_{-N/2}$ SQCD with $F=N$ flavors, which we argued is dual to a single free chiral multiplet, as in \eqref{eq: freechir_dual}. In this dual frame, the $SU(F)$ global symmetry is unfaithful, as no operator transforms under it; more precisely, it only appears through a background Chern--Simons interaction. The only faithful global symmetry is therefore the $U(1)$ topological symmetry, which acts on the free chiral.

We can then gauge this symmetry, taking into account the background Chern--Simons couplings on both sides of the duality. This leads to
\begin{equation}
    SU(N)_{-N/2} \,\, \text{with $F$ $\Box$}
    \quad \longleftrightarrow \quad
    U(1)_{1/2} \,\, \text{with one charge $-1$ chiral} \, .
\end{equation}
On the left-hand side, there is no background Chern--Simons term, so gauging the topological symmetry effectively ungauges the overall $U(1)$ factor.

Starting from this duality, we next gauge the baryonic symmetry on the left-hand side, including an appropriate Chern--Simons level, thereby obtaining a $U(N)_{(-N/2,-3N/2)}$ theory. On the right-hand side, this corresponds to gauging the topological symmetry. Performing this operation carefully yields
\begin{equation}
    U(N)_{(-N/2,-3N/2)} \,\, \text{with $F$ $\Box$}
    \quad \longleftrightarrow \quad
    \begin{array}{c}
        U(1) \times U(1) \,\, \text{with one chiral of charge $(-1,0)$} \\
        \text{and } k=\begin{pmatrix}
            1/2 & -2 \\
            -2 & -1
        \end{pmatrix}
    \end{array} \, .
\end{equation}
We can integrate out the $U(1)_{-1}$ topological sector on the right-hand side, which shifts the effective Chern--Simons level of the remaining $U(1)$ gauge field. This leads to
\begin{equation}\label{eq:N=4enhancementdual}
    U(N)_{(-N/2,-3N/2)} \,\, \text{with $F$ $\Box$}
    \quad \longleftrightarrow \quad
    U(1)_{3/2} \,\, \text{with one charge $-1$ chiral} \, .
\end{equation}

This duality is consistent with a limiting case of Aharony-like dualities (see \cite{Closset_2023a}), and for $N=1$ it reduces to a trivial identity between theories related by time-reversal.

It was observed in \cite{Gang:2018huc} that the theory on the right-hand side exhibits an enhancement to $\mathcal{N}=4$ supersymmetry, with trivial Higgs and Coulomb branches. By virtue of the duality \eqref{eq:N=4enhancementdual}, the theory $U(N)_{(-N/2,-3N/2)}$ with $N$ fundamental chirals must therefore also exhibit $\mathcal{N}=4$ supersymmetry. This enhancement can be further confirmed, for instance, by analyzing the superconformal index using the methods of \cite{Cordova:2016emh,Razamat:2016gzx,Evtikhiev:2017heo}.

\section{Mass Deformations in \texorpdfstring{CS-SQCD}{CS-SQCD} with Fundamental Fields}
\label{sec: mass deformations}
In this section, we study mass deformations in $\NN=2$ $U(N)$ SQCD with CS levels $(k, k+\ell N)$ and $F$ fundamental chiral fields. We introduce a large real mass deformation for one of the fundamentals, implemented through a real mass associated with a $U(1)$ subgroup in the Cartan of the $SU(F)$ flavor symmetry. 
Our goal is to describe the four mass deformations in \eqref{eq:realmass_arrows_schematic}: positive and negative real masses for one of the fundamentals and positive and negative real masses where the gauge group is also Higgsed from $U(N)$ to $U(N-1)$.

We parameterize the real masses associated to the Cartan of $SU(F)$ as $X_i$, $i=1,\dots,F$ subject to the following constraint:
\begin{equation}
	\sum_{i=1}^{F} X_i = 0
\end{equation}
In the presence of these real mass deformations as well as an FI term $\eta$, the equations of motion are \cite{Intriligator_2013}:
\begin{equation}	\label{eq:eom_chiral_UN}
(\sigma_\alpha - X_i ) Q_{\alpha}^i = 0 
,\qquad \alpha = 1,\dots , N, \; i=1,\dots, F
\end{equation}

\begin{equation}	\label{eq:Dterm_chiral_UN}
\sum_{i=1}^F Q_\alpha^{i \, \dagger} Q_i^\beta=
\frac{\delta_{\alpha \beta}}{2 \pi}
\left[
-\eta + k \sigma_\alpha + \ell \sum_{\gamma=1}^N {\sigma}_\gamma+\frac{1}{2} \sum_{i=1}^F\left|{\sigma}_\alpha-{X}_i\right|
\right] 
,\qquad \alpha,\beta = 1,\dots,N
\end{equation}
where $Q_\alpha^i$ are the quark VEVs, and we use gauge freedom to rotate the VEVs for the scalar $\sigma$ in the $\NN=2$ vector multiplet to a diagonal matrix. 

\paragraph{Mass Flows without Higgsing:}
We first turn on a real mass deformation parametrized by the shift:
\begin{equation}	\label{eq:shifts_mass_no_higgs_QCD}
\left\{
\begin{array}{ll}
X_1 \to X_1 \mp (F-1) m, &
\\
X_i \to X_i \pm m, & \qquad 2\leq i \leq F,
\\
\eta \to \eta + \left( \frac{F}{2} \pm( k + \ell N ) \right )m.
\end{array}
\right.
\end{equation}
for large $m$.
Since the quarks transform in the antifundamental representation of the $SU(F)$ flavor group, these shifts correspond to a positive mass (upper sign) and a negative mass (lower sign) for the first quark, respectively.
We consider a vacuum wherein the gauge group is not Higgsed and the last $F-1$ quarks remain massless in the limit $m\to\infty$. This is achieved by giving a large VEV to the real scalars $\sigma$ in the $\NN=2$ vector multiplet:
\begin{equation}	\label{eq:vacuum_mass_no_higgs_QCD}
\sigma \to\pm  \text{diag}(m,m, \dots, m) + \sigma.
\end{equation}
The effective mass for the quarks $Q_\alpha^i$ with $i\geq2$, given by $(\sigma_\alpha - X_i)$ vanish at leading order in $m$, while the effective mass for the first quark $Q_\alpha^1$ is:
\begin{equation}
m_1 = \pm m F + \mathcal{O} (m^0)
\end{equation}
rendering this quark massive in the $m\to \infty$ limit and, hence, can be integrate it out. Up to $\mathcal{O}(m^0)$ terms, the equations of motion \eqref{eq:eom_chiral_UN} reduce to the equations of motion for $U(N)$ with $(F-1)$ fundamentals and CS levels $k \pm \tfrac{1}{2}, k \pm \tfrac{1}{2} + \ell N$. 
The real mass deformations \eqref{eq:shifts_mass_no_higgs_QCD}, together with the choice of vacuum \eqref{eq:vacuum_mass_no_higgs_QCD}, produce the following RG flows:
\begin{equation}
\begin{tikzpicture}[baseline=(current bounding box).center]
\node (start) at (0,0) {$U(N)_{k,k+N\ell} \with F \; \square$ };
\node (mp) at (-2,-2) {$U(N)_{k+\tfrac{1}{2},k+\tfrac{1}{2}+N\ell} \with F - 1 \; \square$ };
\node (mm) at (-2,2) {$U(N)_{k-\tfrac{1}{2},k-\tfrac{1}{2}+N\ell} \with F - 1 \; \square$ };

\draw[->] (start) -- node[midway, left] {$+m$} (mp);
\draw[->] (start) -- node[midway, left] {$-m$} (mm);
\end{tikzpicture}
\end{equation}

In the rest of this paper, we will compactly denote these RG flows as “$+m$" and “$-m$", respectively.

\paragraph{Mass Flows with Higgsing:}
We now consider real mass deformations that simultaneously give mass to one quark and Higgs the gauge group to $U(N-1)$. We consider the following shifts of real mass parameters for large $\mH$:
\begin{equation}	\label{eq:shifts_mass_higgs_QCD}
\left\{
\begin{array}{ll}
X_1 \to X_1 \mp (F-1) \mH, &
\\
X_i \to X_i \pm \mH, & \qquad 2\leq i \leq F
\\
\eta \to \eta + \left( \frac{F}{2} \pm ( k + \ell N - \ell F ) \right ) \mH.
\end{array}
\right.
\end{equation}
and we combine it with a large VEV for the real scalars $\sigma$:
\begin{equation}
\sigma \to 
\pm
\left(
\begin{array}{cc}
-(F-1)  & 0 \\
0 &  \mathbf{1}_{N-1 \times N-1}
\end{array}
\right) \mH
+ \sigma
\end{equation}
which Higgses the gauge group to $U(N-1)\times U(1)$. 
The fundamentals $Q_\alpha^i$ with $\alpha \geq 2$ and $i\geq 2$ remain massless and form $F-1$ fundamentals for the $U(N-1)$ factor, while the component $Q_1^1$ is massless as well and carries a charge $+1$ under the $U(1)$ factor. The other quarks are massive and are integrated out. 
The D-terms for the $U(N-1)$ sector are satisfied up to $\mathcal{O}(\mH^0)$ order and reduce to the equations for $U(N-1)$ with $F-1$ fundamentals. The D-term for the $U(1)$ gauge factor, namely \eqref{eq:Dterm_chiral_UN} with $\alpha=\beta=1$, read:
\begin{equation}
\begin{split}
2\pi |Q_1^1|^2 =& - \eta - \left( \frac{F}{2} \pm ( k + \ell N - \ell F ) \right ) \mH + k \sigma_1 \mp k(F-1) \mH 
\\&
+ \ell \sum_{\gamma=1}^{N} \sigma_\gamma \pm \ell ( -F  + N  )\mH
+ \frac{1}{2} \sum_{i=2}^F |\sigma_1 - X_i \pm ( -F )\mH |
+\frac{1}{2}|\sigma_1-X_1|
\\ =&
\left(
\frac{1}{2} F(F-1) - \frac{1}{2} F
\pm (-  k- \ell N + \ell F - k(F - 1) + \ell(N-F) ) 
\right)\mH + \mathcal{O}(\mH^0)
\\ = &
\left(
\frac{1}{2} F(F-2)
\mp k F
\right)\mH + \mathcal{O}(\mH^0)
\end{split}
\end{equation}
This can be solved by giving a non-zero VEV to $Q_1^1$ such that\footnote{When $F-1\mp k=0$ the D-terms are satisfied with vanishing VEV for the chirals and $Q_1^1$ remains a massless d.o.f. charged under the $U(1)$ factor of the gauge group. This is the case when the SQCD obtained after the mass deformation is marginally chiral. We discuss this case separately in Appendix \ref{app: mass flow S3b}.}:
\begin{equation}
|Q_1^1|^2 = \frac{\mH}{2\pi} \left(
\frac{1}{2} F(F-2)
\mp k F
\right) \geq 0
\end{equation}

\begin{figure}[h!] 
\centering 
\includegraphics[width=0.7\linewidth]{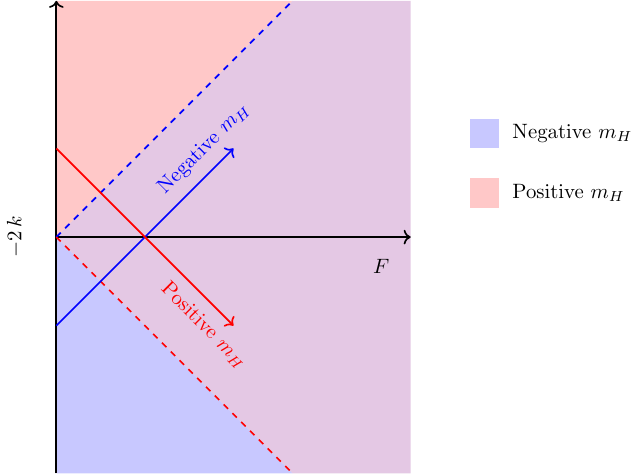} 
\caption{RG-flow trajectories induced by real mass deformations accompanied with Higgsing in the $(F,\,k)$-plane.} \label{fig:range_mass_higgs} \end{figure}

Positivity of this expression determines the range of validity for the two possible deformations: 
\begin{equation} 
    \begin{split} \mH: &\quad F \geq 2(k+1) \, ,\\ -\mH: &\quad F \geq 2(-k+1) \, . 
    \end{split} 
\end{equation} 
With a slight abuse of notation, we can depict the movement in the parameter space induced by the Higgsing deformations as illustrated in Figure \ref{fig:range_mass_higgs}. Indeed, since the rank $N$ changes, the deformation actually moves across two planes with different fixed $N$. 

When the deformation is allowed, the \(U(1)\) factor is fully Higgsed, and the resulting theory is \(U(N-1)\) SQCD with \(F-1\) fundamentals and CS levels \((k \pm \tfrac{1}{2},\, k \pm \tfrac{1}{2} + \ell (N-1))\). Together with the previous non-Higgsing deformations, we obtain four distinct flows from \(U(N)\) SQCD with \(F\) fundamentals to theories with \(F-1\) fundamentals: 

\begin{equation}    \label{eq:defcross_QCD}
\begin{tikzpicture}[baseline=(current bounding box).center]
\node (start) at (0,0) {$U(N)_{k,k+N\ell} \with F \; \square$ };
\node (mp) at (-3.5,-3) {$U(N)_{k+\tfrac{1}{2},k+\tfrac{1}{2}+N\ell} \with F - 1 \; \square$ };
\node (mm) at (-3.5,3) {$U(N)_{k-\tfrac{1}{2},k-\tfrac{1}{2}+N\ell} \with F - 1 \; \square$ };

\node (mpH) at (3.5,-3) {$U(N-1)_{k+\tfrac{1}{2},k+\tfrac{1}{2}+(N-1)\ell} \with F - 1 \; \square$ };
\node (mmH) at (3.5,3) {$U(N-1)_{k-\tfrac{1}{2},k-\tfrac{1}{2}+(N-1)\ell} \with F - 1 \; \square$ };

\draw[->] (start) -- node[midway, left] {$+m$} (mp);
\draw[->] (start) -- node[midway, left] {$-m$} (mm);

\draw[->] (start) -- node[midway, right] {$+\mH$} (mpH);
\draw[->] (start) -- node[midway, right] {$-\mH$} (mmH);

\end{tikzpicture}
\end{equation}

\subsection{Mapping Mass Flows across Aharony-like Dualities}\label{subsec: mass flows Aharony}

We remind the reader that SQCD with $U(N)$ gauge group and CS levels $k,k+\ell N$ and $F$ fundamentals is mapped by Aharony-like duality to $U(\tilde{N})$ SQCD with CS levels $-k,-k+ \ell \tilde{N}$ and $F$ antifundamentals \cite{Aharony:1997gp, Benini_2011a} (see also \cite{Giveon:2008zn, Aharony:2014uya, Closset_2023a} for a comprehensive review). 

For simplicity, we focus on the case $\ell=0$, corresponding to 
vanishing CS level for the diagonal $U(1)$ group.
The general case with $\ell \neq 0$ can be obtained by applying Witten’s $SL(2,\mathbb{Z})$ action on both sides of the duality \cite{witten2003sl2zactionthreedimensionalconformal}. 

The dual rank $\widetilde{N}$ depends on the relative values of $F$ and $k$, dividing the parameter space into three regimes — minimally chiral, marginally chiral, and maximally chiral — characterized as follows:
\begin{equation}
\begin{array}{rl}
\text{minimally chiral: }F < |2k|, & \qquad \Nt= \frac{F}{2} + |k| - N
\\
\text{marginally chiral: }F = |2k|, & \qquad\Nt=\frac{F}{2}  + |k| - N = F - N
\\
\text{maximally chiral: }F > |2k|, & \qquad\Nt=F - N
\end{array}
\end{equation}
In the marginally chiral case there is also an additional gauge singlet in the magnetic theory that flips a monopole:
\begin{equation}
\mathcal{W}^{\tfrac{F}{2}=|k|}_{mag} = T \mathfrak{M}^{\mp}
\end{equation}
where the upper (lower) sign correspond to $k>0$ ($k<0$). 
\\

One can flow between different Aharony-like dualities by consistently turning on real masses on both sides of the duality. We outline the mapping of mass deformations $\pm m$ and $\pm \mH$ across Aharony-like duality; a detailed derivation is provided in Appendix \ref{app: mass flow S3b}.

In the \textit{maximally chiral case}, all four deformations in \eqref{eq:defcross_QCD} are available, as can be checked from Figure \ref{fig:range_mass_higgs}. 
Aharony duality maps positive real masses to negative real masses. Consistency with the value of $\tilde{N}$ implies that mass deformations are mapped as follows:
\begin{equation}    \label{eq:mass_map_aharony_max_chiral}
\begin{cases}
+m \\ - m \\ +\mH \\ - \mH
\end{cases}
\xleftrightarrow[\text{maximally chiral}]{\text{Aharony}}
\begin{cases}
-\widetilde{\mH} \\ + \widetilde{\mH} \\ -\widetilde{m} \\ + \widetilde{m}
\end{cases}
\end{equation}
where $\pm\widetilde{m}$ and $\pm\widetilde{\mH}$ denote mass deformations with or without Higgsing for $U(N)$ SQCD with only antifundamentals, analogous to the deformations discussed above for SQCD with only fundamentals.
\\

In the \textit{minimally} and \textit{marginally chiral} cases, only three of the four deformations are available. The two non-Higgsing deformations, $+m$ and $-m$, are always accessible, whereas only one of the two Higgsing deformations ($\pm\mH$) is allowed, depending on the sign of the Chern–Simons level $k$. For concreteness, we set $k<0$, where the electric theory admits the $\mH$ deformation. The Aharony dual theory has CS level $-k>0$ and one can check that the deformation $-\widetilde{\mH}$ is available. 
The mapping of mass deformations is given by:
\begin{equation}
k<0: \qquad
\begin{cases}
+m \\ - m \\  +\mH
\end{cases}
\xleftrightarrow[\text{marginally or minimally chiral}]{\text{Aharony}}
\begin{cases}
-\widetilde{\mH} \\ + \widetilde{m} \\ -\widetilde{m}
\end{cases}
\end{equation}
similarly for $k>0$:
\begin{equation}
k>0: \qquad
\begin{cases}
+m \\ - m \\  -\mH
\end{cases}
\xleftrightarrow[\text{marginally or minimally chiral}]{\text{Aharony}}
\begin{cases}
-m \\ - \widetilde{\mH} \\ +m
\end{cases}
\end{equation}

\section{Mapping Mass Flows to the Planar Mirror Duals}
\label{sec: planar abelian dual of mass}

In this section, we analyze how real-mass deformations of $\mathcal{N}=2$ SQCD are realized on their planar Abelian duals. We do this by comparing the equations of motion on the electric and mirror sides in the large-mass limit. We then focus on a representative example to elaborate on some features of our proposal.

\subsection{Deformations of the \texorpdfstring{$\NN=4$}{N=4}-descendant SQCD}
\label{subsec: deformations of N4 sqcd}
As a starting point, we consider the planar Abelian dual of $U(N)_{N-\frac{F}{2}}$ SQCD with $F$ fundamental fields, shown in Figure \ref{eq: Dual_N4line_schematic_nosquares} (see Section \ref{subsec: negative mass deformations in mirror}). We remind the reader that this theory corresponds to the $\NN=4$-descendant line in Figure \ref{fig:KF_plane}. 

These theories lie in the maximally chiral regime, where all four deformations $\pm m$ and $\pm \mH$ are available. Across the duality, a real mass for a quark in SQCD maps to an FI term for the gauge nodes contained in a single column of the planar Abelian quiver, while the FI parameter of the SQCD maps to a real mass for one of the two flavor nodes of the Abelian theory.

Consequently, giving a real mass to the $i$-th quark corresponds to turning on a large FI parameter for one (or two) adjacent columns of the Abelian quiver. Although different choices of $i$ are equivalent at the level of the IR-SCFT—being related by Weyl transformations of the $SU(F)$ flavor symmetry—this symmetry is not manifest in the planar quiver representation. As a result, the UV analysis of the deformation depends on the specific choice of $i$, even though the corresponding IR deformations describe the same flow.

For concreteness, we adopt the following conventions for the four possible deformations:
\begin{equation}
\begin{array}{ll}
+m, +\mH: & \text{ large mass for the last quark } Q^F \leftrightarrow X_F \to -\infty
\\
 -\mH,-m: & \text{ large mass for the first quark } Q^1 \leftrightarrow  X_1 \to +\infty
\end{array}
\end{equation}
Where the limits indicate that for the deformations $+m, +\mH$, we have to correspondly take, at the level of the $\mathbf{S}^3_b$ partition function, a large FI limit for the rightmost node in the quiver, while the deformations $-m, -\mH$ correspond to a large FI limit for the leftmost node.
\newpage
\vspace{-1cm}
\begin{figure}[H]
    \centering
    \includegraphics[width=1\linewidth]{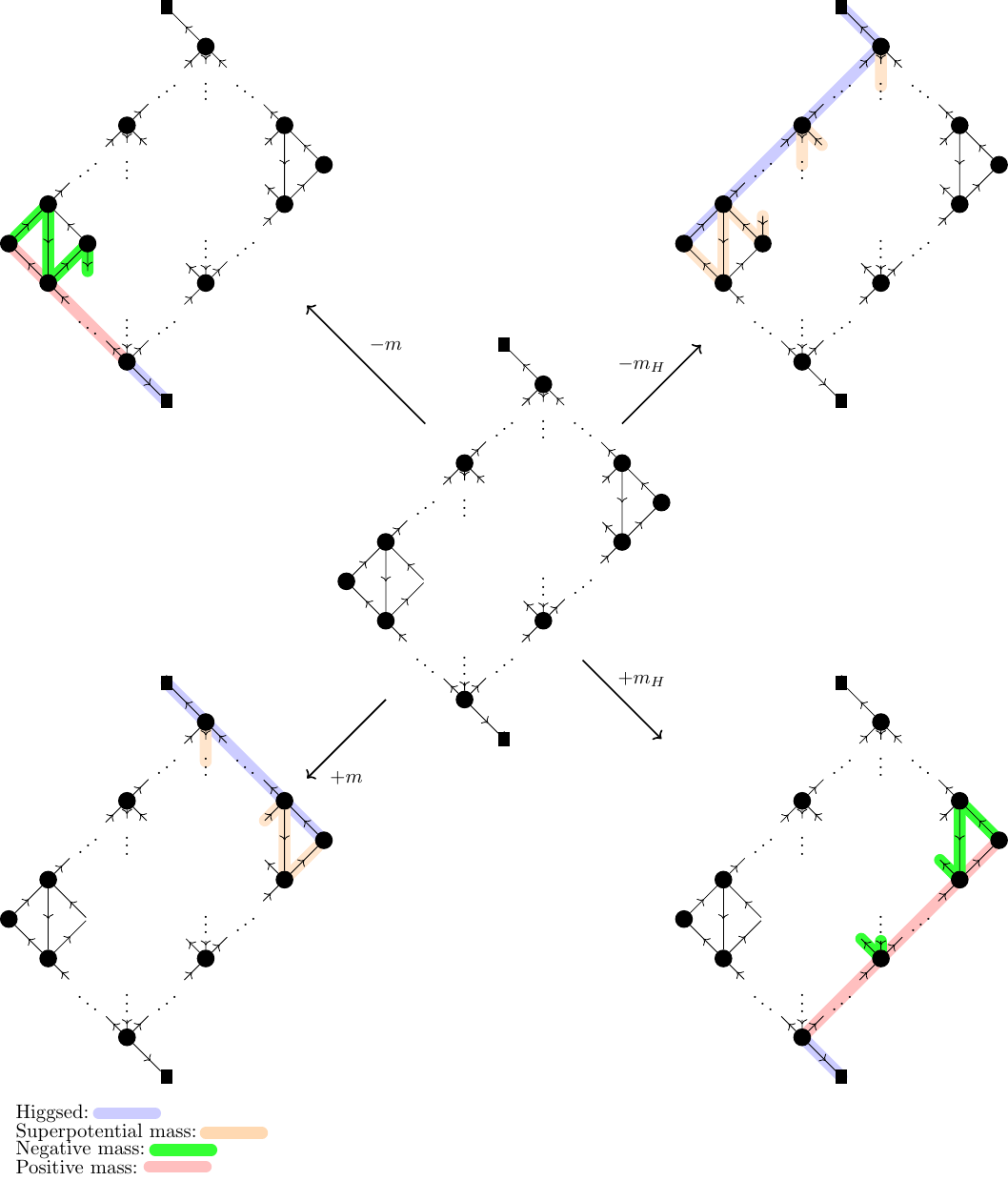}
    \caption{Deformations of the planar Abelian quiver dual to $U(N)_{(N-\tfrac{F}{2})}$ SQCD with $F$ fundamentals.
    Chiral fields highlighted in pink acquire a large positive mass, fields highlighted in green acquire a large negative mass, and fields highlighted in blue acquire a VEV, thereby Higgsing the nodes that they connect to a diagonal $U(1)$. After the Higgsing the fields in orange acquire a complex superpotential mass. The quivers resulting after integrating out the massive fields and performing the Higgsing are shown in Figure \ref{fig:def_planar_final}.
    }
    \label{fig:def_planar_start}
\end{figure}

We propose the following vacua for the four mass deformations:
\begin{itemize}
\item[$+m$: ]  In this deformation, the chiral fields along the top-right diagonal and the chiral field attached to the topmost gauge node acquire non-zero VEVs. These fields are highlighted in blue in the bottom-left panel of Figure \ref{fig:def_planar_start}. The Higgsing identifies all gauge nodes along the top-right diagonal with the adjacent flavor node. Consequently, the planar superpotential reduces to a set of mass terms, allowing one to integrate out the $2(N-1)$ chiral fields highlighted in orange in Figure \ref{fig:def_planar_start}, following the same procedure outlined in \eqref{eq: Higgsing_diag_separate}.

\item[$-m$: ] In this deformation, the real scalars in the vector multiplets associated with the gauge groups along the bottom-left diagonal acquire large VEVs:
	\begin{equation}
		(\sigma_1^{bottom}, \sigma_2^{bottom}, \dots, \sigma_{N-1}^{bottom}) \propto - ((N-1)F m , (N-2)Fm, \dots, Fm )
	\end{equation}
    where $\sigma_i^{bottom}$ is the real scalar of the gauge group associated to the bottom-most node in the $i$-th column.
	As a consequence, the chiral multiplets connecting these nodes acquire large positive real masses, while those connecting them to the rest of the quiver acquire negative real masses. These fields are highlighted in green in Figure \ref{fig:def_planar_start}. Moreover, the chiral field attached to the bottom flavor node is Higgsed.
    
    Integrating out the massive fields yields the quiver shown below:
    \begin{equation}\label{eq:Mm_general_raw}
        \centering
        \includegraphics[width=.9\linewidth]{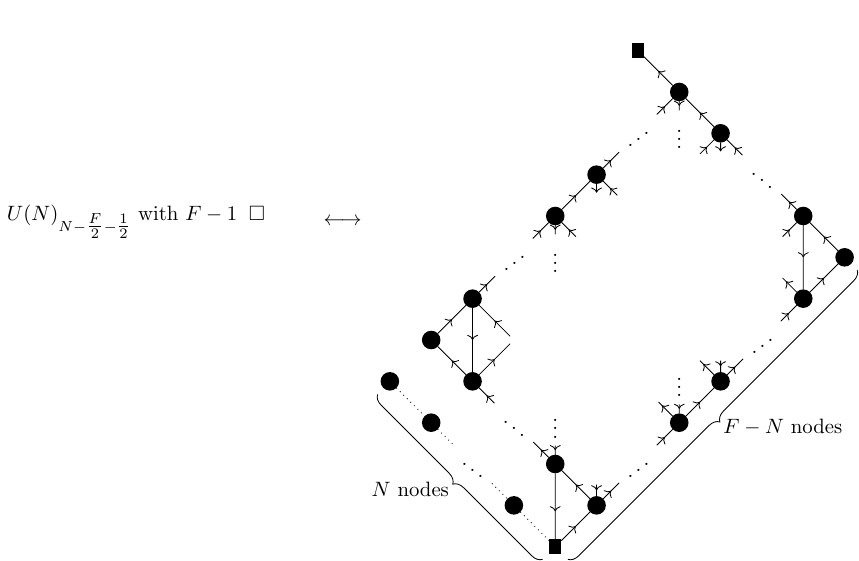}
    \end{equation}
    This intermediate quiver contains a tail of $U(1)$ gauge nodes that interact with the rest of the quiver only through mixed CS couplings. Such a tail can be systematically contracted by sequentially integrating out the last node, which corresponds to an almost trivial $U(1)_1$ Chern–Simons theory. 

    We remind the reader that pure $U(1)_{\pm1}$ Yang-Mills theory is almost trivial and its path integral can be performed exactly \cite{Kapustin:1999ha, witten2003sl2zactionthreedimensionalconformal}:

    \begin{equation}\label{eq:tftintegration}
    \includegraphics[]{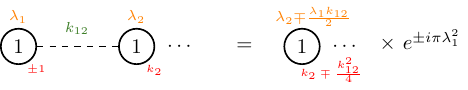}
    \end{equation}

    In the context of Equation \ref{eq:Mm_general_raw}, the contraction of the tail of $U(1)$ gauge nodes proceeds as follows:
    \begin{equation}
    \label{eq: witten sl2z}
    \begin{split}
    &
    \begin{tikzpicture}[baseline=(current bounding box).center]
        \node at (0,0) (g1) [flavor,black] {$1$};
        \draw[CScolor] (g1)++(.45,-.2) node {\tiny$1$};
        \nodeCS(1.5,0)(g2,1,2);
        \nodeCS(5,0)(g3,1,2);
        \nodeCS(6.5,0)(g4,1,1);
        \path (g2) edge[draw=none] node[midway] {$\dots$} (g3);
        \draw[BFline] (g1) -- node[midway,above,BFcolor] {\tiny -2} (g2);
        \draw[BFline] (g2) -- node[midway,above,BFcolor] {\tiny -2} ++(1,0);
        \draw[BFline] (g3)++(-1,0) -- node[midway,above,BFcolor] {\tiny -2} (g3);
        \draw[BFline] (g3) -- node[midway,above,BFcolor] {\tiny -2} (g4);
        \draw[->-] (g1) -- ++(150:1);
        \draw (g1)++(190:1) node {\tiny$\vdots$};
        \draw[-<-] (g1) -- ++(210:1);
    \end{tikzpicture}
    \\
    =&
    \begin{tikzpicture}[baseline=(current bounding box).center]
         \node at (0,0) (g1) [flavor,black] {$1$};
        \draw[CScolor] (g1)++(.45,-.2) node {\tiny$1$};
        \nodeCS(1.5,0)(g2,1,2);
        \nodeCS(5,0)(g3,1,1);
        \path (g2) edge[draw=none] node[midway] {$\dots$} (g3);
        \draw[BFline] (g1) -- node[midway,above,BFcolor] {\tiny -2} (g2);
        \draw[BFline] (g2) -- node[midway,above,BFcolor] {\tiny -2} ++(1,0);
        \draw[BFline] (g3)++(-1,0) -- node[midway,above,BFcolor] {\tiny -2} (g3);
        \draw[->-] (g1) -- ++(150:1);
        \draw (g1)++(190:1) node {\tiny$\vdots$};
        \draw[-<-] (g1) -- ++(210:1);
    \end{tikzpicture}
    \\
    & \vdots
    \\
    =&
    \begin{tikzpicture}[baseline=(current bounding box).center]
         \node at (0,0) (g1) [flavor,black] {$1$};
        \draw[CScolor] (g1)++(.45,-.2) node {\tiny$0$};
        \draw[->-] (g1) -- ++(150:1);
        \draw (g1)++(190:1) node {\tiny$\vdots$};
        \draw[-<-] (g1) -- ++(210:1);
    \end{tikzpicture}
    \end{split}
    \end{equation}
    The resulting duality is then represented schematically as:
    \begin{equation}
        \includegraphics[width=.9\linewidth]{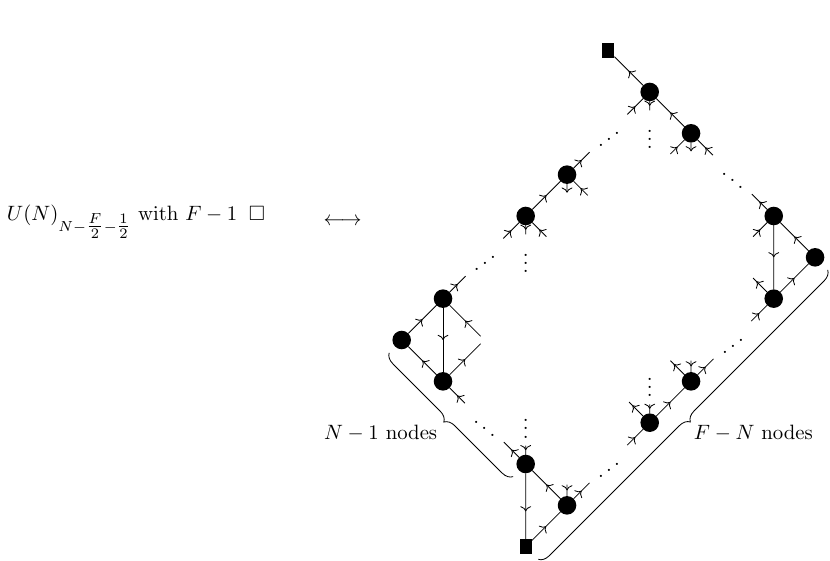}
    \end{equation}
    
\item[$+\mH$: ] In this deformation, the real scalars in the vector multiplets of the gauge groups in the right-bottom-most diagonal acquire non-zero VEVs:
	\begin{equation}
		(\sigma_{F-1}^{bottom}, \sigma_{F-2}^{bottom}, \dots, \sigma_{N+1}^{bottom}) \propto - ((F-N-1)F m , (F-N-2)Fm, \dots, m ).
	\end{equation}
	Consequently, the chiral fields connecting these nodes acquire a large positive mass while the chiral fields connecting these nodes to the rest of the quiver acquire a positive mass, as depicted on the bottom-right panel of Figure \ref{fig:def_planar_start}.
    Furthermore, the chiral field connected to the bottom flavor node is Higgsed.
    Integrating out the massive fields and contracting the resulting tail, analogous to the $+m$ case, results in the following duality:
    \begin{equation}
        \includegraphics[width=1\linewidth]{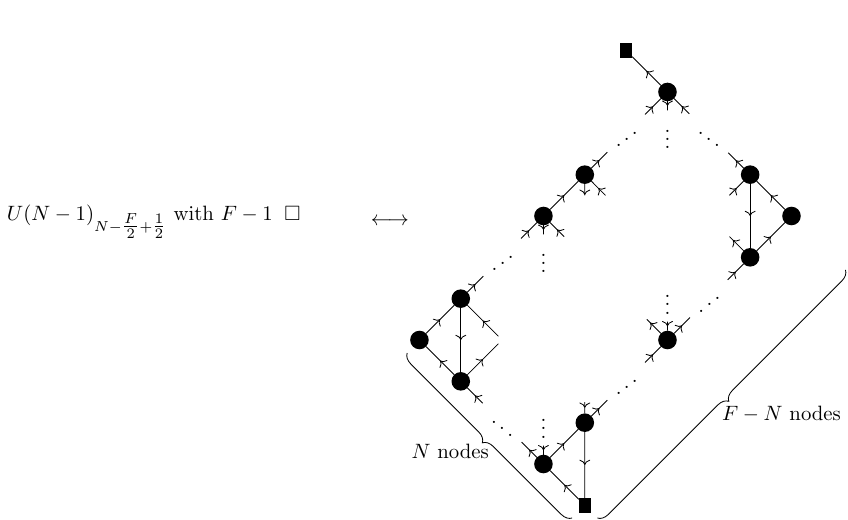}
    \end{equation}
\item[$-\mH$: ] In this deformation, the chiral fields in the top-left diagonal and the chiral field connected to the top-most gauge node acquire a VEV. These fields are highlighted in blue in the top-right panel of Figure \ref{fig:def_planar_start}. As a result, all the gauge nodes in the top-right diagonal are identified with the flavor node. The planar superpotential reduces to mass terms which we use to integrate out $2(N-1)$ chiral fields, analogously to \eqref{eq: Higgsing_diag_separate}.
\end{itemize}

\begin{figure}
    \centering
    \includegraphics[width=1\linewidth]{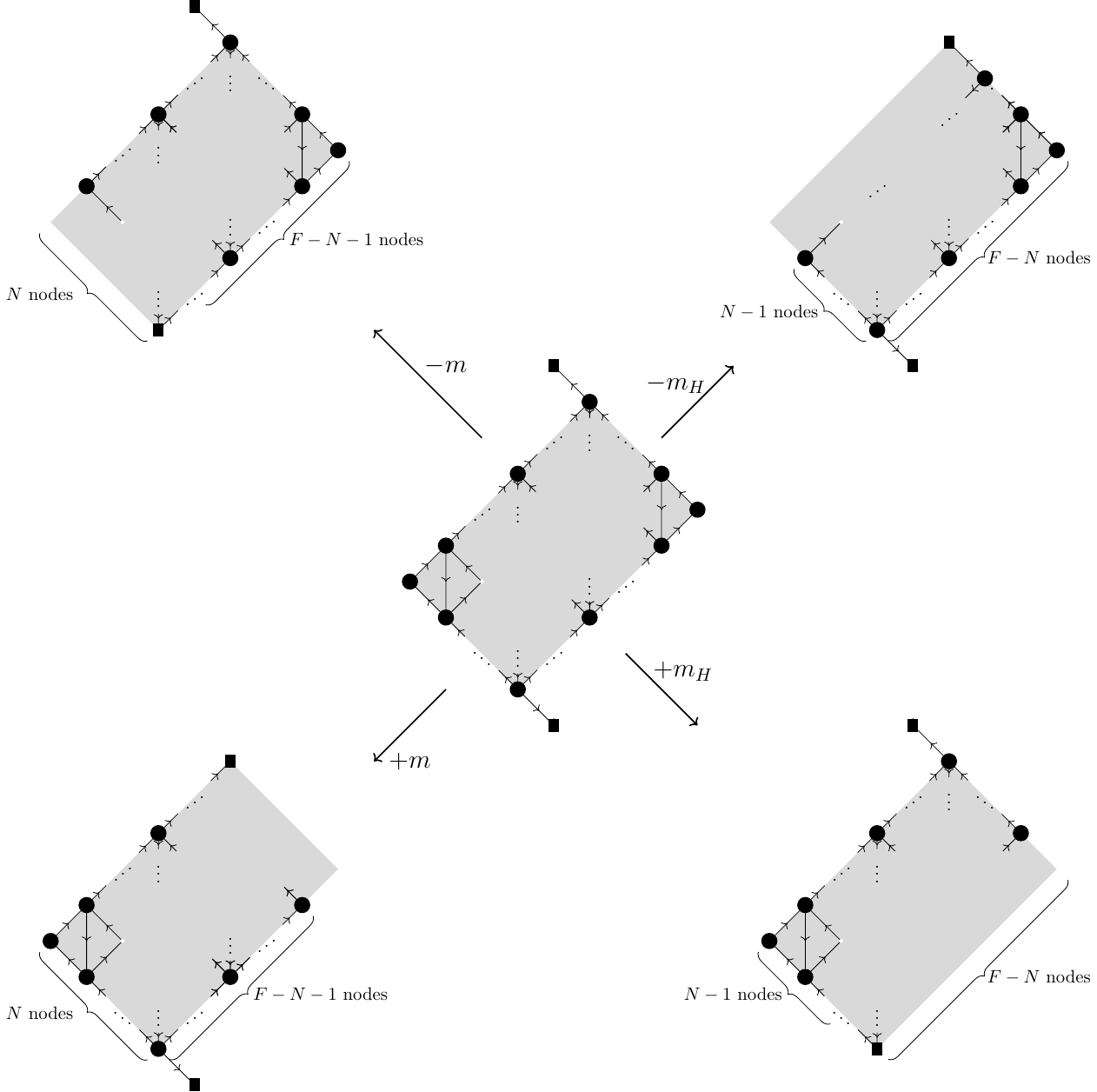}
    \caption{Deformations of the planar quivers dual to the SQCD mass deformations in \eqref{eq:defcross_QCD}. The grey shading highlights which parts of the quiver are modified by each mass deformation and makes explicit how the overall quiver geometry changes relative to the initial configuration.}
    \label{fig:def_planar_final}
\end{figure}

We summarize these mass flows in Figure \ref{fig:def_planar_final}.

The proposal of these deformation arises by comparing the large mass asymptotic behavior of the $\mathbf{S}^3_b$ partition functions of the SQCD and of its planar dual, using the strategies of \cite{Aharony:2013dha,Benini_2011a} (see also Section 2.2.2 of \cite{Benvenuti:2025a}).

One may notice also that, after performing either a $+m$ or a $-\mH$ deformation, the resulting quiver retains the same rectangular structure as the original planar Abelian dual. In fact, the resulting theory is precisely the planar Abelian dual of the SQCD model obtained after the corresponding mass deformation on the electric side. These two deformations therefore provide a nontrivial self-consistency check of the duality originally proposed in \cite{Benvenuti:2024seb,Benvenuti:2025a}.

In contrast, the deformations -$m$ and +$\mH$ yield quivers with a modified geometry, leading to different planar structures. Successive applications of these deformations require a separate analysis, which we undertake in the next section, where we construct and study the planar Abelian duals of SQCD theories obtained through multiple -$m$ or +$\mH$ mass deformations.

\subsection{Exploring Zone 1 via Multiple \texorpdfstring{$+\mH$}{mH} Mass Deformations}
\label{subsec: zone1}

In the previous section, we discussed the mapping of a single positive mass deformation accompanied with Higgsing on the planar Abelian dual of SQCD on the $\NN=4$-descendant line. This resulted in a duality for SQCD theories with $F$ fundamentals and $k=N-\tfrac{F}{2}+1$ CS level, which lie in Zone 1 in the $k/F$ plane (see Figure \ref{fig:KF_plane}), just below the $\NN=4$ descendant (the orange line in Figure \ref{fig:KF_plane}). 

To explore theories in Zone 1 more systematically, we must map the effect of adding multiple positive masses accompanied with Higgsing ($+\mH$) on the planar theory.

We perform the $+\mH$ deformation associated with the $F$-th quark, which is implemented by the real mass deformations in \eqref{eq:shifts_mass_higgs_QCD}. We then proceed with a second $+ \mH$ deformation, which we find to result in the following.
\begin{equation}
    \includegraphics[width=.7\linewidth]{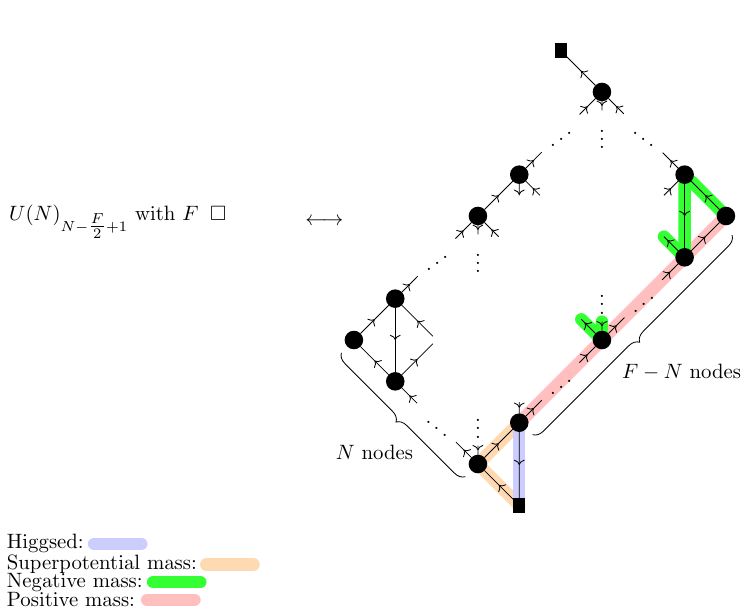}
\end{equation}

The corresponding deformation on the magnetic side is the following:
\begin{itemize}
\item The vertical bifundamental chiral connected to the bottom-most flavor node acquires a VEV, Higgsing the gauge node under which it is charged.
\item The scalars in the vector multiplets of the gauge node in the bottom-right diagonal acquire a VEV:
\begin{equation}
		(\sigma_{F-1}^{bottom}, \sigma_{F-2}^{bottom}, \dots, \sigma_{N+1}^{bottom}) \propto - ((F-N-1)F m , (F-N-2)m, \dots, m )
	\end{equation}
\end{itemize}
Which is a proposal checked that can be followed on the $\mathbf{S}_b^3$ partition function.
Similar to the first mass deformation $+\mH$ considered in Section \ref{subsec: deformations of N4 sqcd}, the gauge groups in the bottom-right diagonal become a tail of CS theories that only interact via mixed CS levels after integrating out the massive fields. This tail can be sequentially contracted as in \eqref{eq: witten sl2z}, resulting in the following duality:
\begin{equation}
    \includegraphics[width=.7\linewidth]{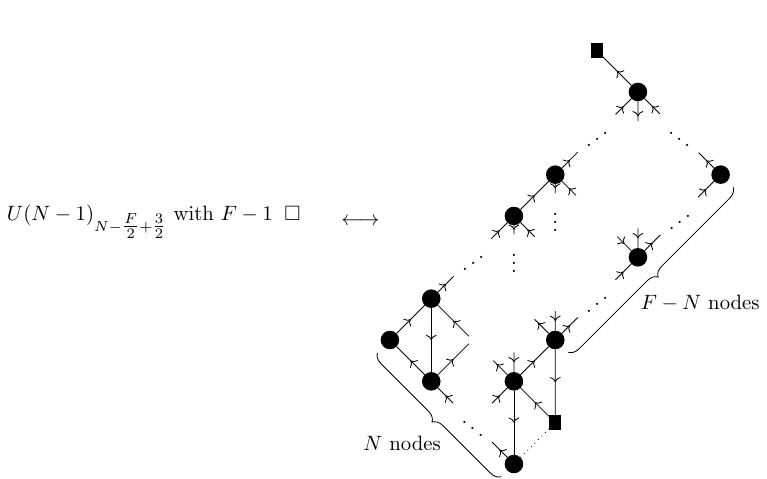}
\end{equation}

Further positive mass deformations accompanied by Higgsing are mapped in an analogous manner. In this case, the real scalars associated with the gauge nodes along the bottom-right diagonal acquire non-zero vacuum expectation values, which in turn generate large masses for the fields charged under them. Simultaneously, the vertical chiral multiplet attached to the bottom flavor node develops a Higgs VEV.
Performing this sequence of deformations yields the planar Abelian dual quiver corresponding to a generic SQCD theory in Zone 1:
\begin{equation}    \label{quiv:Zone1_dual_raw}
    \includegraphics[width=.7\linewidth]{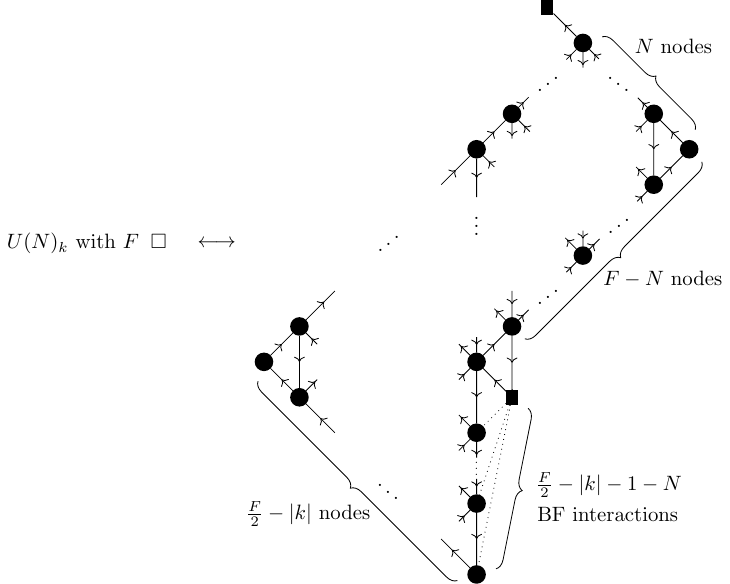}
\end{equation}

The quiver on the r.h.s.~of \eqref{quiv:Zone1_dual_raw} can be systematically simplified by applying local Aharony-like dualities, similar to the analysis of Appendix B of \cite{Benvenuti:2025a}.
Specifically, we apply Aharony duality to all the gauge nodes in a vertical column, proceeding column by column, from left to right and halting at the $(\frac{F}{2}-k)$-th column. The gauge nodes within this column that lie outside the rectangle of dimensions $N\times(F-N)$ confine.
We then repeat the process, restarting from the left and stopping one column earlier each time, until only the leftmost gauge node is dualized. At every iteration, the nodes of the final dualized column that fall outside the $N\times(F-N)$ rectangle again confine. All the dualities utilized in this procedure are described in detail in Appendix \ref{app: rel_duals}.

After performing this complete sequence of local dualizations, we obtain the duality describing SQCD theories with $F>2N+2|k|$, namely those lying in Zone 1 of Figure \ref{fig:KF_plane}:
\begin{equation}
\label{eq: Dual_zone1}
\includegraphics[width=1\linewidth]{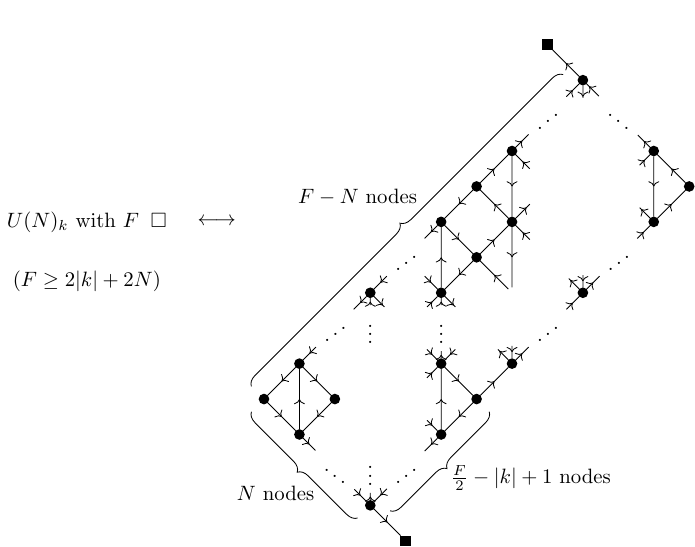}
\end{equation}

The resulting quiver coincides with the dual obtained in Section~\ref{subsec: negative mass deformations in mirror}. Consequently, the duality proposed in Section \ref{subsec: planar abelian dual of U(N)0} for $U(N)_0$ SQCD can be viewed as a descendant of the duality for SQCD on the $\mathcal{N}=4$ line originally discussed in \cite{Benvenuti:2024seb,Benvenuti:2025a}.




\subsection{Exploring Zones \texorpdfstring{$\widetilde{1},2, 3$ and $4$ via Multiple $-m$}{1t, 2, 2t, 3 and 3t via multiple -m} Mass Deformations}\label{subsec: zones1t2t3t}

In Section \ref{subsec: negative mass deformations in mirror}, we briefly summarized the effect of negative mass deformations on the $U(N)_0$ SQCD theory with $F$ fundamental fields and studied the corresponding RG flow in the electric and mirror descriptions. We now provide a more systematic justification of these claims. All these deformations are studied from the point of view of the $\mathbf{S}^3_b$ partition function.

\paragraph{Zones \texorpdfstring{2}{2 and 2t}:} The resulting duality for a generic SQCD theory in Zones $2$ is shown in \eqref{eq: Zone22t_dual_raw}. 

At each step, the $-m$ mass deformation is implemented by giving vacuum expectation values to the real scalars in the vector multiplets associated with the gauge nodes along the bottom-left diagonal:
\begin{equation}
\label{eq: zone22teom}
\left(\sigma_{1}^{\text{bottom}}, \sigma_{2}^{\text{bottom}}, \dots, \sigma_{\tfrac{F}{2} - |k| - 1}^{\text{bottom}}\right)
\propto
\left((\tfrac{F}{2} - |k| - 1) F m,; (\tfrac{F}{2} - |k| - 2) F m,; \dots,; F m \right).
\end{equation}
In addition, the vertical chiral multiplet connected to the bottom flavor node acquires a Higgs VEV.

Starting from the $\mathcal{N}=4$ descendant line, this mapping remains valid for the first $N$ successive negative real mass deformations. This sequence of deformations allows us to fully explore the parameter regions corresponding to Zone~2 in the $k/F$-plane.

\paragraph{Zone \texorpdfstring{$\widetilde{\mathbf{1}}$}{1t}:} The resulting duality for a generic SQCD theory in Zone $\widetilde{1}$ is shown in \eqref{eq: Zone1t_dual_raw}. 

This region of parameter space can be accessed through a sequence of negative real mass deformations starting from \textbf{Zone 2}. Similarly to Section \ref{subsec: deformations of N4 sqcd}, the $+m$ deformations shrink the quiver along the SW-NE directions.

\paragraph{Zones \texorpdfstring{3}{3}:} The resulting dualities for a generic SQCD theory in Zone $3$ are shown in \eqref{eq: Zone33t_a_dual_raw}.

Starting from \textbf{Zone 2}, we perform further $-m$ deformations, each corresponding to a VEV to the real scalars in the $\left(|k|-\tfrac{F}{2}\right)$-th column—namely, the leftmost column still connected to the rest of the quiver by diagonal chiral fields:
\begin{equation}
\sigma^{i}_{|k|-\tfrac{F}{2}} = \sigma^{j}_{|k|-\tfrac{F}{2}}
\qquad \forall i,j
\end{equation}
\begin{equation}
    \sigma^{i}_{|k|-\tfrac{F}{2}}
    -
    \sigma^{i}_{|k|-\tfrac{F}{2}+1}
    \propto Fm
\end{equation}
The vertical chirals in the $\left(|k|-\tfrac{F}{2}\right)$-th column remain massless, whereas the diagonal chirals connecting this column to the next one become massive.
In particular, the diagonals oriented toward the right acquire positive masses, while those pointing left acquire negative masses, as shown in \eqref{eq:columns_howto}.
This mapping remains valid as long as the quiver contains columns of diagonal chiral fields, which corresponds to the condition $F\geq N$ \eqref{eq: Zone33t_a_dual_raw}.
\paragraph{Zones \texorpdfstring{4}{4}:}
In Zone 4, corresponding to $F<N$ \eqref{eq: Zone33t_b_dual_raw}, the negative mass deformation~$-m$ instead corresponds to giving a VEV to the vertical chiral at the top of the $\text{F}^{\text{th}}$ column from the right, together with VEVs for the real scalars in the top-right diagonal:
\begin{equation}
\left(
\sigma_{1}^{top}, \dots, \sigma_{F-1}^{top}
\right) \propto
\left(
(F-1)F m, (F-2)Fm, \dots, Fm
\right)
\end{equation}
where $\sigma_{i}^{top}$ is the real scalar associated to the top node of the $i$-th column from the right.

\section{Relating Planar Abelian Duals to Aharony Duals} \label{sec: aharonyproof}

The $\mathcal{N}=2$ $U(N)_k$ SQCD theory enjoys an Aharony-like duality for arbitrary parameters $(N,k,F)$ (see \cite{Closset_2023a} and references therein). 
In this section, we focus on the case with only fundamental matter fields, whose dual description involves only anti-fundamental fields. 
As noted above, the structure of these dualities changes qualitatively as one moves in the $(k,F)$ plane at fixed $N$ (see Figure~\ref{eq: Dual_k0}).

A natural question is how the planar Abelian dualities proposed in Section~\ref{sec: landscape} relate to Aharony duality. 
Here, we argue that, assuming the validity of the planar Abelian dualities together with certain special instances of Aharony-like duality, specifically those involving Abelian gauge groups in both the electric and magnetic phases, one can derive the general Aharony-like dualities for $U(N)_k$ SQCD with $F$ fundamental chiral fields. 
Consequently, Aharony duality for SQCD with fundamentals can be interpreted as a consequence of mirror symmetry.

Concretely, we consider SQCD in Zone $\tilde{1}$ with $F>2N$ fundamentals and Chern--Simons level $k$, satisfying $2|k|<2N-F$, corresponding to a maximally chiral regime \cite{Benini_2011a}. 
As discussed in detail in Appendix \ref{app: mass flow S3b}, Aharony duality for SQCD theories sitting in the other Zones can be derived from Aharony duality in Zone 1 via real mass deformations, therefore we do not lose generality by focusing on Zone 1.
The Aharony-dual theory is $U(F-N)$ SQCD with $F$ anti-fundamentals and Chern--Simons level $-k$, which lies in (the anti-fundamental analogue of) Zone~1:
\begin{equation}\label{eq:aharony_zone1}
    \text{$U(N)_k$ with $F<2N\;\Box$ }
    \quad \longleftrightarrow \quad
    \text{$U(F-N)_{-k}$ with $F>2(F-N)\;\overline{\Box}$}
\end{equation}

The topological symmetry of the magnetic theory, $U(1)_{\eta_D}$, is mapped to a linear combination of the R-symmetry and minus the topological symmetry of the electric theory, $U(1)_\eta$:
\begin{equation}
    U(1)_{\eta_D} = k\, U(1)_R - U(1)_{\eta}.
\end{equation}
The representations of the fields under the $SU(F) \times U(1)_{\eta} \times U(1)_R$ symmetries are:

\begin{equation}
    \begin{array}{c|cc|ccc}
         & U(N) & U(F-N) & SU(F) & U(1)_R & U(1)_{\eta}  \\ \hline
        Q &  \square & & \overline{\square} & r & 0 \\
        \tilde{q} & & \overline{\square} & \square & 1-r & 0
    \end{array}
\end{equation}

Here, $Q$ denotes the fundamental chiral fields of the electric theory, and $\tilde{q}$ denotes the anti-fundamental chiral fields of the magnetic theory. 
No additional gauge-singlet fields are involved in this duality. 
The only operators charged under $U(1)_\eta$ are the monopole operators in the respective phases.

The corresponding identity between partition functions is given in \eqref{app:aharony_max_chiral_Sb3}, which also encodes shifts for background CS terms for the global symmetries involved in the duality.
\\

\begin{figure}[H]
    \centering
    \includegraphics[width=.8\linewidth]{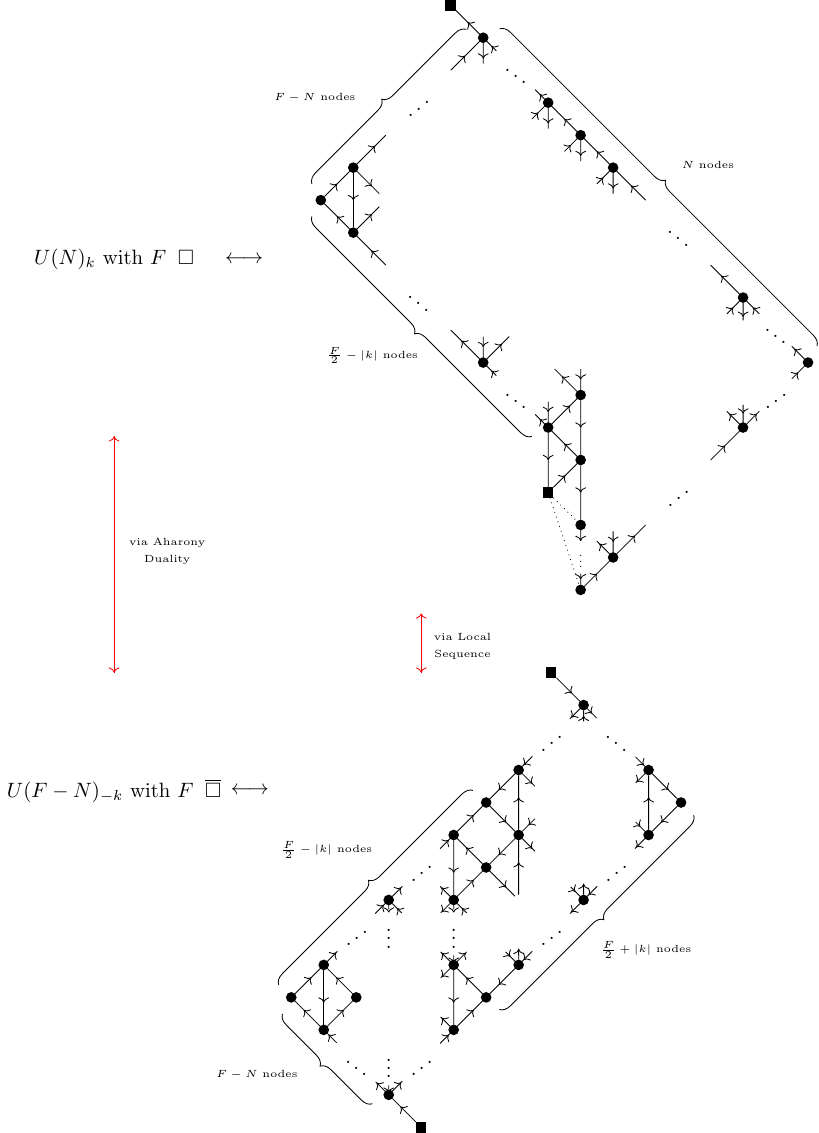}
    \caption{Schematic depiction of the duality web between a theory in Zone $\tilde{1}$ and its Aharony dual in Zone 1.
    The horizontal arrows represent the planar Abelian dualities presented in this paper. 
    The rightmost vertical arrow corresponds to a sequence of local dualizations discussed in the body of the paper.
    The diagram is commutative, therefore Aharony duality can be thought of as a consequence of mirror symmetry.}
    \label{fig: schematic_z1_1t}
\end{figure}

The planar Abelian dual of the electric theory was presented in Section~\ref{sec: planar abelian dual of SQCD}, while the planar Abelian dual of the magnetic theory can be obtained by applying parity and charge conjugation to the dualities discussed in this paper. 
Taken together, this produces four UV Lagrangians that are dual to each other: the two SQCD theories related by Aharony duality, and their respective planar Abelian duals. 
The relations among these theories are summarized schematically in Figure~\ref{fig: schematic_z1_1t}. 

In the remainder of this section, we describe the precise sequence of local dualizations that connects the two planar Abelian quivers shown in Figure~\ref{fig: schematic_z1_1t}.


\paragraph{Electric Theory.}
We begin by massaging the planar Abelian dual of the electric theory \eqref{eq: Zone1t_dual_raw}, which has gauge group $U(N)$, Chern--Simons level $k$, FI parameter $\eta$, and trial R-charge $r$. 
We report the planar Abelian dual of this theory on the left of Figure~\ref{fig: z1_p1_alt}.
We perform a sequence of dualizations analogous to the planar flip--flip duality discussed in \cite{Benvenuti:2025a}. We provide the dualities involved in this procedure in Appendix \ref{app: rel_duals}. 

The dualization proceeds column by column, starting from the rightmost column and stopping at the column immediately to the right of the bottom flavor node. 
The result of this first dualization pass is shown in the central quiver of Figure~\ref{fig: z1_p1_alt}.

After completing this first dualization pass, a column of square nodes appears at position $\tfrac{F}{2}-|k|$. 
Additionally, the flavor node shifts one column to the right, and the bottom-right diagonal of the quiver shortens by one node.

We then perform subsequent dualization passes in a similar manner, starting from the right and ending at the column immediately to the right of the bottom flavor node. 
Each pass further shortens the bottom-right diagonal by one node and requires one fewer dualization. 
The result of the second pass is shown on the right of Figure~\ref{fig: z1_p1_alt}.

After iterating this procedure, we arrive at the quiver depicted on the left of Figure~\ref{fig: z1_p2_alt}, which takes the form of a rectangular lattice of size $N \times (F-N)$. 
Finally, we perform a flip--flip duality \cite{Benvenuti:2025a} on the rightmost triangular subquiver, resulting in the quiver shown on the right of Figure~\ref{fig: z1_p2_alt}. 

\begin{landscape}
\thispagestyle{empty}
\begin{figure}[h]
    \centering
    \makebox[\textwidth][c]{
        \includegraphics[width=1.2\linewidth]{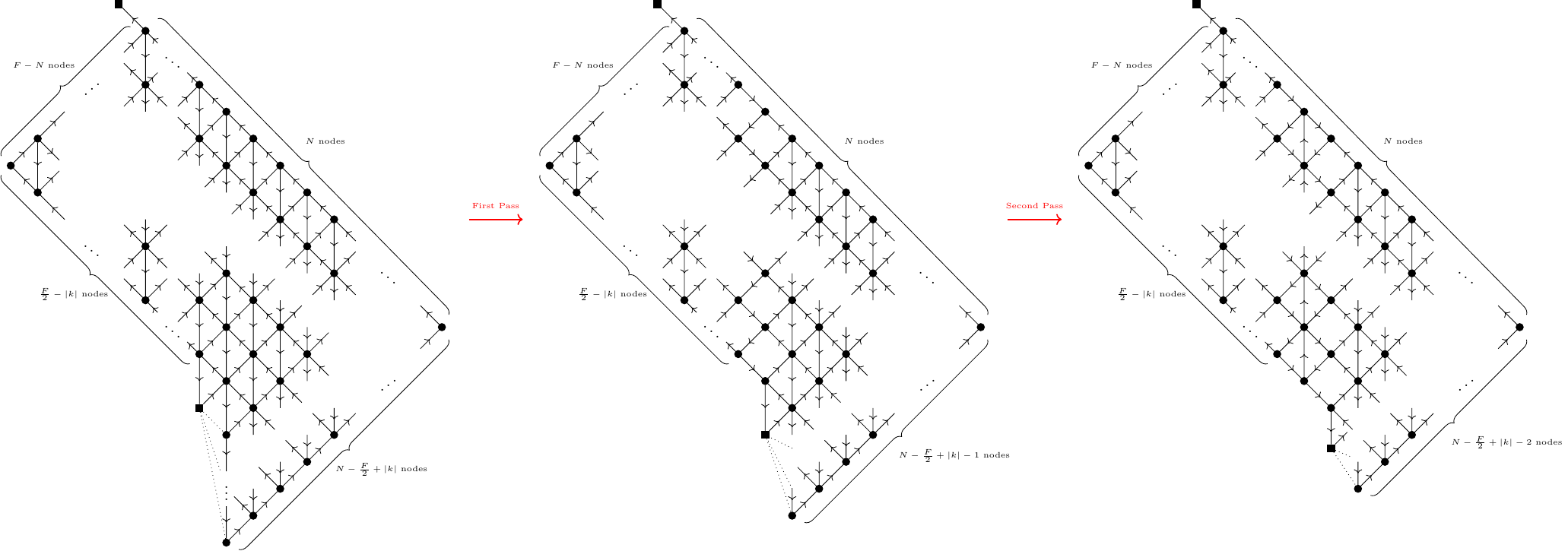}}
    \caption{The planar Abelian dual for $U(N)_k$ SQCD with $F$ fundamentals in Zone $\tilde{1}$ and the result of local dualizations. Here we show the original planar dual presented in \eqref{eq: Zone1t_dual_raw} on the left. In the center and on the right, we show the quivers obtained after one and two dualization passes as discussed in the main text.}
    \label{fig: z1_p1_alt}
\end{figure}
\end{landscape}

\begin{landscape}
\thispagestyle{empty}
\begin{figure}
    \centering
    \makebox[\textwidth][c]{
        \includegraphics[width=1.2\linewidth]{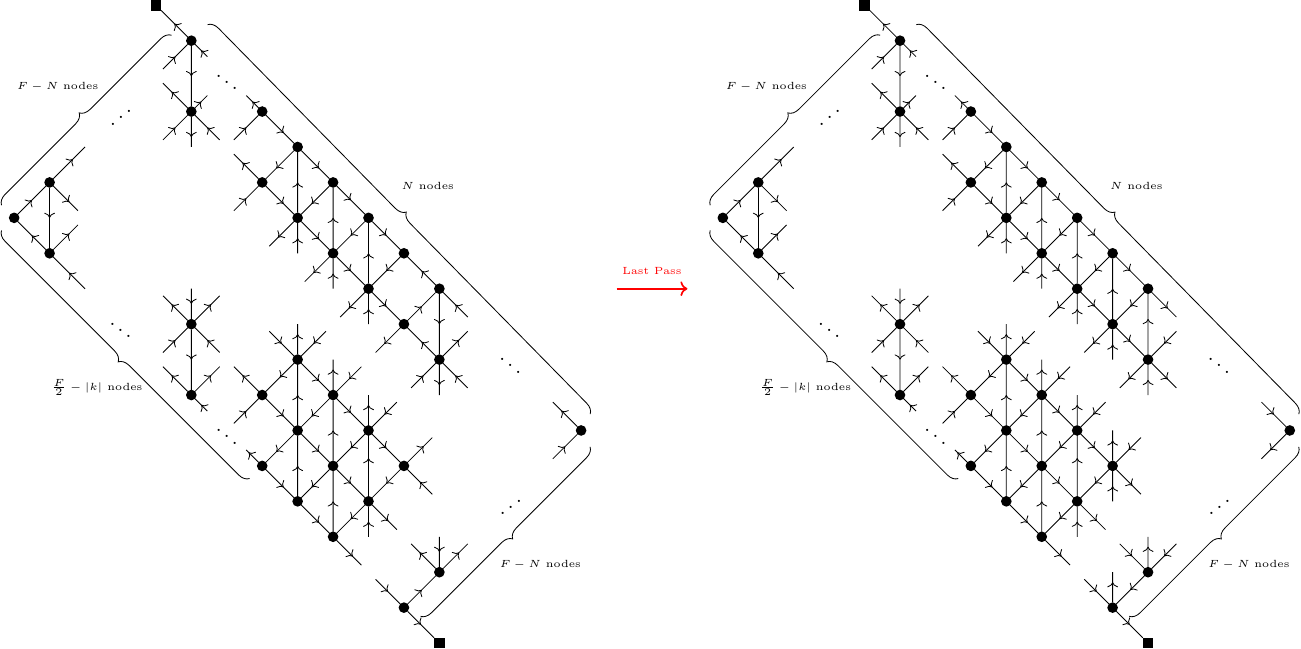}
    }
    \caption{Continued from Figure \ref{fig: z1_p1_alt}.
    The final iterations of the dualization sequence for the Zone $\widetilde{1}$ mirror of $U(N)_k$ with $F$ fundamental fields.}
    \label{fig: z1_p2_alt}
\end{figure}
\end{landscape}

This represents our final expression for the electric theory and provides an alternative presentation of the planar Abelian dual in Zone~$\tilde{1}$. We remind the reader that we use the shorthand $X_{ij} := X_i-X_j$ to parametrize the FI terms of the gauge nodes in the quiver diagram.

\begin{equation}
\label{eq:dual_zone1t_alternative}
    \includegraphics[]{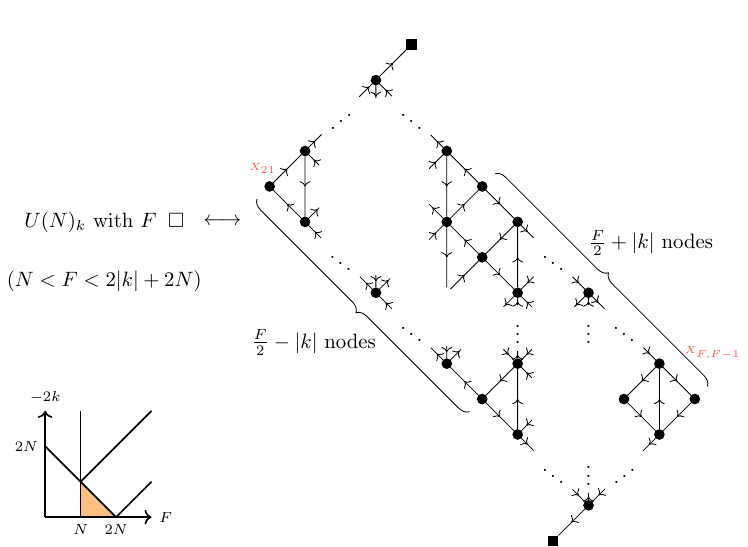}
\end{equation}


\paragraph{Magnetic Theory.}
The planar Abelian dual of SQCD with only anti-fundamental fields can be obtained by applying parity ($P$) and charge conjugation ($C$) to both sides of the dualities presented in this paper. 
In particular, we start from the duality for SQCD in Zone~1 \eqref{eq: Dual_zone1_schematic}, with gauge group $U(N_D)$, Chern--Simons level $k$, FI parameter $\eta - k\omega$, and trial R-charge $1-r$. 
This corresponds to the theory appearing on the right-hand side of the Aharony-like duality \eqref{eq:aharony_zone1}, up to background terms.

Applying the combined $CP$ transformation on the left-hand side produces SQCD with gauge group $U(N_D)$, $F$ anti-fundamental fields, Chern--Simons level $-k$, and FI parameter $k\omega - \eta$. 
On the quiver side, the $CP$ action reverses the direction of all arrows and flips the sign of all Chern--Simons and mixed Chern--Simons terms, including background contributions\footnote{FI terms can be interpreted as mixed Chern--Simons couplings between gauge and topological symmetries and are therefore odd under $P$.}. 


The resulting duality is the following:
\begin{equation}
\label{eq: aharony_mirror_fig1}
    \includegraphics[]{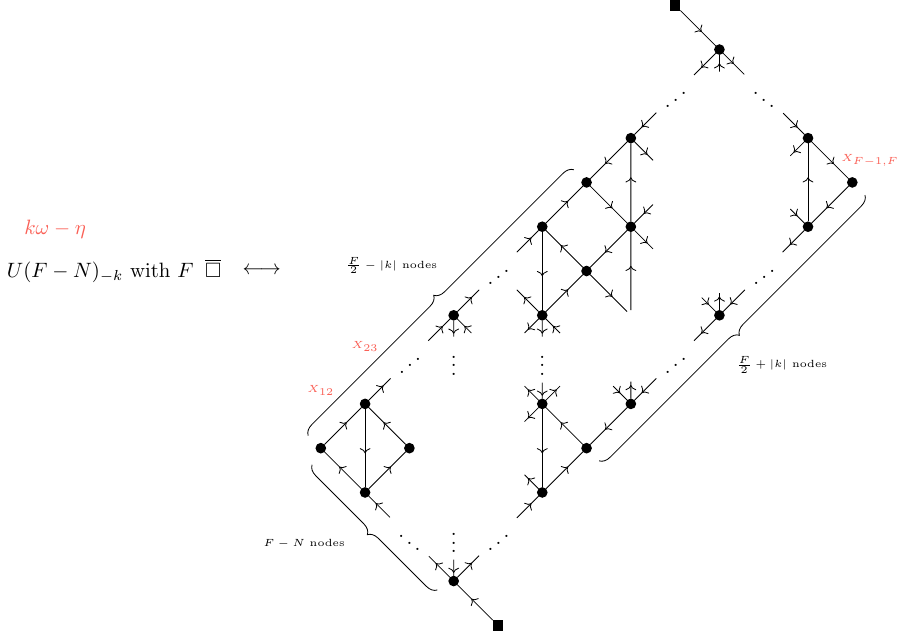}
\end{equation}

We remind the reader that we use the shorthand $X_{ij} := X_i-X_j$ to parametrize the FI terms of the gauge nodes in the quiver diagram. The Chern--Simons levels, mixed Chern--Simons couplings, and trial R-charges can be determined following the rules outlined in Section~\ref{sec: planar abelian dual of SQCD}. 

We now perform a sequence of local dualizations on the quiver on the right-hand side to show that it is equivalent to the right-hand side of \eqref{eq: aharony_mirror_fig1}. All the dualities involved in this dualization are given in Appendix \ref{app: rel_duals}.
While these manipulations can be performed for general values of $N$, $F$, and $k$, it is pedagogical to present them through a concrete example. 
We consider $U(2)_{\frac{1}{2}}$ SQCD with $9$ anti-fundamental fields:

\begin{equation} \label{eq:aharony_example_1}
    U(2)_{\frac{1}{2}} \text{ with 9 } \overline{\square}
    \qquad \leftrightarrow\qquad
    \begin{tikzpicture}[baseline=(current bounding box).center]
         \begin{scope}[scale=.6]
        \node at (0,0) (g11) [gaugefill] {};
        \node at (1,1) (g21) [gaugefill] {};
        \node at (1,-1) (g22) [gaugefill] {};
        \node at (2,-2) (f1) [flavorfill] {};
        \node at (2,2) (g31) [gaugefill] {};
        \node at (2,0) (g32) [gaugefill] {};
        \node at (3,3) (g41) [gaugefill] {};
        \node at (3,1) (g42) [gaugefill] {};
        \node at (4,4) (g51) [gaugefill] {};
        \node at (4,2) (g52) [gaugefill] {};
        \node at (5,5) (g61) [gaugefill] {};
        \node at (5,3) (g62) [gaugefill] {};
        \node at (6,6) (g71) [gaugefill] {};
        \node at (5,7) (f2) [flavorfill] {};
        \node at (6,4) (g72) [gaugefill] {};
        \node at (7,5) (g81) [gaugefill] {};

        \draw[->-] (g11)--(g21);
        \draw[-<-] (g22)--(g21);
        \draw[->-] (g22)--(g11);
        \draw[->-] (f1)--(g22);

        \draw[->-] (g21)--(g31);
        \draw[->-] (g22)--(g32);
        \draw[->-] (g32)--(g21);

        \draw[->-] (g31)--(g41);
        \draw[-<-] (g32)--(g31);
        \draw[->-] (g32)--(g42);
        \draw[->-] (g42)--(g31);

        \draw[-<-] (g41)--(g51);
        \draw[-<-] (g42)--(g52);
        \draw[-<-] (g52)--(g41);

        \draw[-<-] (g51)--(g52);

        \draw[->-] (g51)--(g62);
        \draw[->-] (g62)--(g52);

        \draw[->-] (g61)--(g51);
        
        \draw[-<-] (g61)--(g62);
        \draw[->-] (g71)--(g61);
        \draw[->-] (g72)--(g62);
        \draw[->-] (g61)--(g72);

        \draw[->-] (f2)--(g71);

        \draw[-<-] (g71)--(g72);
        \draw[->-] (g71)--(g81);
        \draw[->-] (g81)--(g72);
        
    \end{scope}
    \end{tikzpicture}
\end{equation}

We begin by performing the ``flip--flip'' dualization sequence \cite{Benvenuti:2025a} on the triangular subquivers at the ends of the quiver. 
In the example above, this corresponds to dualizing the leftmost and rightmost gauge nodes, yielding the following quiver:
\begin{equation}    \label{eq:aharony_example_2}
    \text{R.H.S. of \eqref{eq:aharony_example_1}}
    \qquad \rightarrow\qquad
    \begin{tikzpicture}[baseline=(current bounding box).center]
        
    \begin{scope}[xshift = 5cm,scale=.6]
        \node at (0,0) (g11) [gaugefill] {};
        \node at (1,1) (g21) [gaugefill] {};
        \node at (1,-1) (g22) [gaugefill] {};
        \node at (2,-2) (f1) [flavorfill] {};
        \node at (2,2) (g31) [gaugefill] {};
        \node at (2,0) (g32) [gaugefill] {};
        \node at (3,3) (g41) [gaugefill] {};
        \node at (3,1) (g42) [gaugefill] {};
        \node at (4,4) (g51) [gaugefill] {};
        \node at (4,2) (g52) [gaugefill] {};
        \node at (5,5) (g61) [gaugefill] {};
        \node at (5,3) (g62) [gaugefill] {};
        \node at (6,6) (g71) [gaugefill] {};
        \node at (5,7) (f2) [flavorfill] {};
        \node at (6,4) (g72) [gaugefill] {};
        \node at (7,5) (g81) [gaugefill] {};

        \draw[-<-] (g11)--(g21);
        \draw[-<-] (g22)--(g11);
        \draw[->-] (f1)--(g22);

        \draw[->-] (g21)--(g31);
        \draw[->-] (g22)--(g32);
        \draw[->-] (g32)--(g21);

        \draw[->-] (g31)--(g41);
        \draw[-<-] (g32)--(g31);
        \draw[->-] (g32)--(g42);
        \draw[->-] (g42)--(g31);

        \draw[-<-] (g41)--(g51);
        \draw[-<-] (g42)--(g52);
        \draw[-<-] (g52)--(g41);

        \draw[-<-] (g51)--(g52);

        \draw[->-] (g51)--(g62);
        \draw[->-] (g62)--(g52);

        \draw[->-] (g61)--(g51);
        
        \draw[-<-] (g61)--(g62);
        \draw[->-] (g71)--(g61);
        \draw[->-] (g72)--(g62);
        \draw[->-] (g61)--(g72);

        \draw[->-] (f2)--(g71);

        \draw[-<-] (g71)--(g81);
        \draw[-<-] (g81)--(g72);
        
    \end{scope}

    \end{tikzpicture}
\end{equation}

Next, we dualize the column of nodes corresponding to the column of squares in the bulk of the quiver, located at position $\tfrac{F}{2}-|k|$. 
We then proceed to dualize the column immediately to its left and continue dualizing columns from right to left until reaching the column attached to the flavor node, thereby completing the first pass. 
In our example, this corresponds to dualizing column~4, followed by column~3. 
The resulting quiver is:

\begin{equation}    \label{eq:aharony_example_3}
    \text{R.H.S. of \eqref{eq:aharony_example_2}}
    \qquad \rightarrow\qquad
    \begin{tikzpicture}[baseline=(current bounding box).center]
        
    \begin{scope}[xshift = 10cm,scale=.6]
        \node at (0,0) (g11) [gaugefill] {};
        \node at (1,1) (g21) [gaugefill] {};
        \node at (1,-1) (g22) [gaugefill] {};
        \node at (2,-2) (f1) [flavorfill] {};
        \node at (2,2) (g31) [gaugefill] {};
        \node at (2,0) (g32) [gaugefill] {};
        \node at (3,3) (g41) [gaugefill] {};
        \node at (3,1) (g42) [gaugefill] {};
        \node at (4,4) (g51) [gaugefill] {};
        \node at (4,2) (g52) [gaugefill] {};
        \node at (5,5) (g61) [gaugefill] {};
        \node at (5,3) (g62) [gaugefill] {};
        \node at (6,6) (g71) [gaugefill] {};
        \node at (5,7) (f2) [flavorfill] {};
        \node at (6,4) (g72) [gaugefill] {};
        \node at (7,5) (g81) [gaugefill] {};

        \draw[-<-] (g11)--(g21);
        \draw[-<-] (g22)--(g11);
        \draw[->-] (f1)--(g22);
        \draw[->-] (g22)--(g21);
        
        \draw[-<-] (g21)--(g31);
        \draw[-<-] (g22)--(g32);
        \draw[-<-] (g32)--(g21);

        \draw[-<-] (g31)--(g41);
        \draw[->-] (g32)--(g42);
        \draw[->-] (g42)--(g31);
        
        \draw[->-] (g41)--(g42);
        \draw[->-] (g41)--(g51);
        \draw[->-] (g42)--(g52);
        \draw[->-] (g52)--(g41);

        \draw[->-] (g51)--(g62);
        \draw[->-] (g62)--(g52);

        \draw[->-] (g61)--(g51);
        
        \draw[-<-] (g61)--(g62);
        \draw[->-] (g71)--(g61);
        \draw[->-] (g72)--(g62);
        \draw[->-] (g61)--(g72);

        \draw[->-] (f2)--(g71);

        \draw[-<-] (g71)--(g81);
        \draw[-<-] (g81)--(g72);
        
    \end{scope}
    
    \end{tikzpicture}
\end{equation}

We then repeat the procedure, starting from the column at position $\tfrac{F}{2}-|k|+1$, which is now a column of squares. 
We dualize columns from right to left until we reach the opposite column of squares, completing the second pass. 
This process is iterated until the rightmost flavor node is reached. 

In our example, two additional passes are required:

\begin{equation}    \label{eq:aharony_example_4}
    \text{R.H.S. of \eqref{eq:aharony_example_3}}
    \qquad \rightarrow\qquad
    \begin{tikzpicture}[baseline=(current bounding box).center]
        
   \begin{scope}[xshift = 0cm,yshift = -7cm,scale=.6]
        \node at (0,0) (g11) [gaugefill] {};
        \node at (1,1) (g21) [gaugefill] {};
        \node at (1,-1) (g22) [gaugefill] {};
        \node at (2,-2) (f1) [flavorfill] {};
        \node at (2,2) (g31) [gaugefill] {};
        \node at (2,0) (g32) [gaugefill] {};
        \node at (3,3) (g41) [gaugefill] {};
        \node at (3,1) (g42) [gaugefill] {};
        \node at (4,4) (g51) [gaugefill] {};
        \node at (4,2) (g52) [gaugefill] {};
        \node at (5,5) (g61) [gaugefill] {};
        \node at (5,3) (g62) [gaugefill] {};
        \node at (6,6) (g71) [gaugefill] {};
        \node at (5,7) (f2) [flavorfill] {};
        \node at (6,4) (g72) [gaugefill] {};
        \node at (7,5) (g81) [gaugefill] {};

        \draw[-<-] (g11)--(g21);
        \draw[-<-] (g22)--(g11);
        \draw[->-] (f1)--(g22);
        \draw[->-] (g22)--(g21);
        
        \draw[-<-] (g21)--(g31);
        \draw[-<-] (g22)--(g32);
        \draw[-<-] (g32)--(g21);

        \draw[->-] (g32)--(g31);
        \draw[-<-] (g31)--(g41);
        \draw[-<-] (g32)--(g42);
        \draw[-<-] (g42)--(g31);
        
        \draw[->-] (g41)--(g51);
        \draw[->-] (g42)--(g52);
        \draw[->-] (g52)--(g41);

        \draw[->-] (g51)--(g52);
        \draw[-<-] (g51)--(g62);
        \draw[-<-] (g62)--(g52);

        \draw[-<-] (g61)--(g51);
        \draw[->-] (g71)--(g61);
        \draw[->-] (g72)--(g62);
        \draw[->-] (g61)--(g72);

        \draw[->-] (f2)--(g71);

        \draw[-<-] (g71)--(g81);
        \draw[-<-] (g81)--(g72);
        
    \end{scope}
    \end{tikzpicture}
    \qquad \rightarrow\qquad
    \begin{tikzpicture}[baseline=(current bounding box).center]
        
    \begin{scope}[xshift = 5cm,yshift = -7cm,scale=.6]
        \node at (0,0) (g11) [gaugefill] {};
        \node at (1,1) (g21) [gaugefill] {};
        \node at (1,-1) (g22) [gaugefill] {};
        \node at (2,-2) (f1) [flavorfill] {};
        \node at (2,2) (g31) [gaugefill] {};
        \node at (2,0) (g32) [gaugefill] {};
        \node at (3,3) (g41) [gaugefill] {};
        \node at (3,1) (g42) [gaugefill] {};
        \node at (4,4) (g51) [gaugefill] {};
        \node at (4,2) (g52) [gaugefill] {};
        \node at (5,5) (g61) [gaugefill] {};
        \node at (5,3) (g62) [gaugefill] {};
        \node at (6,6) (g71) [gaugefill] {};
        \node at (5,7) (f2) [flavorfill] {};
        \node at (6,4) (g72) [gaugefill] {};
        \node at (7,5) (g81) [gaugefill] {};

        \draw[-<-] (g11)--(g21);
        \draw[-<-] (g22)--(g11);
        \draw[->-] (f1)--(g22);
        \draw[->-] (g22)--(g21);
        
        \draw[-<-] (g21)--(g31);
        \draw[-<-] (g22)--(g32);
        \draw[-<-] (g32)--(g21);

        \draw[->-] (g32)--(g31);
        \draw[-<-] (g31)--(g41);
        \draw[-<-] (g32)--(g42);
        \draw[-<-] (g42)--(g31);
        \draw[-<-] (g41)--(g42);
        
        \draw[-<-] (g41)--(g51);
        \draw[-<-] (g42)--(g52);
        \draw[-<-] (g52)--(g41);

        \draw[-<-] (g51)--(g62);
        \draw[-<-] (g62)--(g52);

        \draw[->-] (g61)--(g62);
        \draw[-<-] (g61)--(g51);
        \draw[-<-] (g71)--(g61);
        \draw[-<-] (g72)--(g62);
        \draw[-<-] (g61)--(g72);

        \draw[->-] (f2)--(g71);
        \draw[->-] (g71)--(g72);
        \draw[-<-] (g71)--(g81);
        \draw[-<-] (g81)--(g72);
        
    \end{scope}
    \end{tikzpicture}
\end{equation}

In general, the resulting quiver takes the form: 
\begin{equation}
\label{eq: aharony_mirror_fig2}
    \includegraphics[]{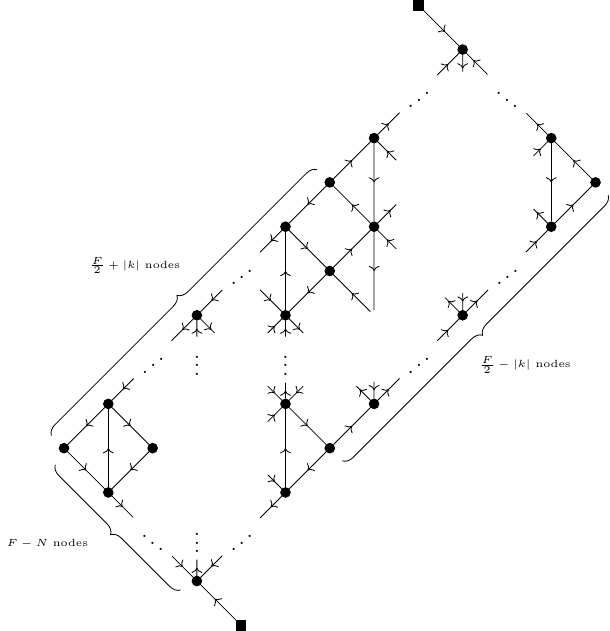}
\end{equation}

The final step consists of modifying the two chiral fields connected to the flavor nodes. 
This manipulation is implemented via a sequence of Witten $SL(2,\mathbb{Z})$ actions \cite{witten2003sl2zactionthreedimensionalconformal} on the $U(1)_{\eta}$ symmetry of the quiver, which together combine into a trivial action. 
To make this operation nontrivial, we use the fact that the quiver has a single $U(1)$ flavor symmetry, which can be associated either with the left or the right flavor node; the two choices are gauge-equivalent.

Concretely, we perform a $T^{-1}ST^{-1}$ transformation on the left flavor node and a $TS^{-1}T$ transformation on the right flavor node. 
Although these combine into a trivial $SL(2;\mathbb{Z})$ action, the intermediate steps induce a non-trivial reorganization of the theory, which remains IR dual to the original quiver. 
The $T$ transformation shifts the Chern--Simons level by $+1$, while the $S$ transformation gauges the symmetry, introducing a new $U(1)$ topological symmetry\footnote{At the level of the $\mathbf S^3_b$ partition function, the $T$ and $S$ transformations act as
\begin{equation}
\begin{split}
    T:& \quad Z(\mu) \to e^{-i\pi \mu^{2}} Z(\mu)\,, \\
    S:& \quad Z(\mu) \to \int d\mu \; e^{-2 i\pi \mu \lambda} Z(\mu)\,,
\end{split}
\end{equation}
where $\mu$ and $\lambda$ are the fugacities associated with the original and new $U(1)$ symmetries, respectively.}. 

After these transformations, both nodes are confined to a chiral field, corresponding to a monopole operator of the associated gauge node. 

The resulting quiver is shown in Equation \ref{eq: aharony_mirror_fig3}. We remind the reader that we use the shorthand $X_{ij} := X_i-X_j$ to parametrize the FI terms of the gauge nodes in the quiver diagram.
\newpage
\begin{equation}
\label{eq: aharony_mirror_fig3}
    \includegraphics[]{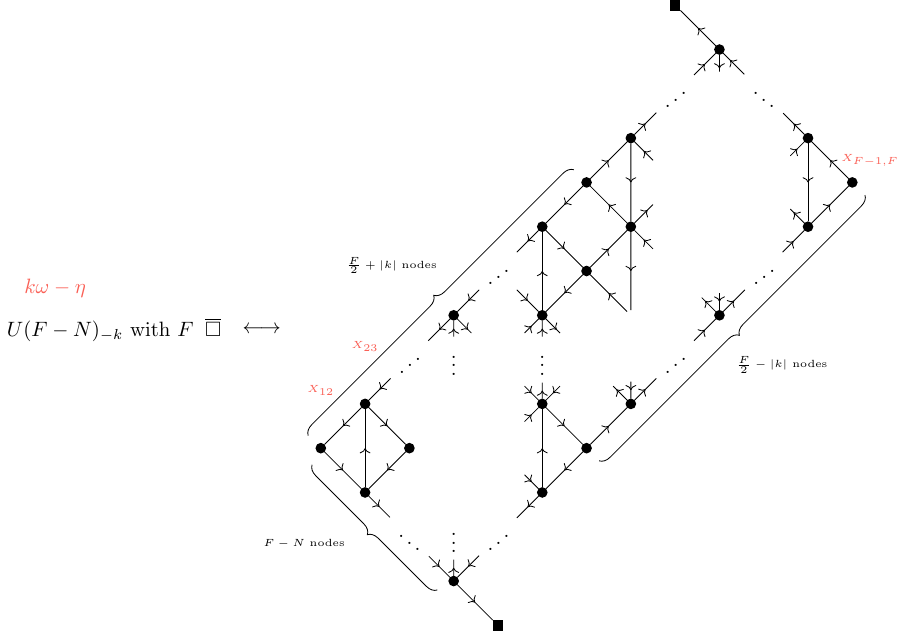}
\end{equation}

Notice that the planar quiver obtained this way is equivalent to the planar Abelian dual of Zone $\tilde 1$ in \eqref{eq:dual_zone1t_alternative}, up to a horizontal reflection and a permutation of the $X_i$ fugacities.

This concludes the argument that Aharony duality for $U(N)_k$ SQCD with $F$ fundamental chiral fields follows from the planar Abelian dualities discussed in Section \ref{sec: landscape}, together with their massive RG flows and the assumption of Aharony duality for Abelian gauge theories. 
We expect that a similar argument extends to more general $U(N)_k$ theories with both fundamental and anti-fundamental matter, although we leave this extension for future work.


\acknowledgments

We are grateful to Sara Pasquetti and Vittorio Cagioni for their comments on the draft and for their collaboration on related topics. AS thanks Davide Bason, Leonardo Goller, and Marina Moleti for helpful discussions, and Davide Morgante for help with making the diagrams (created with TikZ \cite{Ellis_2017}). SB is partially supported by the MUR-PRIN grant No. 2022NY2MXY. SR is supported by the MUR-PRIN grant No. 2022NY2MXY and the POC grant No. 41355/GRFVG. RC is supported by the STFC grant ST/X000575/1.

\appendix

\section{The Fate of the Monopole}
We now review several properties of monopole operators in three-dimensional supersymmetric gauge theories that are repeatedly used in the main text.
\subsubsection*{Monopoles in \texorpdfstring{$U(N)$}{U(N)} Chern-Simons Theories}

We begin by considering a generic $U(N)_{(k,k+\ell N)}$ gauge theory with $F$ fundamental and $A$ anti-fundamental chiral multiplets. This is shown in the quiver gauge theory below:
\begin{center}
\begin{tikzpicture}  
    \node at (0,0) (g1) [gauge, black] {$N$};
    \node at (2,0) (f1) [flavor,black] {$A$};
    \node at (-2,0) (f2) [flavor,black] {$F$};
    \draw[-<-] (0.3,0) -- (1.7,0);
    \draw[-<-] (-1.7,0) -- (-0.3,0);
    \draw[red](0.5,-0.5) node {$_{(k,k+\ell N)}$};
    \draw (-2,-0.7) node {$\vec{Y}$};
    \draw (2,-0.7) node {$\vec{X}$};
    \draw (0,0.6) node {$\vec{Z}$};
\end{tikzpicture}
\end{center}

Here $\vec{X}$ and $\vec{Y}$ denote fugacities for the $S[U(A)\times U(F)]$ flavor symmetry, while $\vec{Z}$ are the gauge fugacities associated with the Cartan of $U(N)$.

\paragraph{Quantum Numbers of Supersymmetric Monopoles:} A bare monopole operator $\mon^{\vec{m}}$, labeled by a magnetic flux vector
\(
\vec{m}=(m_1,\dots,m_N),
\)
has R-charge given by \cite{Aharony:1997bx,Borokhov:2002cg,Benini_2011}
\begin{equation}
 R[\mon^{\vec{m}}]
    =
    \Lambda \sum_{i=1}^N m_i
    - \frac{1}{2}
    \left[
        \sum_{i \neq j} |m_i - m_j|
        + F (r_f - 1) \sum_{i=1}^N |m_i|
        + A (r_a - 1) \sum_{i=1}^N |m_i|
    \right],
    \label{monfor}
\end{equation}
where $\Lambda$ parametrizes the mixing between the R-symmetry and the topological symmetry and is proportional to the imaginary part of the FI parameter.
$(r_f,r_a)$ are the R-charges of the fundamental and anti-fundamental chiral multiplets, respectively. T

The representation of the monopole under the gauge symmetry is encoded in the polynomial
\begin{equation}
    Q_{\text{gauge}}[\mon^{\vec{m}}]
    =
    -\sum_{i=1}^N \left(k Z_i + \ell \sum_{j=1}^N Z_j \right) m_i
    + \frac{1}{2}\sum_{i=1}^N (F - A) Z_i |m_i|,
    \label{eq:gauge_charge_monopole}
\end{equation}
which determines the highest weight of the gauge representation carried by the bare monopole.

A generic magnetic flux $\vec{m}$ breaks the gauge symmetry as
\[
U(N) \;\longrightarrow\; \prod_{p=1}^L U(M_p),
\qquad
\sum_{p=1}^L M_p = N,
\]
where the decomposition corresponds to blocks of equal magnetic flux. The Cartan generators are accordingly grouped as
\[
Z_{1,\dots,M_1},\;
Z_{M_1+1,\dots,M_1+M_2},\;
\dots .
\]

\subsubsection*{Monopoles in Abelian Quiver Gauge Theories}

In the main text, an important role is played by Abelian quiver gauge theories with mixed Chern--Simons interactions and matter in bifundamental representations. We therefore summarize here the corresponding monopole charge formulas.

We consider an Abelian quiver theory specified by:
\begin{itemize}
    \item $n_g$ Abelian gauge nodes with gauge parameters $u_i$, $i=1,\dots,n_g$;
    \item $n_f$ flavor nodes;
    \item A global symmetry $U(1)^r$, excluding topological symmetries, with fugacities $y_a$, $a=1,\dots,r$;
    \item Chern-Simons levels $k_i$, BF couplings $k_{ij}$, and FI parameters $\lambda_i$;
    \item Bifundamental chiral fields $\alpha_{ij}$, of charge $-1$ under node $i$ and $+1$ under node $j$;
    \item Fundamental fields $\beta_{ai}$ and anti-fundamental fields $\tilde{\beta}_{ia}$.
\end{itemize}

We consider a monopole operator with magnetic flux
\(
\vec{m}=(m_1,\dots,m_{n_g})
\)
under the gauge nodes and vanishing flux under global symmetries.

The charges of the monopole under gauge and global symmetries are compactly encoded in the polynomial
\begin{equation}
\begin{split}
\mathcal{M}(\vec{m})
=&
-\frac{1}{2}
\Bigg[
    \sum_{\alpha_{ij}} |m_i - m_j|
    \left(
        (R[\alpha_{ij}]-1)\omega
        - u_i + u_j
        + \sum_{a=1}^{r} Q_a[\alpha_{ij}]\, y_a
    \right)
\\&\qquad
    +
    \sum_{\tilde{\beta}_{ia}} |m_i|
    \left(
        (R[\tilde{\beta}_{ia}]-1)\omega
        - u_i
        + \sum_{a=1}^{r} Q_a[\tilde{\beta}_{ia}]\, y_a
    \right)
\\&\qquad
    +
    \sum_{\beta_{ai}} |m_i|
    \left(
        (R[\beta_{ai}]-1)\omega
        + u_i
        + \sum_{a=1}^{r} Q_a[\beta_{ai}]\, y_a
    \right)
\Bigg]
\\&
    - \sum_{i=1}^{n_g} k_i m_i u_i
    + \sum_{i=1}^{n_g} \lambda_i m_i
    - \frac{1}{2} \sum_{i<j} k_{ij} (m_i u_j + m_j u_i) .
\end{split}
\label{eq:monopole_charge}
\end{equation}

The first line captures contributions from bifundamental fields, the second and third from anti-fundamental and fundamental matter, respectively. The final line accounts for Chern--Simons terms, FI couplings, and BF interactions. The parameters $\lambda_i$ encode the mixing between topological symmetries and the R-symmetry.

The charges of the monopole under the various symmetries are read off as the coefficients of the corresponding fugacities. The R-charge is given by the coefficient of $\omega$ in $\mathcal{M}(\vec{m})$.

\section{From the Basic \texorpdfstring{$\mathcal{N}=4$ Abelian Confining Dualities to the $\mathcal{N}=2$}{N=4 to N=2} Abelian Dualities}\label{app: rel_duals}
In this section we show how to prove the self-dualities for $U(1)_0$ with $[2,2]$ flavors, for $U(1)_{1/2}$ with $[2,1]$ flavors and $U(1)_{1}$ with $[1,1]$ flavors assuming only that $U(1)_0$ with $[1,1]$ flavors confines.

Starting from the basic $\mathcal{N}=4$ Abelian mirror duality, we present the Aharony-like dualities needed to dualize the planar quivers presented in the paper. We start from the basic $\mathcal{N}=4$ Abelian mirror dual pair (Equation \ref{eq: basic_mirror_n=4}) which relates a $U(1)$ gauge theory with one hypermultiplet to a free hypermultiplet.
\begin{equation} \label{eq: basic_mirror_n=4}
\begin{tikzpicture}[baseline=(current bounding box).center]
    \node at (0,0) [gauge,black] {$1$};
    \node at (1.5,0) [flavor,black] {$1$};
    \draw[->, thick] (0.25,.1)--(1.2,.1);
    \draw[->, thick] (1.2,-.1)--(.25,-.1);
    \node at (.75,.3) {\tiny$Q$};
    \node at (.75,-.3) {\tiny$\tilde{Q}$};
    \node at (0.75,0) {\huge$\times$};
    \node at (.75,-.7) {$\mathcal{W}=Flip[Q\tilde{Q}]$};

    \node at (2.5,0) {$\leftrightarrow$};

    \node at (3.5,0) [flavor,black] {$1$};
    \node at (5,0) [flavor,black] {$1$};
    \node at (4.25,.3) {\tiny$X$};
    \node at (4.25,-.3) {\tiny$\tilde{X}$};
    \draw[->,thick] (3.8,.1) -- (4.7,.1);
    \draw[->,thick] (4.7,-.1)--(3.8,-.1);
    \node at (4.25,-.7) {$\mathcal{W}=0$};
\end{tikzpicture}
\end{equation}
The operator map is shown below:
\begin{equation}
    \begin{aligned}
        \mathfrak{M}^+ \, & \leftrightarrow\, X\\ 
        \mathfrak{M}^- \, & \leftrightarrow\, \tilde{X}.
    \end{aligned}
\end{equation}

\subsection*{\texorpdfstring{$U(1)_0$ with $[1,1]$}{U10 with 1,1} chiral fields:} Analogous to the $\mathcal{N}=4$ mirror pair (Equation \ref{eq: basic_mirror_n=4}), the fundamental $\mathcal{N}=2$ mirror pair is shown in Equation \ref{eq: basic_mirror_n=2}:

\begin{equation} \label{eq: basic_mirror_n=2}
\begin{tikzpicture}[baseline=(current bounding box).center]
    \node at (0,0) [gauge,black] {$1$};
    \node at (1.5,0) [flavor,black] {$1$};
    \draw[->, thick] (0.25,.1)--(1.2,.1);
    \draw[->, thick] (1.2,-.1)--(.25,-.1);
    \node at (.75,.3) {\tiny$Q$};
    \node at (.75,-.3) {\tiny$\tilde{Q}$};
    \draw[red] node at (.2,-.5) {\tiny$0$};
    \draw[FIcolor] node at (0,.5) {\tiny$\lambda$};       
    \node at (.75,-1) {$\mathcal{W}=0$};

    \node at (2.5,0) {$\leftrightarrow$};

    \node at (3.5,0) [flavor,black] {$1$};
    \node at (5,0) [flavor,black] {$1$};
    \node at (4.25,.3) {\tiny$X$};
    \node at (4.25,-.3) {\tiny$\tilde{X}$};
    \node at (4.25,0) {\huge$\times$};
    \draw[->,thick] (3.8,.1) -- (4.7,.1);
    \draw[->,thick] (4.7,-.1)--(3.8,-.1);
    \node at (4.25,-1) {$\mathcal{W}=Flip[X\tilde{X}]$};
\end{tikzpicture}
\end{equation}
with the operator map:
\begin{equation}
    \begin{aligned}
        Q\tilde{Q}\, & \leftrightarrow\, \mathcal{F}[X\tilde{X}]\\ 
        \mathfrak{M}^+\, & \leftrightarrow \, X \\ 
        \mathfrak{M}^- \, & \leftrightarrow \, \tilde{X}.
    \end{aligned}
\end{equation}

\subsection*{Self-Dual SQED}
By composing the duality operation shown in Equation \ref{eq: basic_mirror_n=4}, one can derive mirror dual pairs of SQED with a generic number of hypermultiplets \textit{a la} Kapustin-Strassler \cite{Kapustin:1999ha}. We provide a brief summary of the piecewise dualization procedure needed to find the mirror of $\mathcal{N}=2$ SQED with $F$ flavors \cite{Benvenuti:2016wet}. Each flavor is dualized using Equation \ref{eq: basic_mirror_n=2} to obtain a $U(1)^{F+1}$ with $2F$ charged fields, $F$ \textit{flippers} (and a corresponding superpotential) and $2F$ mixed BF interactions connecting the original $U(1)$ gauge group with the $2F$ new gauge factors. At this point, the original $U(1)$ gauge group has no flavors and we can perform the functional integral over it obtaining a functional delta function. Implementing the delta reduces the number of gauge groups to $F-1$, which is indeed the correct mirror dual.

To dualize the planar quivers in this paper, we need some additional self-dualities of $\mathcal{N}=2$ SQED with $F=2$. The starting point of this is the $\mathcal{N}=2$ self-dual (\textit{modulo} flips) pair shown in Equation \ref{eq: basic_self_dual_mirror_n=2}:
\begin{equation} 
\label{eq: basic_self_dual_mirror_n=2}
\begin{tikzpicture}[baseline=(current bounding box).center]
   \node at (-1.5,0) [flavor,black] {$1$};
   \node at (0,0) [gauge,black] {$1$};
   \node at (1.5,0) [flavor,black] {$1$};
   \draw[->,thick] (-1.2,.1) -- (-.3,.1);
   \draw[->,thick] (-.3,-.1) --(-1.2,-.1);
   \draw[->,thick] (.3,.1)--(1.2,.1);
   \draw[->,thick] (1.2,-.1)--(.3,-.1);
   \node at (-.75,.35) {\tiny$\tilde{Q}_1$};
   \node at (.75,.35) {\tiny$Q_2$};
   \node at (-.75,-.35) {\tiny$Q_1$};
   \node at (.75,-.35) {\tiny$\tilde{Q}_2$};
    \node at (0,-1) {$\mathcal{W}=0$};
   \node at (3,0) {$\leftrightarrow$};
   \begin{scope}[xshift=6cm]
    \node at (-1.5,0) [flavor,black] {$1$};
   \node at (0,0) [gauge,black] {$1$};
   \node at (1.5,0) [flavor,black] {$1$};
   \draw[->,thick] (-1.2,.1) -- (-.3,.1);
   \draw[->,thick] (-.3,-.1) --(-1.2,-.1);
   \draw[->,thick] (.3,.1)--(1.2,.1);
   \draw[->,thick] (1.2,-.1)--(.3,-.1);
  \node at (-.75,0) {\huge$\times$};
   \node at (.75,0) {\huge$\times$};
   \node at (-.75,.35) {\tiny$\tilde{X}_1$};
   \node at (.75,.35) {\tiny$X_2$};
   \node at (-.75,-.35) {\tiny$X_1$};
   \node at (.75,-.35) {\tiny$\tilde{X}_2$};
   \node at (0,-1) {$\mathcal{W}=Flip[X_1\tilde{X}_1, X_2\tilde{X}_2]$};
   \end{scope}
\end{tikzpicture}
\end{equation}
with the operator map:
\begin{equation}
    \begin{aligned}
        \begin{pmatrix} 
        Q_1\tilde{Q}_1 & Q_1\tilde{Q}_2 \\ Q_2\tilde{Q}_1 & Q_2\tilde{Q}_2
        \end{pmatrix} \, & \leftrightarrow \, 
        \begin{pmatrix}
            \mathcal{F}[X_1\tilde{X}_1] & \mathfrak{M}^+ \\ \mathfrak{M}^- & \mathcal{F}[X_2\tilde{X}_2]    
        \end{pmatrix}\\ \mathfrak{M}^+ \,& \leftrightarrow \, X_2\tilde{X}_1 \\ \mathfrak{M}^- \, & \leftrightarrow\, X_1\tilde{X}_2
    \end{aligned}
\end{equation}
We can apply the same duality with a different choice of mesons that are flipped on the dual side, which amounts to a relabeling of the fields $X_{1,2}$ and $\widetilde{X}_{1,2}$. This provides a third mirror dual frame (Equation \ref{eq: basic_self_dual_mirror_triality_n=2}):
\begin{equation} 
\label{eq: basic_self_dual_mirror_triality_n=2}
\begin{tikzpicture}[baseline=(current bounding box).center]
    \begin{scope}[xshift=-4cm]
        \node at (-1,1) (f1) [flavor,black] {$1$};
        \node at (0,0) (g1) [gauge,black] {$1$};
        \node at (1,1) (f2) [flavor,black] {$1$};
        \node at (-1,-1) (f3) [flavor,black] {$1$};
        \node at (1,-1) (f4) [flavor,black] {$1$};

        \draw[->-] (f1)--(g1);
        \draw[->-] (g1)--(f2);
        \draw[->-] (f4)--(g1);
        \draw[->-] (g1)--(f3);
        \draw[->-] (f3)--(f1);
        \draw[->-] (f2)--(f4);

        \draw (f1)++(.7,-.3) node {\tiny$\tilde{X}_1$};
        \draw (f2)++(-.7,-.3) node {\tiny$X_2$};
        \draw (f3)++(.7,.3) node {\tiny$X_1$};
        \draw (f4)++(-.7,.3) node {\tiny$\tilde{X}_2$};

        \node at (0,-1.7) {$\mathcal{W}=\sum_{i=1}^{2}Flip[X_i\tilde{X}_i]$};

        \node at (2,0) {$\leftrightarrow$};
  \end{scope}
  \begin{scope}
        \node at (-1,1) (f1) [flavor,black] {$1$};
        \node at (0,0) (g1) [gauge,black] {$1$};
        \node at (1,1) (f2) [flavor,black] {$1$};
        \node at (-1,-1) (f3) [flavor,black] {$1$};
        \node at (1,-1) (f4) [flavor,black] {$1$};

        \draw[->-] (f1)--(g1);
        \draw[->-] (g1)--(f2);
        \draw[->-] (f4)--(g1);
        \draw[->-] (g1)--(f3);

        \draw (f1)++(.7,-.3) node {\tiny$\tilde{Q}_1$};
        \draw (f2)++(-.7,-.3) node {\tiny$Q_2$};
        \draw (f3)++(.7,.3) node {\tiny$Q_1$};
        \draw (f4)++(-.7,.3) node {\tiny$\tilde{Q}_2$};

        \node at (0,-1.7) {$\mathcal{W}=0$};

        \node at (2,0) {$\leftrightarrow$};
  \end{scope}
  \begin{scope}[xshift=4cm]
        \node at (-1,1) (f1) [flavor,black] {$1$};
        \node at (0,0) (g1) [gauge,black] {$1$};
        \node at (1,1) (f2) [flavor,black] {$1$};
        \node at (-1,-1) (f3) [flavor,black] {$1$};
        \node at (1,-1) (f4) [flavor,black] {$1$};

        \draw[->-] (f1)--(g1);
        \draw[->-] (g1)--(f2);
        \draw[->-] (f4)--(g1);
        \draw[->-] (g1)--(f3);
        \draw[->-] (f2)--(f1);
        \draw[->-] (f3)--(f4);
        \draw (f1)++(.7,-.3) node {\tiny$\tilde{A}_1$};
        \draw (f2)++(-.7,-.3) node {\tiny$A_2$};
        \draw (f3)++(.7,.3) node {\tiny$A_1$};
        \draw (f4)++(-.7,.3) node {\tiny$\tilde{A}_2$};

        \node at (0,-1.7) {$\mathcal{W}=\sum_{i\neq j}^{2}Flip[A_i\tilde{A}_j]$};
  \end{scope}
\end{tikzpicture}
\end{equation}

where the mapping of operators across the triality is given by:
\begin{equation}
    \begin{aligned}
       X_1\tilde{X}_2\, &\leftrightarrow\, \mathfrak{M}^- \, \leftrightarrow \, A_2\tilde{A}_2\\
       X_2\tilde{X}_1\, & \leftrightarrow\, \mathfrak{M}^+ \, \leftrightarrow\, A_1\tilde{A}_1 \\
       \begin{pmatrix}
           \mathcal{F}[X_1\tilde{X}_1] & \mathfrak{M}^+ \\ \mathfrak{M}^- & \mathcal{F}[X_2\tilde{X}_2]
       \end{pmatrix} \, & \leftrightarrow Q\tilde{Q}\, \leftrightarrow\, \begin{pmatrix}
           \mathfrak{M}^+ & \mathcal{F}[A_2\tilde{A}_1]\\ \mathcal{F}[A_1\tilde{A}_2] & \mathfrak{M}^-
       \end{pmatrix}.
    \end{aligned}
\end{equation}  
We arrive at an analogous Seiberg-like duality by observing the first and third frame of Equation \ref{eq: basic_self_dual_mirror_triality_n=2}:
\begin{equation} 
\label{eq: basic_self_dual_seiberg_n=2}
\begin{tikzpicture}[baseline=(current bounding box).center]
    \begin{scope}[xshift=-4cm]
        \node at (-1,1) (f1) [flavor,black] {$1$};
        \node at (0,0) (g1) [gauge,black] {$1$};
        \node at (1,1) (f2) [flavor,black] {$1$};
        \node at (-1,-1) (f3) [flavor,black] {$1$};
        \node at (1,-1) (f4) [flavor,black] {$1$};

        \draw[->-] (f1)--(g1);
        \draw[->-] (g1)--(f2);
        \draw[->-] (f4)--(g1);
        \draw[->-] (g1)--(f3);
        \draw[->-] (f3)--(f1);
        \draw[->-] (f2)--(f4);

        \draw (f1)++(.7,-.3) node {\tiny$\tilde{X}_1$};
        \draw (f2)++(-.7,-.3) node {\tiny$X_2$};
        \draw (f3)++(.7,.3) node {\tiny$X_1$};
        \draw (f4)++(-.7,.3) node {\tiny$\tilde{X}_2$};

        \node at (0,-1.7) {$\mathcal{W}=\sum_{i=1}^2Flip[X_i\tilde{X}_i]$};

        \node at (2.5,0) {$\leftrightarrow$};
  \end{scope}
 
  \begin{scope}[xshift=1cm]
\node at (-1,1) (f1) [flavor,black] {$1$};
        \node at (0,0) (g1) [gauge,black] {$1$};
        \node at (1,1) (f2) [flavor,black] {$1$};
        \node at (-1,-1) (f3) [flavor,black] {$1$};
        \node at (1,-1) (f4) [flavor,black] {$1$};

        \draw[->-] (f1)--(g1);
        \draw[->-] (g1)--(f2);
        \draw[->-] (f4)--(g1);
        \draw[->-] (g1)--(f3);
        \draw[->-] (f2)--(f1);
        \draw[->-] (f3)--(f4);
        \draw (f1)++(.7,-.3) node {\tiny$\tilde{A}_1$};
        \draw (f2)++(-.7,-.3) node {\tiny$A_2$};
        \draw (f3)++(.7,.3) node {\tiny$A_1$};
        \draw (f4)++(-.7,.3) node {\tiny$\tilde{A}_2$};

        \node at (0,-1.7) {$\mathcal{W}=\sum_{i\neq j}^{2}Flip[A_i\tilde{A}_j]$};
  \end{scope}
\end{tikzpicture}
\end{equation}
We can \textit{flip} the flippers in the quiver theory shown on the l.h.s.~ of Equation \ref{eq: basic_self_dual_seiberg_n=2} to obtain a Seiberg-like duality:

\begin{equation} 
\label{eq: basic_self_dual_seiberg_d_n=2}
\begin{tikzpicture}[baseline=(current bounding box).center]
    \begin{scope}[xshift=-4cm]
        \node at (-1,1) (f1) [flavor,black] {$1$};
        \node at (0,0) (g1) [gauge,black] {$1$};
        \node at (1,1) (f2) [flavor,black] {$1$};
        \node at (-1,-1) (f3) [flavor,black] {$1$};
        \node at (1,-1) (f4) [flavor,black] {$1$};

        \draw[->-] (f1)--(g1);
        \draw[->-] (g1)--(f2);
        \draw[->-] (f4)--(g1);
        \draw[->-] (g1)--(f3);

        \draw (f1)++(.7,-.3) node {\tiny$\tilde{X}_1$};
        \draw (f2)++(-.7,-.3) node {\tiny$X_2$};
        \draw (f3)++(.7,.3) node {\tiny$X_1$};
        \draw (f4)++(-.7,.3) node {\tiny$\tilde{X}_2$};

        \node at (0,-2) {$\mathcal{W}=0$};

        \node at (2.5,0) {$\leftrightarrow$};
  \end{scope}
 
  \begin{scope}[xshift=1.5cm]
        \node at (-1,1) (f1) [flavor,black] {$1$};
        \node at (0,0) (g1) [gauge,black] {$1$};
        \node at (1,1) (f2) [flavor,black] {$1$};
        \node at (-1,-1) (f3) [flavor,black] {$1$};
        \node at (1,-1) (f4) [flavor,black] {$1$};

        \draw[->-] (f1)--(g1);
        \draw[->-] (g1)--(f2);
        \draw[->-] (f4)--(g1);
        \draw[->-] (g1)--(f3);
        \draw[->-] (f2)--(f1);
        \draw[->-,red] (f1) to [out=20, in=160](f2);
        \draw[->-,red] (f4) to [out=-160, in=-20](f3);
        \draw[->-] (f3)--(f4);
        \draw (f1)++(.7,-.3) node {\tiny$\tilde{A}_1$};
        \draw (f2)++(-.7,-.3) node {\tiny$A_2$};
        \draw (f3)++(.7,.3) node {\tiny$A_1$};
        \draw (f4)++(-.7,.3) node {\tiny$\tilde{A}_2$};

        \node at (0,-2) {$\mathcal{W}=\sum_{i\neq j}^2Flip[A_i\tilde{A}_j]+Flip[\mathfrak{M}^{\pm}]$};
  \end{scope}
\end{tikzpicture}
\end{equation}
with the operator map:
\begin{equation}
    \begin{aligned}
      \mathfrak{M}^+ \, & \leftrightarrow\, \mathcal{F}[A_2\tilde{A}_1]\\ 
      \mathfrak{M}^- \, & \leftrightarrow\, \mathcal{F}[A_1\tilde{A}_2]\\
      X\tilde{X}\, & \leftrightarrow \, \begin{pmatrix}
          \mathcal{F}[\mathfrak{M}^+] & A_2\tilde{A}_2\\
          A_1\tilde{A}_1 & \mathcal{F}[\mathfrak{M}^-]
      \end{pmatrix}.
    \end{aligned}
\end{equation}  

We are now able to establish a fourth mirror dual frame (Equation \ref{eq: basic_self_dual_mirror_triality_2_n=2}):
\begin{equation} 
\label{eq: basic_self_dual_mirror_triality_2_n=2}
\begin{tikzpicture}[baseline=(current bounding box).center]
    \begin{scope}[xshift=-5cm]
        \node at (-1,1) (f1) [flavor,black] {$1$};
        \node at (0,0) (g1) [gauge,black] {$1$};
        \node at (1,1) (f2) [flavor,black] {$1$};
        \node at (-1,-1) (f3) [flavor,black] {$1$};
        \node at (1,-1) (f4) [flavor,black] {$1$};

        \draw[->-] (f1)--(g1);
        \draw[->-] (g1)--(f2);
        \draw[->-] (f4)--(g1);
        \draw[->-] (g1)--(f3);
        \draw[->-] (f2)--(f1);
        \draw[->-,red] (f1) to [out=20, in=160](f2);
        \draw[->-,red] (f4) to [out=-160, in=-20](f3);
        \draw[->-] (f3)--(f4);
        \draw (f1)++(.7,-.3) node {\tiny$\tilde{A}_1$};
        \draw (f2)++(-.7,-.3) node {\tiny$A_2$};
        \draw (f3)++(.7,.3) node {\tiny$A_1$};
        \draw (f4)++(-.7,.3) node {\tiny$\tilde{A}_2$};

        \node at (0,-2) {$\mathcal{W}=\sum_{i\neq j}^2Flip[A_i\tilde{A}_j]+Flip[\mathfrak{M}^{\pm}]$};
        \node at (2,0) {$\leftrightarrow$};
  \end{scope}
  \begin{scope}
        \node at (-1,1) (f1) [flavor,black] {$1$};
        \node at (0,0) (g1) [gauge,black] {$1$};
        \node at (1,1) (f2) [flavor,black] {$1$};
        \node at (-1,-1) (f3) [flavor,black] {$1$};
        \node at (1,-1) (f4) [flavor,black] {$1$};

        \draw[->-] (f1)--(g1);
        \draw[->-] (g1)--(f2);
        \draw[->-] (f4)--(g1);
        \draw[->-] (g1)--(f3);

        \draw (f1)++(.7,-.3) node {\tiny$\tilde{Q}_1$};
        \draw (f2)++(-.7,-.3) node {\tiny$Q_2$};
        \draw (f3)++(.7,.3) node {\tiny$Q_1$};
        \draw (f4)++(-.7,.3) node {\tiny$\tilde{Q}_2$};

        \node at (0,-2) {$\mathcal{W}=0$};

        \node at (2,0) {$\leftrightarrow$};
  \end{scope}
  \begin{scope}[xshift=4cm]
        \node at (-1,1) (f1) [flavor,black] {$1$};
        \node at (0,0) (g1) [gauge,black] {$1$};
        \node at (1,1) (f2) [flavor,black] {$1$};
        \node at (-1,-1) (f3) [flavor,black] {$1$};
        \node at (1,-1) (f4) [flavor,black] {$1$};

        \draw[->-] (f1)--(g1);
        \draw[->-] (g1)--(f2);
        \draw[->-] (f4)--(g1);
        \draw[->-] (g1)--(f3);
        \draw[->-] (f2)--(f1);
        \draw[->-] (f3)--(f4);
        \draw (f1)++(.7,-.3) node {\tiny$\tilde{B}_1$};
        \draw (f2)++(-.7,-.3) node {\tiny$B_1$};
        \draw (f3)++(.7,.3) node {\tiny$B_2$};
        \draw (f4)++(-.7,.3) node {\tiny$\tilde{B}_2$};

        \node at (0,-2) {$\mathcal{W}=\sum_{i=1}^2Flip[B_i\tilde{B}_i]$};
  \end{scope}
\end{tikzpicture}
\end{equation}
with corresponding operator map:
\begin{equation}
    \begin{aligned}
        \mathcal{F}[A_2\tilde{A}_1]\,&\leftrightarrow\, \mathfrak{M}^+\, \leftrightarrow \, B_2\tilde{B}_1 \\
        \mathcal{F}[A_1\tilde{A}_2]\, & \leftrightarrow\, \mathfrak{M}^-\, \leftrightarrow\, B_1\tilde{B}_2\\
        \begin{pmatrix}
            \mathcal{F}[\mathfrak{M}^+] & A_2\tilde{A}_2\\ A_1\tilde{A}_1 & \mathfrak{M}^-
        \end{pmatrix} \, & \leftrightarrow\, Q\tilde{Q}\, \leftrightarrow\, \begin{pmatrix}
            \mathfrak{M}^+ & \mathcal{F}[B_1\tilde{B}_1]\\ \mathcal{F}[B_2\tilde{B}_2] & \mathfrak{M}^-
        \end{pmatrix}.
    \end{aligned}
\end{equation}

We arrive at an analogous Seiberg-like duality by observing the first and third frame of Equation \ref{eq: basic_self_dual_mirror_triality_2_n=2}:
\begin{equation} 
\label{eq: basic_self_dual_mirror_d_2_n=2}
\begin{tikzpicture}[baseline=(current bounding box).center]
    \begin{scope}[xshift=-4cm]
        \node at (-1,1) (f1) [flavor,black] {$1$};
        \node at (0,0) (g1) [gauge,black] {$1$};
        \node at (1,1) (f2) [flavor,black] {$1$};
        \node at (-1,-1) (f3) [flavor,black] {$1$};
        \node at (1,-1) (f4) [flavor,black] {$1$};

        \draw[->-] (f1)--(g1);
        \draw[->-] (g1)--(f2);
        \draw[->-] (f4)--(g1);
        \draw[->-] (g1)--(f3);
        \draw[->-] (f2)--(f1);
        \draw[->-,red] (f1) to [out=20, in=160](f2);
        \draw[->-,red] (f4) to [out=-160, in=-20](f3);
        \draw[->-] (f3)--(f4);
        \draw (f1)++(.7,-.3) node {\tiny$\tilde{A}_1$};
        \draw (f2)++(-.7,-.3) node {\tiny$A_2$};
        \draw (f3)++(.7,.3) node {\tiny$A_1$};
        \draw (f4)++(-.7,.3) node {\tiny$\tilde{A}_2$};

        \node at (0,-2) {$\mathcal{W}=\sum_{i\neq j}^2Flip[A_i\tilde{A}_j]+Flip[\mathfrak{M}^{\pm}]$};
        \node at (2.5,0) {$\leftrightarrow$};
  \end{scope}

  \begin{scope}[xshift=1cm]
\node at (-1,1) (f1) [flavor,black] {$1$};
        \node at (0,0) (g1) [gauge,black] {$1$};
        \node at (1,1) (f2) [flavor,black] {$1$};
        \node at (-1,-1) (f3) [flavor,black] {$1$};
        \node at (1,-1) (f4) [flavor,black] {$1$};

        \draw[->-] (f1)--(g1);
        \draw[->-] (g1)--(f2);
        \draw[->-] (f4)--(g1);
        \draw[->-] (g1)--(f3);
        \draw[->-] (f2)--(f1);
        \draw[->-] (f3)--(f4);
        \draw (f1)++(.7,-.3) node {\tiny$\tilde{B}_1$};
        \draw (f2)++(-.7,-.3) node {\tiny$B_1$};
        \draw (f3)++(.7,.3) node {\tiny$B_2$};
        \draw (f4)++(-.7,.3) node {\tiny$\tilde{B}_2$};

        \node at (0,-2) {$\mathcal{W}=\sum_{i=1}^2Flip[B_i\tilde{B}_i]$};;
  \end{scope}
\end{tikzpicture}
\end{equation}
We can once again \textit{flip} the flippers in the quiver theory shown on the r.h.s.~ of Equation \ref{eq: basic_self_dual_mirror_d_2_n=2} to obtain a Seiberg-like duality:
\begin{equation} 
\label{eq: basic_self_dual_mirror_d_fin_n=2}
\begin{tikzpicture}[baseline=(current bounding box).center]
    \begin{scope}[xshift=-4cm]
        \node at (-1,1) (f1) [flavor,black] {$1$};
        \node at (0,0) (g1) [gauge,black] {$1$};
        \node at (1,1) (f2) [flavor,black] {$1$};
        \node at (-1,-1) (f3) [flavor,black] {$1$};
        \node at (1,-1) (f4) [flavor,black] {$1$};

        \draw[->-] (f1)--(g1);
        \draw[->-] (g1)--(f2);
        \draw[->-] (f4)--(g1);
        \draw[->-] (g1)--(f3);
        \draw[->-] (f2)--(f1);
        \draw[->-] (f3)--(f1);
        \draw[->-] (f2)--(f4);
         \draw[->-,red] (f1) to [out=20, in=160](f2);
        \draw[->-,red] (f4) to [out=-160, in=-20](f3);
        \draw[->-] (f3)--(f4);
        \draw (f1)++(.7,-.3) node {\tiny$\tilde{q}_1$};
        \draw (f2)++(-.7,-.3) node {\tiny$q_2$};
        \draw (f3)++(.7,.3) node {\tiny$q_1$};
        \draw (f4)++(-.7,.3) node {\tiny$\tilde{q}_2$};

        \node at (0,-2) {$\mathcal{W}=\sum_{i,j}Flip[q_i\tilde{q}_j]+Flip[\mathfrak{M}^{\pm}]$};
        
        \node at (2.5,0) {$\leftrightarrow$};
  \end{scope}

  \begin{scope}[xshift=1cm]
        \node at (-1,1) (f1) [flavor,black] {$1$};
        \node at (0,0) (g1) [gauge,black] {$1$};
        \node at (1,1) (f2) [flavor,black] {$1$};
        \node at (-1,-1) (f3) [flavor,black] {$1$};
        \node at (1,-1) (f4) [flavor,black] {$1$};

        \draw[->-] (f1)--(g1);
        \draw[->-] (g1)--(f2);
        \draw[->-] (f4)--(g1);
        \draw[->-] (g1)--(f3);
        
        \draw (f1)++(.7,-.3) node {\tiny$\tilde{Q}_1$};
        \draw (f2)++(-.7,-.3) node {\tiny$Q_2$};
        \draw (f3)++(.7,.3) node {\tiny$Q_1$};
        \draw (f4)++(-.7,.3) node {\tiny$\tilde{Q}_2$};

        \node at (0,-2) {$\mathcal{W}=0$};
  \end{scope}
\end{tikzpicture}
\end{equation}
with operator map:

\begin{equation}
    \begin{aligned}
        \mathcal{F}[\mathfrak{M}^{\pm}] \, &\leftrightarrow\,  \mathfrak{M}^{\pm}\\
        \begin{pmatrix}
            \mathcal{F}[q_2\tilde{q}_1] & \mathcal{F}[q_1\tilde{q}_1]\\ \mathcal{F}[q_2\tilde{q}_2] & \mathcal{F}[q_1\tilde{q}_2]
        \end{pmatrix} \, & \leftrightarrow\, \begin{pmatrix}
            Q_1\tilde{Q}_1 & Q_1\tilde{Q}_2\\ Q_2\tilde{Q}_1 & Q_2\tilde{Q}_2
        \end{pmatrix}.
    \end{aligned}
\end{equation}

Thus, we have recovered the known Aharony self-duality \textit{modulo flips} of $U(1)$ with 2 flavors \cite{Aharony:1997gp} by only invoking basic $\mathcal{N}=4$ Abelian mirror duality (Equation \ref{eq: basic_mirror_n=4}). Starting from the basic $\mathcal{N}=2$ mirror pair (Equation \ref{eq: basic_self_dual_mirror_n=2}), one can keep track of the CS and mixed BF interactions in the theory, which would yield the quiver description: 

\begin{equation} 
\label{eq: basic_self_dual_mirror_d_fin_v2_n=2}
\begin{tikzpicture}[baseline=(current bounding box).center]
    \begin{scope}[xshift=-4cm]
        \node at (-1,1) (f1) [flavor,black] {$1$};
        \node at (0,0) (g1) [gauge,black] {$1$};
        \node at (1,1) (f2) [flavor,black] {$1$};
        \node at (-1,-1) (f3) [flavor,black] {$1$};
        \node at (1,-1) (f4) [flavor,black] {$1$};

        \draw[->-] (f1)--(g1);
        \draw[->-] (g1)--(f2);
        \draw[->-] (f4)--(g1);
        \draw[->-] (g1)--(f3);
        \draw[->-] (f2)--(f1);
        \draw[->-] (f3)--(f1);
        \draw[->-] (f2)--(f4);
         \draw[->-,red] (f1) to [out=20, in=160](f2);
        \draw[->-,red] (f4) to [out=-160, in=-20](f3);
        \draw[->-] (f3)--(f4);

        \draw[CScolor] (g1)++(.2,-.4) node {\tiny$0$};
        \draw[FIcolor] (g1)++(0,.5) node {\tiny$-\lambda$};
        \draw[BFcolor] (f1)++(1,-.2) node {\tiny$0$};
        \draw[BFcolor] (f3)++(1,.2) node {\tiny$0$};
        \draw[BFcolor] (f1)++(-.3,-1) node {\tiny$-1$};
        \draw[BFcolor] (f2)++(.3,-1) node {\tiny$1$};
        \draw[BFcolor] (f1)++(.3,-.6) node {\tiny$1$};
        \draw[BFcolor] (f2)++(-.3,-.6) node {\tiny$-1$};
        \draw[BFcolor] (f3)++(.3,.6) node {\tiny$1$};
        \draw[BFcolor] (f4)++(-.3,.6) node {\tiny$-1$};

        \node at (0,-2) {$\mathcal{W}=\mathcal{W}_{planar}+Flip[\mathfrak{M}^{\pm}]$};
        
        \node at (2.5,0) {$\leftrightarrow$};
  \end{scope}

  \begin{scope}[xshift=1cm]
        \node at (-1,1) (f1) [flavor,black] {$1$};
        \node at (0,0) (g1) [gauge,black] {$1$};
        \node at (1,1) (f2) [flavor,black] {$1$};
        \node at (-1,-1) (f3) [flavor,black] {$1$};
        \node at (1,-1) (f4) [flavor,black] {$1$};

        \draw[-<-] (f1)--(g1);
        \draw[-<-] (g1)--(f2);
        \draw[-<-] (f4)--(g1);
        \draw[-<-] (g1)--(f3);

        \draw[CScolor] (g1)++(.2,-.4) node {\tiny$0$};
        \draw[CScolor] (f1)++(.5,-.2) node {\tiny$\frac{1}{2}$};
        \draw[CScolor] (f2)++(.5,-.2) node {\tiny$-\frac{1}{2}$};
        \draw[CScolor] (f3)++(.5,-.2) node {\tiny$\frac{1}{2}$};
        \draw[CScolor] (f4)++(.5,-.2) node {\tiny$-\frac{1}{2}$};
        \draw[FIcolor] (g1)++(0,.5) node {\tiny$\lambda$};
        \draw[FIcolor] (f1)++(0,.5) node {\tiny$-\frac{\lambda}{2}$};
        \draw[FIcolor] (f2)++(0,.5) node {\tiny$-\frac{\lambda}{2}$};
        \draw[FIcolor] (f3)++(0,-.5) node {\tiny$-\frac{\lambda}{2}$};
        \draw[FIcolor] (f4)++(0,-.5) node {\tiny$-\frac{\lambda}{2}$};
     
        \draw[BFcolor] (f1)++(.3,-.6) node {\tiny$-1$};
        \draw[BFcolor] (f2)++(-.3,-.6) node {\tiny$1$};
        \draw[BFcolor] (f3)++(.3,.6) node {\tiny$-1$};
        \draw[BFcolor] (f4)++(-.3,.6) node {\tiny$1$};

        \node at (0,-2) {$\mathcal{W}=0$};
  \end{scope}
\end{tikzpicture}
\end{equation}

\subsubsection*{Mass Deformations of Known Aharony Duals}
We can obtain other pairs of Aharony duals by adding suitable real mass deformations to the dual pairs discussed above.

\subsubsection*{\texorpdfstring{$U(1)_{-\frac{1}{2}}$ with $[1,0]$}{U112 with 1,0} chiral fields:} The mirror pair shown in Equation \ref{eq: basic_mirror_n=2_Cs} can be obtained by adding a real mass deformation to the mirror pair shown in Equation \ref{eq: basic_mirror_n=2} and integrating out 1 chiral field:

\begin{equation} \label{eq: basic_mirror_n=2_Cs}
\begin{tikzpicture}[baseline=(current bounding box).center]
    \node at (0,0) (g1) [gauge,black] {$1$};
    \node at (1.5,0) (f1) [flavor,black] {$1$};
    \draw[->-, thick] (g1)--(f1);

    \node at (.75,.3) {\tiny$Q$};
   
    \draw[red] node at (.2,-.5) {\tiny$-\frac{1}{2}$};
    \draw[FIcolor] node at (0,.5) {\tiny$\lambda$};       
    \node at (.75,-1) {$\mathcal{W}=0$};

    \node at (2.5,0) {$\leftrightarrow$};

    \node at (3.5,0) (a1) [flavor,black] {$1$};
    \draw[red] node at (3.8,-.5) {\tiny$\frac{1}{2}$};
    \node at (4,-.2) {$,$};
    \node at (5.2,0) {$X\text{ (singlet)}$};

    \node at (4.05,-1) {$\mathcal{W}=0$};
\end{tikzpicture}
\end{equation}

with the operator map:
\begin{equation}
    \mathfrak{M}^+\, \leftrightarrow\, X.
\end{equation}
Mirror symmetry maps topological symmetries to flavor symmetries - in this case, the topological symmetry of the electric CS-SQED theory ($U(1)_{\lambda}$) is mapped to a flavor symmetry in the dual theory, and the singlet $X$ carries a charge $+1$ under it.

\subsubsection*{\texorpdfstring{$U(1)_{-\frac{1}{2}}$ with $[1,2]$}{U112 with 1,2} chiral fields:} 
Starting from the dual pair shown in Equation \ref{eq: basic_self_dual_mirror_d_fin_v2_n=2}, we can add a suitable real mass deformation and integrate out one chiral field on the r.h.s.~ of the duality, which generates a CS level for the $U(1)$ gauge group. Upon mapping this deformation to the l.h.s.~, we flow to the duality shown in Equation \ref{eq: U_1_1/2_w2_1} (note that we have assigned a trial R-charge $r$ to the fundamentals $Q$ and anti-fundamentals $\tilde{Q}$ of the theory shown on the r.h.s.~):

\begin{equation} 
\label{eq: U_1_1/2_w2_1}
\begin{tikzpicture}[baseline=(current bounding box).center]
    \begin{scope}[xshift=-4cm]
        \node at (0,0) (g1) [gauge,black] {$1$};
        \node at (1,1) (f2) [flavor,black] {$1$};
        \node at (-1,-1) (f3) [flavor,black] {$1$};
        \node at (1,-1) (f4) [flavor,black] {$1$};

        \draw[->-] (g1)--(f2);
        \draw[->-] (f4)--(g1);
        \draw[->-] (g1)--(f3);
        \draw[->-] (f2)--(f4);
        \draw[->-,red] (f4) to [out=-160, in=-20](f3);
        \draw[->-] (f3)--(f4);

        \draw[CScolor] (g1)++(.15,-.5) node {\tiny$\frac{1}{2}$};
        \draw[FIcolor] (g1)++(-.1,.5) node {\tiny$-\lambda-\frac{1}{2}$};
        \draw[BFcolor] (f3)++(1,.2) node {\tiny$0$};
        \draw[BFcolor] (f2)++(.3,-1) node {\tiny$1$};
        \draw[BFcolor] (f2)++(-.3,-.6) node {\tiny$-1$};
        \draw[BFcolor] (f3)++(.3,.6) node {\tiny$1$};
        \draw[BFcolor] (f4)++(-.3,.6) node {\tiny$-1$};

        \node at (0,-2) {$\mathcal{W}=\mathcal{W}_{planar}+Flip[\mathfrak{M}^{+}]$};
        
        \node at (2.5,0) {$\leftrightarrow$};
  \end{scope}

  \begin{scope}[xshift=1cm]
        \node at (0,0) (g1) [gauge,black] {$1$};
        \node at (1,1) (f2) [flavor,black] {$1$};
        \node at (-1,-1) (f3) [flavor,black] {$1$};
        \node at (1,-1) (f4) [flavor,black] {$1$};

        \draw[-<-] (g1)--(f2);
        \draw[-<-] (f4)--(g1);
        \draw[-<-] (g1)--(f3);

        \draw[CScolor] (g1)++(.1,-.5) node {\tiny$-\frac{1}{2}$};
        \draw[CScolor] (f2)++(.5,-.2) node {\tiny$-\frac{1}{2}$};
        \draw[CScolor] (f3)++(.5,-.2) node {\tiny$1$};
        \draw[CScolor] (f4)++(.5,-.2) node {\tiny$-\frac{1}{2}$};
        \draw[FIcolor] (g1)++(0,.5) node {\tiny$\lambda$};
        \draw[FIcolor] (f2)++(0,.5) node {\tiny$-\lambda - \frac{1-r}{2}$};
        \draw[FIcolor] (f3)++(0,-.5) node {\tiny$-\frac{\lambda}{2}-\frac{3r-1}{4}$};
        \draw[FIcolor] (f4)++(0,-.5) node {\tiny$-\frac{\lambda}{2}-\frac{1-r}{4}$};
     
        \draw[BFcolor] (f2)++(-.3,-.6) node {\tiny$1$};
        \draw[BFcolor] (f3)++(.3,.6) node {\tiny$-1$};
        \draw[BFcolor] (f4)++(-.3,.6) node {\tiny$1$};

        \node at (0,-2) {$\mathcal{W}=0$};
  \end{scope}
\end{tikzpicture}
\end{equation}
with the map of operators:
\begin{equation}
    \begin{aligned}
        \mathcal{F}[\mathfrak{M}^+]\, & \leftrightarrow\, \mathfrak{M}^- \\ 
        \text{flippers}\, & \leftrightarrow\, Q\tilde{Q}\,.
    \end{aligned}
\end{equation}

\subsubsection*{\texorpdfstring{$U(1)_{-1}$ with $[1,1]$}{U11 with 1,1} chiral fields:} 
Starting from the dual pair shown in Equation \ref{eq: U_1_1/2_w2_1}, we can add another suitable real mass deformation and integrate out one antichiral field on the r.h.s.~ of the duality. Upon mapping this deformation to the l.h.s.~, we flow to the duality shown in Equation \ref{eq: U_1_1_w1_1}:

\begin{equation} 
\label{eq: U_1_1_w1_1}
\begin{tikzpicture}[baseline=(current bounding box).center]
    \begin{scope}[xshift=-4cm]
        \node at (0,0) (g1) [gauge,black] {$1$};
        \node at (1,1) (f2) [flavor,black] {$1$};
        \node at (1,-1) (f4) [flavor,black] {$1$};

        \draw[->-] (g1)--(f2);
        \draw[->-] (f4)--(g1);
        \draw[->-] (f2)--(f4);

        \draw[CScolor] (g1)++(.15,-.5) node {\tiny$1$};
        \draw[FIcolor] (g1)++(-.1,.5) node {\tiny$-\lambda$};
        \draw[BFcolor] (f2)++(.3,-1) node {\tiny$1$};
        \draw[BFcolor] (f2)++(-.3,-.6) node {\tiny$-1$};
        \draw[BFcolor] (f4)++(-.3,.6) node {\tiny$-1$};

        \node at (.5,-2) {$\mathcal{W}=\mathcal{W}_{planar}$};
        
        \node at (2.5,0) {$\leftrightarrow$};
  \end{scope}

  \begin{scope}[xshift=-.5cm]
        \node at (0,0) (g1) [gauge,black] {$1$};
        \node at (1,1) (f2) [flavor,black] {$1$};
        \node at (1,-1) (f4) [flavor,black] {$1$};

        \draw[-<-] (g1)--(f2);
        \draw[-<-] (f4)--(g1);

        \draw[CScolor] (g1)++(.1,-.5) node {\tiny$-1$};
        \draw[CScolor] (f2)++(.5,-.2) node {\tiny$-\frac{1}{2}$};
        \draw[CScolor] (f4)++(.5,-.2) node {\tiny$-\frac{1}{2}$};
        \draw[FIcolor] (g1)++(0,.5) node {\tiny$\lambda$};
        \draw[FIcolor] (f2)++(0,.5) node {\tiny$-\lambda$};
        \draw[FIcolor] (f4)++(0,-.5) node {\tiny$-\lambda$};
     
        \draw[BFcolor] (f2)++(-.3,-.6) node {\tiny$1$};
        \draw[BFcolor] (f4)++(-.3,.6) node {\tiny$1$};

        \node at (.5,-2) {$\mathcal{W}=0$};
  \end{scope}
\end{tikzpicture}
\end{equation}
with the flipper of the l.h.s.~ being mapped to the gauge invariant meson $Q\tilde{Q}$ on the r.h.s.~.

\subsection{An Application: Recovering the Symmetric Bifundamental Block from a Planar Abelian Quiver Gauge Theory}
\begin{figure}[]
    \centering
    \includegraphics[width=0.4\linewidth]{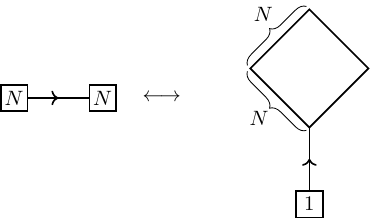}
    \caption{The duality between the $[N,N]$ bifundamental block and a planar quiver gauge theory is shown here.}
    \label{fig: symmetric bifundamental block}
\end{figure}

In \cite{Benvenuti:2025a}, it was demonstrated that $N$ free chiral fields are mirror dual to a planar quiver gauge theory. This is schematically shown in Figure \ref{fig: symmetric bifundamental block}. 

Our aim in this appendix is to sequentially confine the planar quiver using only the Aharony dualities discussed previously. 



For clarity, we fix \( N = 3 \). We begin with the planar mirror of the \([3,3]\) bifundamental\footnote{We follow the same conventions for the Chern–Simons, mixed Chern–Simons, and FI interactions as before, and will only mention them explicitly when they differ from these standard assignments.  
The FI term associated with the gauge node of fugacity $u_{3,3}$ receives an additional contribution of $-\tfrac{\omega}{2}$ due to mixing with the $U(1)_R$ symmetry which has been suppressed for brevity.
}:
\begin{equation}
    \includegraphics[width=0.35\linewidth]{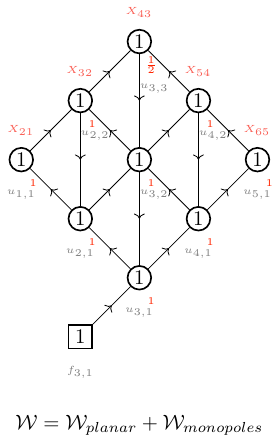}
\end{equation}

\begin{figure}[H]
    \centering
    \includegraphics[width=1\linewidth]{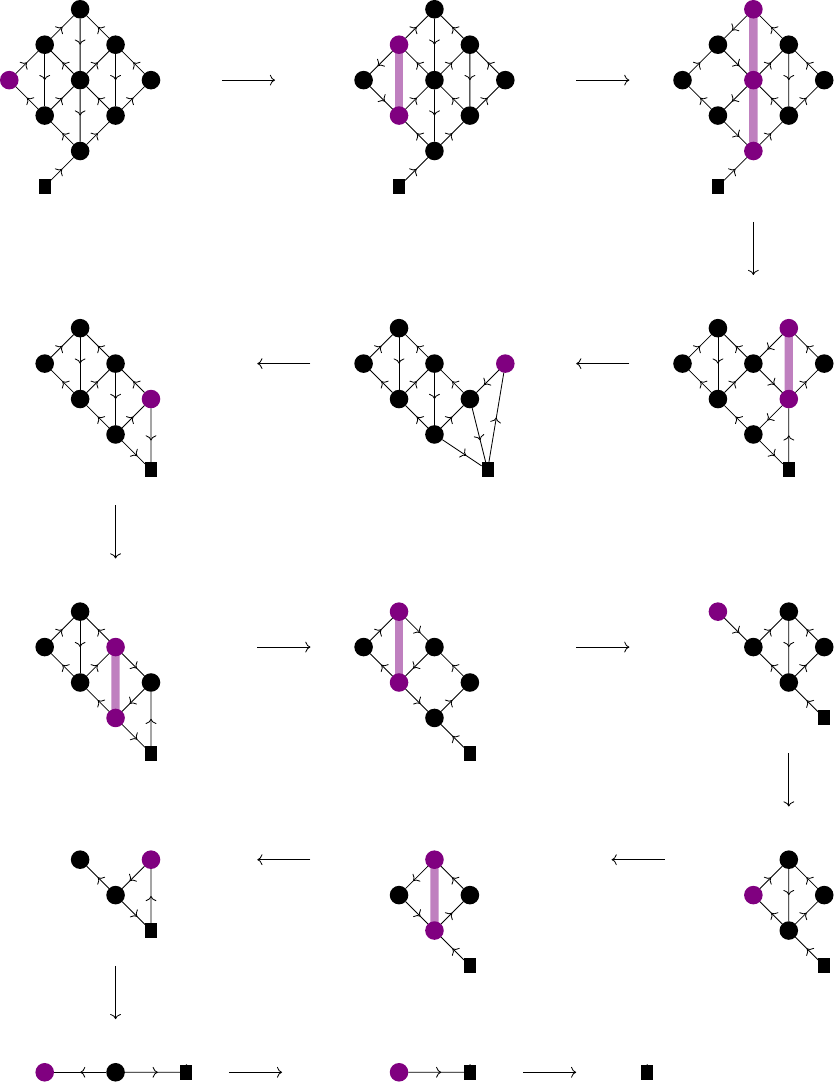}
    \caption{Schematic summary of the dualization sequence leading to the confinement of the $[3,3]$ symmetric block. In each frame, the columns being dualized are highlighted in violet, meaning that we perform Aharony duality on all the nodes in the column in any order. Each dualization will change the CS and FI data, these can be tracked by looking at the detailed dualities at the beginning of the section.}
    \label{fig: aharony dual sym}
\end{figure}

We now perform a sequence of local Aharony dualizations, column by column. We provide a schematic summary of the same in Figure \ref{fig: aharony dual sym}.
\newpage

By generalizing this procedure, one can show that any \([N,N]\) bifundamental block can be systematically confined to yield $N^2$ free chirals. 

\section{Real Mass Flows Across Aharony-like Duality on the \texorpdfstring{$\mathbf{S}_b^3$}{S3b} Partition Function}\label{app: mass flow S3b}

In this Appendix, we consider some of the real mass flows described in Section \ref{subsec: mass flows Aharony} from the point of view of the $\mathbf{S}_b^3$ partition function. 
Our main goal is to show that, assuming only the “pseudo-self-duality":
\begin{equation}    \label{app:aharony_pseudo_self_dual}
U(N)_{0} \with 2N \square \dualto U(N)_{0} \with 2N \overline{\square}
\end{equation}
for any $N$, one can derive all Aharony dual pairs for SQCD with only (anti-) fundamentals by performing mass deformations.
This derivation is split into three steps, each of which can be followed on the $\mathbf{S}_b^3$ partition function:
\begin{itemize}
\item Real mass flows from maximally chiral theories to maximally chiral theories, which are mapped across Aharony duality as in  \eqref{eq:mass_map_aharony_max_chiral}.
\item Real mass flows from maximally chiral SQCD with $F=2|k| + 1$ fundamentals to marginally chiral SQCD with $F=2|k|$ fundamentals. These flows were already discussed in \cite{Amariti:2014lla} and we review them here for completeness.
\item Real mass flows from marginally chiral SQCD to minimally chiral SQCD.
\end{itemize}

\subsection{Flows between maximally chiral theories}\label{app: max_chiral}
We start from Aharony-like duality between maximally chiral SQCD with only (anti-) fundamental fields:
\begin{equation}    \label{app:aharony_max_chiral}
U(N)_{k,k+N \ell} \with F \square \dualto U(\Nt)_{-k, -k + \ell \Nt} \with F \overline{\square}
,\qquad F> 2|k|
\end{equation}

The corresponding partition function identity, including background terms, is given by:
\begin{equation}	\label{app:aharony_max_chiral_Sb3}
Z_{U(N)_{k,k+\ell N}}^{[F,0]} \left( X, \cdot; \lambda \right)
=
e^{\frac{\pi i}{2} \phi}
e^{\frac{\pi i}{2} ( -2k \sum_{a=1}^{F} X_a^2 + 4 (F-N) \omega \lambda ) } \;
Z_{U(\Nt)_{-k,-k+\ell \Nt}}^{[0,F]} \left( \cdot,X-\omega; (k - \ell \Nt)\omega -  \lambda \right)
\end{equation}
where:
\begin{equation}
\begin{split}
Z_{U(N)_{k,k+\ell N}}^{[n_f,n_a]} \left( X, Y; \lambda \right)
=&
\frac{1}{N!} \int \prod_{i=1}^{N} \frac{du_i}{2\pi i }
e^{\pi i \left( -k \sum_{i=1}^{N} u_i^2 - \ell \left( \sum_{i=1}^{N} u_i \right)^2 + 2 \lambda \sum_{i=1}^{N} u_i \right)}
\\
&\times \frac{\prod_{i=1}^{N} \prod_{a=1}^{n_f} \Gamma(u_i - X_a) \prod_{b=1}^{n_a} \Gamma(-u_i + Y_b)}
	{\prod_{i<j}^N \Gamma(\pm(u_i-u_j)) }
\end{split}
\end{equation}
is the $\mathbf{S}_b^3$ partition function of $\NN=2$ SQCD with gauge group $U(N)$, CS levels $k,k+\ell N$, $n_f$ fundamentals, and $n_a$ antifundamentals. $\vec{X}$ and $\vec{Y}$ are real masses for the $S(U(n_f)\times U(n_a))$ flavor symmetries and $\lambda$ is the FI term and $\phi$ is a phase that only depends on the squashing parameter $\omega$:
\begin{equation}
\phi = 2 k \omega^2 (F-N)
\end{equation}
Here $\Gamma(x) \equiv \Gamma(x; \omega_1, \omega_2)$ is the Hyperbolic Gamma function \cite{rains2005, vandebult}.

Let us consider the real mass flows without Higgsing $\pm m$ for the electric theory as described in Section \ref{sec: mass deformations}. 
Here, we assume that the SQCD obtained after integrating out the massive fields is also maximally chiral.
These real mass deformations are implemented by the shifts in flavor fugacities:
\begin{equation}	\label{app:real_mass_pmm}
\left\{
\begin{array}{ll}
X_1 \to X_1 \mp (F-1) s &
\\
X_i \to X_i \pm s & \qquad 2\leq i \leq F
\\
\eta \to \eta + \left( \frac{F}{2} \pm( k + \ell N ) \right )s
\end{array}
\right.
\end{equation}
combined with the large shifts in gauge fugacities $u_i$ for the electric theory:
\begin{equation}	\label{app:real_mass_pmm_ele_shift}
u_i \to u_i \pm s
\end{equation}
We then take the $s\to \infty$ limit and use the asymptotic behavior of the hyperbolic gamma function:
\begin{equation}
\lim_{x\to \pm\infty} \zeta^{\text{sign}(x)}e^{-\left(\frac{\pi i}{2 \omega_1 \omega_2} \text{sign}(x) (x-\omega)^2 \right)}   \Gamma (x) = 1
,\qquad
\zeta \equiv \exp \left(\frac{\pi i\left(\omega_1^2+\omega_2^2\right)}{24 \omega_1 \omega_2}\right)
\end{equation}
Then, up to a highly oscillating phase that factorizes outside the integral, the l.h.s. of \eqref{app:aharony_max_chiral_Sb3} reduces to the partition function of $U(N)$ SQCD with CS levels $k \pm \tfrac{1}{2}, k \pm \tfrac{1}{2} + \ell N$ and $F-1$ fundamentals:

\begin{equation}
\left. 
Z_{U(N)_{k,k+\ell N}}^{[F,0]} \left( X, \cdot; \lambda \right)
\right|_{\eqref{app:real_mass_pmm},\eqref{app:real_mass_pmm_ele_shift}}
\to
e^{\frac{\pi i}{2} \phi_{ele}}
Z_{U(N)_{k\pm \tfrac{1}{2},k\pm \tfrac{1}{2}+\ell N}}^{[F-1,0]} \left( \{ X_a \}_{a=2}^{F}, \cdot; \lambda \right)
\end{equation}

On the magnetic side, following the analysis of the equation of motion \eqref{eq:eom_chiral_UN}, we want to explore a vacuum where both the scalars in the vector multiplet $\tilde{\sigma}$ and one component of the quarks $Q_1^1$ take a large VEV. The VEV for the real scalars can be implemented by shifting the gauge fugacities of the magnetic theory $\tilde{u}_i$:
\begin{equation}	\label{app:real_mass_pmm_mag_shift}
\begin{cases}
\tilde{u}_1 \to \tilde{u}_1 \mp (F-1)s
\\
\tilde{u}_i \to \tilde{u}_i \pm s \qquad i=2,\dots, \Nt
\end{cases}
\end{equation}
The VEV for the quark $Q_1^1$ breaks the $U(1)$ symmetry under which the quark is charged, which imposes the constraint $\tilde{u}_1 = X_1 - \omega$ in the partition function. 

Subject to this constraint, the Gamma function corresponding to the quark $Q_1^1$ reduces to $\Gamma(0)$ and diverges. We expect this VEV can be implemented in the partition function by substituting the Gamma function with $\delta(-\tilde{u}_1 + X_1 - \omega)$. 
More formally, we implement the VEV for the quark $Q_1^1$ by taking the limit:
\begin{equation}	\label{app:real_mass_pmm_higgs}
e^{-2\pi i s (-u_1 + X_1 - \omega)} \Gamma(-u_1 + X_1 - \omega) \to \delta(-u_1 + X_1 - \omega)
\end{equation}
and then use the $\delta$ to perform the integral over $\tilde{u}_1$. 
After taking the $s\to\infty$ limit the r.h.s. of \eqref{app:aharony_max_chiral_Sb3} reduces to the partition function of $U(\Nt - 1)$ SQCD with CS level $-k\mp\tfrac{1}{2},-k\mp\tfrac{1}{2} + \ell \Nt$ with $F-1$ antifundamentals, and a divergent prefactor. 
\begin{equation}
\begin{split}
\left.
e^{\frac{\pi i}{2} ( -2k \sum_{a=1}^{F} X_a^2 - 4 N \omega \lambda ) } \;
Z_{U(\Nt)_{-k,-k+\ell \Nt}}^{[0,F]} \left( \cdot,\omega - X; (k - \ell \Nt)\omega -  \lambda \right)
\right|_{\eqref{app:real_mass_pmm}, \eqref{app:real_mass_pmm_mag_shift}, \eqref{app:real_mass_pmm_higgs} }
\\
\to
e^{\frac{\pi i}{2}  \phi_{mag}}
e^{\frac{\pi i}{2} ( -2k \sum_{a=2}^{F} X_a^2 - 4 N \omega \lambda ) }
Z_{U(\Nt - 1)_{-k\mp\tfrac{1}{2},-k\mp\tfrac{1}{2}+\ell \Nt}}^{[0,F-1]} \left( \cdot,\omega - \{ X_a \}_{a=2}^{F}; (k - \ell (\Nt-1))\omega -  \lambda \right)
\end{split}
\end{equation}
We checked that the divergent prefactors cancel between the l.h.s. and the r.h.s., i.e.: $\phi_{ele}=\phi_{mag}$, and the resulting identity corresponds to Aharony duality for the resulting theories with $F-1$ charged fields.
Thus, this provides a check of the mapping of real masses \eqref{eq:mass_map_aharony_max_chiral} at the level of the $\mathbf{S}_b^3$ partition function.
In particular, one can obtain any Aharony-like duality for maximally chiral SQCD, namely with $F>2|k|$, with only (anti-) fundamentals by starting from the pseudo-self-duality \eqref{app:aharony_pseudo_self_dual} and performing a sequence of mass deformations.

\subsection{From maximally chiral SQCD to marginally chiral SQCD}\label{app: max_to_marg}
Here, we discuss the real mass flow from maximally chiral SQCD with only fundamentals to a marginally chiral SQCD theory.
This is a special case of a flow studied in \cite{Amariti:2014lla} and we report it here for completeness.

For simplicity, we consider the case $k>0$ and $\ell = 0$. The case $k<0$ is analogous, and the cases with $\ell \neq 0$ can be obtained by applying Witten's $SL(2,\zz)$ action \cite{witten2003sl2zactionthreedimensionalconformal} as discussed in \cite{Benvenuti:2025a}. 

We start from Aharony-like duality for maximally chiral SQCD \eqref{app:aharony_max_chiral} with $k= \tfrac{F}{2}-1$ and $\ell=0$. The corresponding partition function identity is a special case of \eqref{app:aharony_max_chiral_Sb3}. 
We turn on a positive real mass for one fundamental in the electric theory by considering the real masses:
\begin{equation}	\label{app:real_mass_n}
\left\{
\begin{array}{ll}
X_1 \to X_1 - (F-1) s &
\\
X_i \to X_i + s & \qquad 2\leq i \leq F
\\
\eta \to \eta + \left(F-1 \right )s
\end{array}
\right.
\end{equation}
combined with the large shift in electric gauge fugacities $u_i$:
\begin{equation}
u_i \to u_i + s
\end{equation}
The first fundamental takes a positive mass, and we integrate it out, leaving us with the marginally chiral $U(N)_{\tfrac{F-1}{2}}$ SQCD with $F'=F-1$ fundamentals.
On the magnetic side, we consider the following shift in gauge fugacities $\widetilde{u}_i$:
\begin{equation}
\begin{cases}
\tilde{u}_1 \to \tilde{u}_1 - (F-1)s
\\
\tilde{u}_i \to \tilde{u}_i + s \qquad i=2,\dots, \Nt
\end{cases}
\end{equation}
which breaks the gauge group to $U(\Nt-1) \times U(1)$. The massless component of the quarks form $F'=F-1$ antifundamentals of $U(\Nt-1)$ and one chiral with charge $-1$ under the $U(1)$ gauge group. Furthermore, due to the Polyakov mechanism \cite{Polyakov1977} observed during adjoint Higgsing, there is a linear monopole superpotential involving a monopole with GNO fluxes $-1/+1$ under the $U(\Nt-1)$ and $U(1)$ gauge groups, respectively. 
The $U(1)$ gauge group has CS level $\tfrac{1}{2}$, and, hence, is dual to a free chiral $t$ corresponding to its monopole with flux $+1$. Dualizing this gauge factor, we obtain the magnetic side of Aharony-like duality for marginally chiral SQCD:
\begin{equation}    \label{app:aharony_marg_chiral}
U(N)_{\frac{F'}{2}} \with F' \square \dualto U(\Nt)_{-\frac{F'}{2}} \with F' \overline{\square} \text{ and } \mathcal{W} = t \mathfrak{M}^{-}
\end{equation}
We checked that the divergent prefactors in the $\mathbf{S}_b^3$ partition function cancel between the electric and magnetic sides. The resulting identity, corresponding to the duality \eqref{app:aharony_marg_chiral}, reads:
\begin{equation}	\label{app:aharony_marg_chiral_Sb3}
\begin{split}
Z_{U(N)_{\tfrac{F'}{2}}}^{[F',0]} \left( X, \cdot; \lambda \right)
=&
e^{\frac{\pi i}{2} \phi}
e^{\frac{\pi i}{2} ( -F' \sum_{a=1}^{F} X_a^2 +(3F-2N) \omega \lambda  + \lambda^2) } \;
\Gamma\left(-\lambda + \omega \left( \frac{F'}{2} - N +1 \right)\right)
\\&
Z_{U(\Nt)_{-\frac{F'}{2}}}^{[0,F']} \left( \cdot,\omega - X; -\frac{F'}{2} \omega -  \lambda \right)
\end{split}
\end{equation}

Combining this real mass flow with the flows discussed in Appendix \ref{app:aharony_pseudo_self_dual} one can obtain Aharony-like duality for marginally chiral SQCD starting from the pseudo-self-dual case \eqref{app:aharony_pseudo_self_dual} and performing a series of mass deformations.

\subsection{From marginally chiral SQCD to minimally chiral SQCD}
We start from Aharony-like duality for marginally chiral SQCD with only (anti-) fundamental fields \eqref{app:aharony_marg_chiral}. Without loss of generality, we consider the case $k=\frac{F}{2}>0$ and $\ell = 0$.
The corresponding partition function identity is given by \eqref{app:aharony_marg_chiral_Sb3}.

On the electric side, we give a positive mass to $F'-F$ fundamentals by turning on the real masses:
\begin{equation}
\begin{cases}
X_a \to X_a + s \qquad a=1,\dots,F'
\\
X_b \to X_b - \frac{F'}{F-F'} s \qquad b=F'+1, \dots F
\\
\lambda \to \lambda + s F
\end{cases}
\end{equation}
combined with a large shift for the electric gauge fugacities $u_i$:
\begin{equation}
	u_i \to u_i +s
\end{equation}
The first $F'$ quarks remain massless while the last $F-F'$ acquire a positive mass, and are integrated out. 

On the magnetic side, we perform the same real mass combined with a large shift for the magnetic gauge fugacities $\widetilde{u}_i$:
\begin{equation}
\widetilde{u} \to \widetilde{u} + s
\end{equation}
The monopole flipper and the last $F-F'$ anti-fundamentals acquire a mass, and are integrated out. We checked that the divergent prefactors in the $\mathbf{S}_b^3$ partition function cancel between the electric and magnetic sides, and we obtain the identity:
\begin{equation}	\label{app:aharony_min_chiral_Sb3}
\begin{split}
Z_{U(N)_{k}}^{[F',0]} \left( X, \cdot; \lambda \right)
=
e^{\frac{\pi i}{2} \phi}
&
e^{\frac{\pi i}{2} 
	( -(k+\frac{F'}{2}) \sum_{a=1}^{F} X_a^2 
	+2F' \omega \lambda  + 2\lambda^2) } \;
Z_{U(\Nt)_{-k}}^{[0,F']} \left( \cdot,\omega - X; -\frac{F'}{2} \omega -  \lambda \right)
\\
& k = F - \frac{F'}{2} > \frac{F'}{2}
\end{split}
\end{equation}
where:
\begin{equation}
\Nt  =|k| + \frac{F'}{2} - N
\end{equation}
which corresponds to Aharony-like duality for minimally chiral SQCD with $F'$ fundamentals:
\begin{equation}    \label{app:aharony_min_chiral}
U(N)_{k} \with F' \square \dualto U(\Nt)_{-k} \with F' \overline{\square}
,\qquad F'< 2|k|
\end{equation}

Combining this real mass flow with the flows discussed in Appendices \ref{app: max_chiral} and \ref{app: max_to_marg}, one can obtain any Aharony-like duality for SQCD with only fundamentals starting from the pseudo-self-dual case \eqref{app:aharony_pseudo_self_dual} and performing a series of mass deformations.

\bibliographystyle{JHEP}
\bibliography{References}

\end{document}